\documentclass[aps,prd,superscriptaddress,onecolumn, nofootinbib]{revtex4-2}

\usepackage{blindtext}
\usepackage{cancel}
\usepackage{mathtools, mathalfa, amssymb, amsmath, amsfonts, amsthm, bm}
\usepackage{graphicx}
\usepackage{tikz}
\usepackage{multirow}
\usepackage[table, dvipsnames]{xcolor}
\usepackage{color}
\usepackage{framed}
\usepackage{aas_macros}

\usepackage{amsthm} 

\usepackage{appendix}
\usepackage{float}
\usepackage{subcaption}
\usepackage{xcolor}
\usepackage{orcidlink}
\usepackage{algorithm}
\usepackage{algpseudocode}
\usepackage{circledsteps}
\pgfkeys{/csteps/fill color=yellow}

\usepackage{hyperref}
\hypersetup{
    colorlinks = true,
    citecolor  = blue,
    linkcolor = blue,
    urlcolor  = blue,
    linkbordercolor = white
}

\graphicspath{{./figures/}}
\usepackage[bottom]{footmisc}
\usepackage{cleveref}

\begin{document}
    \title{Which Saddles Contribute? \\ The South-East Rule for Multidimensional Integrals}
    
    \author{In\^{e}s Aniceto 
    \orcidlink{0000-0002-5390-5175}} 
    \email{I.Aniceto@soton.ac.uk}
    \affiliation{Mathematical Sciences, University of Southampton, Highfield, Southampton, SO17 1BJ, United Kingdom}

    \author{Job Feldbrugge
    \orcidlink{0000-0003-2414-8707}}
    \email{Job.Feldbrugge@ed.ac.uk}
    \affiliation{Higgs Centre for Theoretical Physics, University of Edinburgh, James Clerk Maxwell Building, Edinburgh EH9 3FD, UK}

    \author{Christopher J. Howls
    \orcidlink{0000-0001-7989-7807}} 
	\email{C.J.Howls@soton.ac.uk}
	\affiliation{Mathematical Sciences, University of Southampton, Highfield, Southampton, SO17 1BJ, United Kingdom}

	\begin{abstract}
	\end{abstract}

    \begin{abstract}
        In this paper, we introduce and demonstrate a simple geometric algorithm to determine which critical points, both complex as well as real, contribute to the asymptotic evaluation of multiple integrals with exponential integrands of the form $e^{ikf(\bm{x})}$ over $\mathbb R^d$, for finite $d\ge 1$ and $f$ is analytic.  In so doing, the algorithm removes the need to compute the flows of $-\text{Re} (i\nabla f)$ in $\mathbb C^d$ that is required to identify such {\it relevant} critical points in Picard-Lefschetz approaches to the derivation of such asymptotic expansions. By contrast, our algorithm relies on the combination of three simple features: the values of $f$ at all the critical points plotted in the complex Borel plane, the concept of adjacency between such points derived from algebraic resurgence/hyperasymptotic approaches and the new result here of a geometric ``South-East" rule.  The algorithm incorporates functions $f$ that remain bounded or unbounded on $\mathbb R^d$. We illustrate this new approach with both pedagogical and advanced examples, and draw conclusions as to its importance for resolving issues associated with Wick rotations and its implications for path integrals. This is a significant step towards a systematic way of identifying instanton contributions in real-time path integrals.

    \end{abstract}
    
    \maketitle

    
    \section{Introduction}\label{sec:introduction}
    Oscillatory integrals of the form 
    \begin{align}
        \Psi = \left(\frac{k}{2\pi i}\right)^{d/2}\int_{\mathbb{R}^d} e^{i k f(\bm{x})}\mathrm{d}\bm{x}\,,\label{eq:integral}
    \end{align}
    where $f:\mathbb{R}^d\to \mathbb{R}$ is a real (locally) analytic function and $k$ is a positive number, play a central role in both mathematics and theoretical physics. They arise naturally in problems of wave propagation in acoustics and optics \cite{Schneider:1992, Feldbrugge:2023b, Feldbrugge:2025PhRvD.111f3061B, Feldbrugge:2020arXiv200801154F}, laser physics \cite{Balian:1978, salieres:2001, Pisanty:2020, Milosevic:2024}, as well as in the Feynman path integral formulation of real-time quantum mechanics and quantum field theory \cite{Feynman:1948,Feynman:1965}. 
    
    Their study lies at the intersection of analysis, catastrophe theory, and resurgence, revealing a remarkably rich mathematical structure. At the same time, these integrals are notoriously subtle. Their definition is often delicate, and their numerical evaluation is costly, as convergence is conditional and relies on intricate cancellations of rapidly oscillating contributions. 
    
    In the limit $|k| \to \infty$, oscillatory integrals are classically analysed using Laplace, stationary-phase, and steepest-descent methods. Their asymptotic behaviour is governed by the critical (or stationary) points of the phase function, while contributions from non-critical regions are suppressed by destructive interference. 
   
    Jones and Kline \cite{JonesKline:1958}, Bleistein and Handelsman \cite{Bleistein:1969}, and Ursell \cite{Ursell:1980} developed powerful methods for constructing asymptotic expansions associated with isolated or coalescing critical points.   In the Russian and singularity-theoretic literature, Fedoryuk’s saddle-point method, \cite{Fedoryuk:1977}, together with the work of Vasiliev \cite{Vasiliev:1995} and the Arnold–Varchenko– Gusein-Zade school \cite{ArnoldVG:1985, ArnoldVG:1988}, connected oscillatory integral asymptotics with complex geometry, vanishing cycles, monodromy, and the classification of critical-point singularities.  McClure and Wong \cite{Wong1989} and Benaissa and Rogers \cite{Benaissa:2001,Benaissa:2013} extended these ideas to oscillatory integrals possessing non-isolated critical manifolds\footnote{We focus here on integrals of the form \eqref{eq:integral} and will cover non-isolated critical points in a future publication.}.   
    
    However, much of this classical work was primarily local or canonical.  It described how to expand the contribution of a critical point, or a cluster of coalescing critical points, in a specified number of dimensions (usually 2 or 3), {\it once} the relevant contour geometry was understood.  What was not provided was a general practical criterion for determining which real {\it and/or} complex critical points contribute to a prescribed original integration cycle. Since asymptotic approximations may generally include contributions from both real and complex critical points, identifying the critical points that are genuinely {\it relevant} to the integral remains a fundamental challenge across a wide range of disciplines.

    In $d=1$, there is a simple geometric approach to identifying the contributing, or {\it relevant} saddles.  This requires the plotting the steepest descent contours through each critical point $x_n$ given by Im$[ik(f(x)-f(x_n))]=0$ and then identifying the chain of such contours that run continuously between the endpoints of the integral (in asymptotic valleys of Re$[ik(f(x)-f(x_n))]<0$ as $|x|\rightarrow +\infty$), \cite{deBruijn1958, Copson1965, Wong1989, Berry:1991, Bennett:2018}.

    For higher dimensions $d>1$ a Picard-Lefschetz approach is taken, whereby the real plane $\mathbb{R}^d$ (or more generally any initial integration domain) is deformed into the complex space $\mathbb{C}^d$ and onto a set of steepest descent manifolds (see for example \cite{Pham1965FormulesDP, Fedoryuk:1977, Howls:1997, Delabaere:2002, Feldbrugge:2017}).   

    Due to the single constraint Im$[ik(f(x)-f(x_n))]=0$, the deformed steepest descent manifold is of (real) co-dimension $2d-1$ in $\mathbb{C}^d$ and is so not unique.   That this is not a problem may be understood by converting the deformed integrals into representations in the single complex dimensional Borel plane:  effectively using $if(x)$ as the new (action) variable. Following Pham \cite{Pham1965FormulesDP}, Howls \cite{Howls:1997} demonstrated how this was possible. 

    The asymptotic expansions about such critical points as $|k|\rightarrow +\infty$ may then be derived, and in general, these will be divergent. Historically, the divergence of these expansions has been a contentious subject. While the full asymptotic series diverges, the initial convergent part often converges more quickly than many convergent series and has proven instrumental in many studies in mathematics and physics. Indeed, the Feynman diagram expansion in quantum electrodynamics is a famous example of an asymptotic series \cite{Dyson:1952}. This is a general property of the Feynman expansion in quantum field theory. 
    
    Over the past century, the pioneering work of Borel, Dingle, Écalle, and later Berry, Howls, and Olde Daalhuis has provided a rigorous framework for the analysis of asymptotic series, known as the theory of resurgence \cite{Dingle:1973, Ecalle:1981, Berry:1990, Ecalle:1993, Berry:1991,Daalhuis:1996, Howls:1997, Daalhuis:1998, Daalhuis:1999}. Their contributions reveal how asymptotic series can be resummed, yielding exponentially accurate approximations to the underlying integrals or, more generally, solutions to differential equations. In particular, exact remainder terms may be obtained, and better-than exponential accuracy obtained through hyperasymptotic methods \cite{Berry:1991, Howls:1992, Howls:1997, Delabaere:2002}.  

    The work of Howls and later Delabaere \& Howls on multidimensional integrals \cite{Howls:1997, Delabaere:2002} again assumed that all the {\it relevant} critical points had been identified {\it a priori}.  However, the procedure for identifying the relevant points in $d>1$ was far from trivial and was not discussed in those papers.  In principle, this would involve solving the downward flow partial differential equation, see, for example, Kaminski \cite{Kaminsk:i1994, Weber:2025, Shoji:2026}:
    \begin{align}
        \dot{\bm{x}}(\lambda)= -\nabla {\rm Re}[i k f(\bm{x})], \qquad \bm{x}(0)=\bm{x}_0\in \mathbb{R}^d\,,
    \end{align}
    to identify the manifolds that run between each initial point $\bm{x}_0 \in\mathbb{R}^d$ to each relevant critical point in $d$-complex dimensions.  Whilst feasible for $d=2$ \cite{Kaminsk:i1994, Weber:2025, Shoji:2026}, this rapidly become inconvenient or intractable as $d$ increases.  

    The goal of this paper is to present a simple approach to identifying the relevant critical points in arbitrary dimensions $d$. The method combines the geometry of the locations of the critical points in the Borel plane, with the algebraically derived {\it adjacency} of these critical points, arising from the Riemann sheet structure of the Borel plane.  The latter algebraic resolving of the Riemann sheet structure may be achieved through methods of Olde Daalhuis \cite{Daalhuis:1998b, Daalhuis:1998c, Daalhuis:1999}, Howls \cite{Howls:1997} or from a Borel-Padé approach, see \cite{Aniceto:2018uik,lustri2025borelpade}, as well as works of Costin and Dunne \cite{Costin:2019xql,costin2021conformal}. In \cite{Serone:2017} the authors addressed a similar problem in a quantum mechanical setting, where a quantum deformation of the original potential was used instead of the adjacency of critical points discussed here.

    Recent work by Assier, Shanin and collaborators \cite{Assier:2022, Shanin:2024} has developed powerful methods for determining relevant contributions in $\mathbb{C}^2$ integrals and related diffraction problems. These methods exploit detailed properties of the singularities in the integrand in two complex dimensions and associated integration surfaces to determine the relevant asymptotic contributions. Our paper instead seeks a dimension-independent criterion for determining the critical points contributing to the exponential asymptotic expansion of oscillatory integrals of the form \eqref{eq:integral}.
     
    Alternative approaches for multidimensional Mellin–Barnes integral representations using Newton polyhedra were developed by Kaminski and Paris \cite{KaminskiParis:1997a,KaminskiParis:1997b}. 

    In section \ref{sec:Crit}, we examine the structure of the integrals  \eqref{eq:integral} and formally define the concept of relevance of critical points.  In section \ref{sec:resurgence}, we explain how such integrals may be reduced to the single Borel plane.  In section \ref{sec:Adjacency}, we explain the concept of adjacency.  This leads us in section \ref{sec:Relevance} to derive a simple, practical, algebro-geometric algorithm to identify the relevance from adjacency, in arbitrary integer $d>1$, which we have called the``South-East rule".  We discover that the cases where $f$ is bounded from below or not need slightly different treatments.  We illustrate the use of this algorithm practically with several examples in section \ref{sec:Examples}, for integrals of dimension $d\ge1$.  Finally, in section \ref{sec:Discussion}, we conclude with a discussion addressing the extension of the South-East rule to path integrals and its potential to address issues arising from Wick rotation. 

    \section{The Relevance of Critical Points to the Saddle Point Approximation}\label{sec:Crit}

    We shall here study analytical integrals of type \eqref{eq:integral}, where the dominant contributions emerge from neighbourhoods of an (assumed) isolated critical point of the exponent $f$ located at finite $\bm{x}_j$, satisfying

    \begin{align}
        \nabla f(\bm{x}_j)=\bm{0}\,, \qquad \mathcal{H} f(\bm{x}_j)\ne 0.
    \end{align}
    We exclude consideration of functions $f$ which have critical points at infinite values of $f$.
    
    Given the critical points $\bm{x}_j$, the saddle-point approximation expands the exponent $f$ to second-order around the points $\bm{x}_j$ as
    \begin{align}
        f(\bm{x}) \approx f(\bm{x}_j) + \frac{1}{2}(\bm{x}-\bm{x}_j)^T\mathcal{H} f(\bm{x}_j)(\bm{x}-\bm{x}_j)\,.
    \end{align} 
    Substituting this approximation in \cref{eq:integral}, we approximate the integral as a Gaussian integral of the form
    \begin{align} 
            \left(\frac{k}{2\pi i}\right)^{d/2}\int e^{i k \left[f(\bm{x}_j) + \frac{1}{2} (\bm{x} - \bm{x}_j)^T\mathcal{H}f(\bm{x}_j) (\bm{x}-\bm{x}_j)\right]}\mathrm{d}\bm{x}
             = 
            \frac{e^{i k f(\bm{x}_j)}}{\sqrt{\det \mathcal{H} f(\bm{x}_j)}}\,.\label{eq:sadlptapprox}
    \end{align}
    The leading-order saddle-point approximation consists of a sum of these Gaussian contributions. While the real critical points give a good approximation of this integral for large $k$, the accuracy typically improves when including complex critical points, which may contribute additional damped but oscillatory terms (as well as higher-order corrections in each of the saddlepoint expansions). 
    
    Note that the saddle point approximation \eqref{eq:sadlptapprox} diverges for degenerate critical points, \textit{i.e.}, when $\det \mathcal{H}f(\bm{x}_j) = 0$, which is why we excluded consideration of such cases above.  (For a method to treat degenerate contributions in $d=1$, see \cite{Bennett:2018}.)
    
    Integral \eqref{eq:integral} is highly oscillatory, depends on delicate cancellations, and converges only conditionally. Under different regulation schemes, the integral typically evaluates to different numbers. In this paper, we define the conditionally convergent integral using analyticity, which can be efficiently implemented with smooth analytic regulators (for more details, see \cite{Feldbrugge:2023}). The choice to define the integral with analyticity directly leads to the Picard-Lefschetz representation of the integral and ultimately the application of resurgence theory.
   
    Picard-Lefschetz theory \cite{Pham1965FormulesDP, pham2011singularities} beautifully generalises the saddle point approximation to an exact non-oscillatory representation of integral \eqref{eq:integral}. Using a multidimensional version of Cauchy's integral theorem, we analytically continue the exponent $f$ into the complex plane $\mathbb{C}^d$ and deform the original integration domain into the complex plane $\mathbb{C}^d$ while avoiding singularities. Explicitly, we deform the original integration domain  $\mathbb{R}^d$ with the downward flow onto a set of steepest descent manifolds $\mathcal{J}_j \subset \mathbb{C}^d$ associated with the critical point $\bm{x}_j$ with respect to the real part of the exponent, 
    \begin{align}
        h(\bm{x}) = \text{Re}[i f(\bm{x})]\,,
    \end{align}    
    leading to the Picard-Lefschetz formula\footnote{
        For completeness, we explicitly define the downward flow, the steepest descent and ascent manifolds, and discuss the Picard-Lefschetz deformation.
        \begin{itemize}
            \item \textit{Downward flow:} We define the downward flow $\frac{\mathrm{d} \bm{u}^i(\lambda)}{\mathrm{d} \lambda} = -g^{ij} \frac{\partial h(\bm{u}(\lambda))}{\partial \bm{u}^j}$ with the real and imaginary parts $\bm{x}(\lambda)= \bm{u}^1(\lambda) + i \bm{u}^2(\lambda)$ of a point starting at $\bm{u}^1(0) + i \bm{u}^2(0) = \bm{x}_0 $. For convenience, we use the diagonal Riemannian metric $g_{ij}$ on the complex plane $\mathbb{C}^d$ given by $g=\begin{pmatrix}1 & 0 \\ 0 & 1 \end{pmatrix}$. 
            \item \textit{Steepest descent manifold:} The steepest descent manifold $\mathcal{J}_j$ consists of the points $\bm{x}_0$ for which the reverse flow reaches the critical point $\bm{x}_j$ in the limit $\lambda \to -\infty$.
            \item \textit{Steepest ascent manifold:} Mirroring the definition of the descent manifold, the steepest ascent manifold $\mathcal{K}_j$ consists of the point $\bm{x}_0$ for which the downward flow reaches the critical point $\bm{x}_j$ in the limit $\lambda \to \infty$.
            \item \textit{The Picard-Lefschetz deformation:} When acting on the original integration domain, the downward flow deforms the original integration domain onto a set of steepest descent manifolds in the limit $\lambda \to \infty$, \textit{i.e.}, $\mathbb{R}^d \simeq \sum_j n_j \mathcal{J}_j$ up to homology equivalence for some integers $n_j$. The coefficient $n_j$ counts the intersections of the original integration domain with the steepest ascent manifold, $n_j = \langle\mathbb{R}^d, \mathcal{J}_j\rangle$.
        \end{itemize}
    }
    \begin{align} \label{eq:PLdecom}
        \Psi = \sum_{j}  n_j\,\Psi^{(j)}(k)=\left(\frac{k}{2\pi i}\right)^{d/2}\sum_{j}  n_j \int_{\mathcal{J}_j} e^{i k f(\bm{x})}\mathrm{d}\bm{x}\,.
    \end{align}
    The sum ranges over both the real and complex critical points of exponent, and the number $n_j \in \mathbb{Z}$ weights the contribution of the steepest descent manifold $\mathcal{J}_j$ to the deformation.

    We are now able to define {\it relevant} critical points as those critical points for which $n_j\ne 0$.  If $n_j=0$, the critical point does not contribute to the deformation and hence to the eventual asymptotic expansion and is said to be {\it irrelevant} \cite{Howls:2004}. 
    
    Remarkably, the Picard-Lefschetz formula expresses the weight $n_j$ as the number of times the often overlooked steepest {\it ascent} manifold $\mathcal{K}_j$ intersects the original integration domain $\mathbb{R}^d$. This is sometimes formally written as $n_j = \langle \mathcal{K}_j, \mathbb{R}^d\rangle$. 
    
    It follows that the real critical points are always relevant, as they lie on the original integration domain. A complex critical point $\bm{x}_j$ can only be relevant when $h(\bm{x}_j) \leq 0$, as the $h$-function vanishes on the original integration domain and the steepest ascent $\mathcal{K}_j$ cannot intersect the original integration domain when $h(\bm{x}_j) > 0$. 
    
    By the multidimensional generalisation of Cauchy-Riemann equations, the imaginary part of the exponential $H(\bm{x}) = \text{Im}[i f(\bm{x})]$ is constant along both the steepest descent and ascent manifolds $\mathcal{J}_j$, $\mathcal{K}_j$. Consequently, the deformed integral does not oscillate and rapidly decays along the descent manifolds. Colloquially, the integrand determines the optimal integration cycle along which the integrand decays exponentially, and the integral converges absolutely. 
    
    While the saddle point approximation is valid in the limit of large $|k|$ and only applies to non-degenerate critical points (for which the Hessian $\mathcal{H}f$ is non-singular), the Picard-Lefschetz representation is exact for general $k$ and also applies to situations where the exponent $f$ includes degenerate critical points. 
    
    In the limit of large $|k|$ when the function $f$ has only non-degenerate critical points, the Picard-Lefschetz formula reduces to the leading-order saddle point approximation 
    \begin{align}
        \Psi \approx \sum_j  \frac{n_j\, e^{i k f(\bm{x}_j)}}{\sqrt{\det \mathcal{H} f(\bm{x}_j)}}\,.\label{eq:saddlePointApproximation}
    \end{align}
    The saddle point approximation inherits the intersection number $n_j$ from the Picard-Lefschetz formula. The saddle point approximation is known as the Eikonal approximation in optics and the Wentzel-Kramers-Brillouin (WKB) approximation in quantum physics.

    To see the Picard-Lefschetz formula in action, let us consider the Pearcey integral 
    \begin{align}
        \Psi(y,z) =\sqrt{\frac{k}{2\pi i}} \int_{-\infty}^\infty e^{i k\left[x^4 + y x^2+z  x\right]}\mathrm{d}x\,,\label{eq:Pearcey}
    \end{align}
    with the external parameters $y,\,z$ (where the notational translation between \eqref{eq:Pearcey} and equation 36.2.10 of the DLMF \cite{DLMF36E10} is given by $\Psi(y, z)=\frac{1}{\sqrt{2\pi i}}\Psi_2(z,y;k)$).
    
\begin{figure}
        \centering
        \begin{subfigure}[b]{0.49\textwidth}
            \resizebox{\textwidth}{!}{%
            \begin{tikzpicture}
                \node[anchor=south west, inner sep=0pt, outer sep=0pt] (image) at (0,0) {\includegraphics{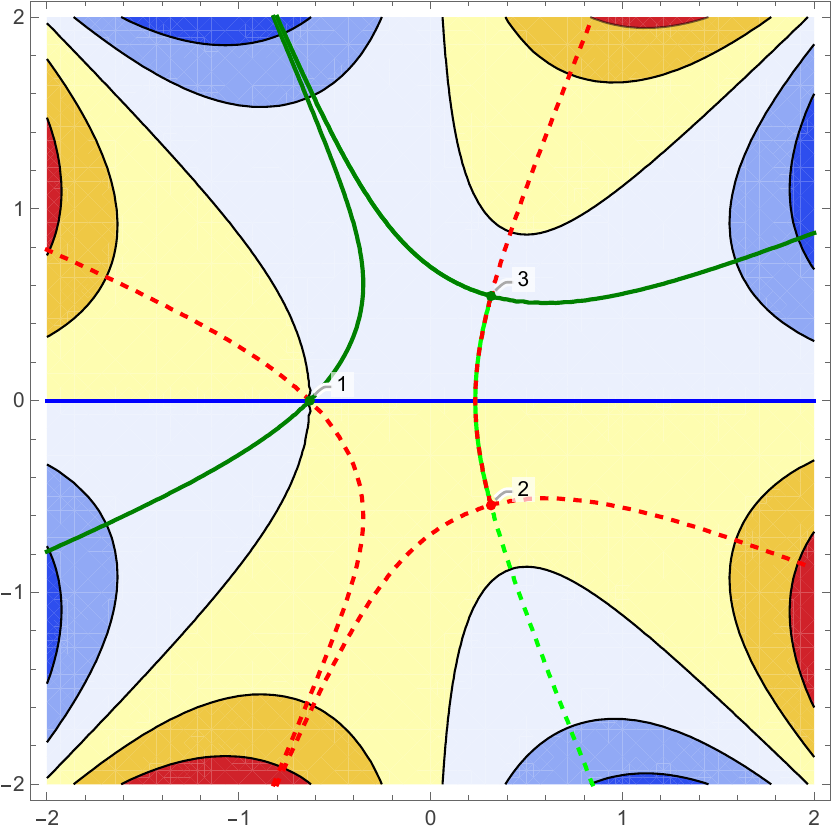}};
                \node[below right, font=\huge] at (image.south) {Re$[x]$};
                \node[above left, font=\huge, rotate = 90] at (image.west)  {Im$[x]$};
            \end{tikzpicture}
            }
            \caption{The complex $x$-plane}
        \end{subfigure}
        \begin{subfigure}[b]{0.49\textwidth}
            \includegraphics[width=\textwidth]{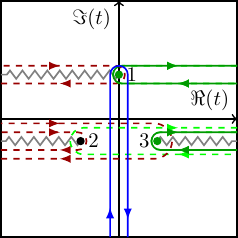}
            \caption{The complex $t$-plane}
        \end{subfigure}
        \caption{Pearcey integral for $y=0$ and $z=1$. \textit{Left:} The Picard-Lefschetz deformation (the dark green curve) of the original integration domain (the blue line) in the complex $x$-plane. The exponent has three critical points, of which points $1$ and $2$ are relevant (the green points) and $3$ is irrelevant (the red point). The descent contours are plotted in green, and the ascent contours are plotted in red. \textit{Right:} The associated branch point singularities and steepest ascent and descent contours in the Borel plane $t=-i f(\bm{x})$. The original integration domain maps to a contour on the imaginary axes wrapping around branch point $1$. The branch points associated with the relevant (irrelevant) critical points are plotted in green (black).}\label{fig:Pearcey}
    \end{figure}

    The integral has three critical points, solving $4x^3 + 2yx+z=0$. For $y=0$ and $z=1$, one critical point is real, and the remaining critical points form a complex conjugate pair (see the left panel of \cref{fig:Pearcey}). The original integral starts in the valley on the lower left at $x = \infty e^{-7i\pi/8}$ and terminates in the valley in the upper right at $x = \infty e^{i \pi/8}$. The real line $\mathbb{R}$ is deformed onto the steepest descent manifold of the real critical point $1$ and the complex critical point $3$ without crossing singularities. Observe that the steepest ascent manifold of critical point $3$ intersects the real line at a point. Critical point $2$ is irrelevant, as its steepest ascent manifold misses the real line. Along the steepest descent manifolds, the integrand decays exponentially away from the critical points.

    \section{Resurgence}\label{sec:resurgence}
    
    In this paper, we relate the intersection numbers of Picard-Lefschetz theory, governing the relevance of critical points, to the {\it adjacency} relations of resurgence theory. For consistency, we summarise the central parts of resurgence theory, focusing on the adjacency relations and introduce the notation used throughout. For a detailed exposition of resurgence in the context of integrals, we refer to \cite{Berry:1990,Berry:1991, Howls:1997}. A more general review of resurgence methods, including a transseries and Borel plane perspective, can be found in \cite{Aniceto:2018bis}.

    \subsection{The transseries}
    
    In the derivation of the saddle point method, we expand the exponent $f$ to second order around the critical points, reducing the oscillatory integral to a Gaussian one. When including higher-order terms,
    \begin{align}
        f(\bm{x}) = f(\bm{x}_j) + \frac{1}{2} (\bm{x}- \bm{x}_j)^T \mathcal{H}f(\bm{x}_j)(\bm{x}-\bm{x}_j) 
        + \sum_{3 \leq |\alpha| } \frac{(\bm{x}-\bm{x}_j)^\alpha}{\alpha!}  \partial^\alpha f(\bm{x}_j)\,,
    \end{align}
    using multi-index notation $\alpha\in \mathbb{Z}_{\geq 0}^d$, the saddle point approximation naturally expands as
    \begin{align}
        \Psi 
        &= \left(\frac{k}{2\pi i}\right)^{d/2}\sum_{j}  n_j \int_{\mathcal{J}_j} e^{i k f(\bm{x})}\mathrm{d}\bm{x}\\
        &= \left(\frac{k}{2\pi i}\right)^{d/2}\sum_{j}  n_j e^{ik f(\bm{x}_j)}\int_{\mathcal{J}_j} e^{  \frac{i k}{2} (\bm{x}- \bm{x}_j)^T \mathcal{H}f(\bm{x}_j)(\bm{x}-\bm{x}_j) 
        }\sum_{m=0}^\infty \frac{(i k)^m}{m!}\left[ \sum_{3 \leq |\alpha| } \frac{(\bm{x}-\bm{x}_j)^\alpha}{\alpha!}  \partial^\alpha f(\bm{x}_j)\right]^m\mathrm{d}\bm{x}\\
        &\sim  \sum_j \frac{n_j e^{i k f(\bm{x}_j)}}{{\sqrt{\det \mathcal{H} f(\bm{x}_j)}}} \sum_{m=0}^\infty \frac{U_m^{(j)}}{k^m}\\
        &=  \sum_j n_j e^{i k f(\bm{x}_j)} \sum_{m=0}^\infty \frac{T_m^{(j)}}{k^m}\label{eq:Useries}
        \,,
    \end{align}
    where the series coefficients $U_m^{(j)}$ (or equivalently the rescaled coefficients $T_m^{(j)} = {U_m^{(j)}/\sqrt{\det \mathcal{H} f(\bm{x}_j)}}$) are associated with critical point $\bm{x}_j$ \cite{Dingle:1973, Nemes:2013:GFSN, Feldbrugge:2026}. This series of asymptotic expansions over critical points $j$ is known as a {\it transseries}.  The contribution from each critical point $\bm{x}_j$ in the transseries is called a transseries sector, identified as $\Psi^{(j)}(k)$ in \eqref{eq:PLdecom}. To zeroth order in $k$, the transseries reduces to the saddle point approximation of the integral as $U_0^{(j)} = 1$.
    
    At first sight, the transseries seems like a straightforward generalisation of the saddle point approximation \eqref{eq:saddlePointApproximation}. However, the transseries is strictly formal as the infinite sum involving the $T_m^{(j)}$ for each $j$ diverge. Transseries and the divergence of asymptotic series arising from integrals are extensively studied in resurgence theory. See \cite{Aniceto:2018bis} and references therein.
    
    For each transseries sector $j$ in \eqref{eq:Useries}, the terms  $U_m^{(j)}$ of the corresponding $U-$series typically first decrease in magnitude before diverging factorially to infinity.  Truncation at suitable finite orders may, nevertheless, lead to exponentially accurate approximations.  The divergence of transseries sectors can in particular be controlled using the hyperasymptotic expansion, which in turn can lead to successfully increasing better-than-exponential accuracy \cite{Berry:1990, Berry:1991}.
    An alternative approach of controlling divergences and re-summing the transseries involves taking an integral (Borel) transform of each transseries sector and will be discussed below.
    As an example, we provide the explicit derivation of the terms in the transseries for the canonical diffraction integrals associated with seven elementary catastrophes in appendix \ref{ap:asymptotic}.  
    
    \subsection{Adjacency relations}
    
    As it turns out, the (assumed simple) critical points are intimately related through their asymptotic series. Resurgence theory relates the behaviour of the coefficients $T_m^{(j)}$ for large $m$ to the critical values $f(\bm{x}_j)$ and the nature of the critical points. Specifically, by Darboux's theorem (see, for example, \cite{Dingle:1973,GauntGutmann1974,Henrici1977}), the late terms of a power series encode the singularities of its analytic continuation, \textit{i.e.}, the \textit{resurgence relation} states (for the quadratic critical points here) that 
    \begin{align}
        T_N^{(j)} \sim \frac{1}{2 \pi i} \sum_{k} \sum_{m=0}^{\infty} K_{jk}\frac{(N-m-1)!}{F_{jk}^{N-m}}T_m^{(k)}=\sum_{k}\,\frac{K_{jk}}{2\pi i}\,\frac{(N-1)!}{(F_{jk})^{N}} \sum_{m=0}^{\infty} \frac{(N-m-1)!}{(N-1)!}\,(F_{jk})^{m}\,T_m^{(k)}.
        \label{eq:resurgence}
    \end{align}
    The $F_{jk}$ are known as the ``singulant'' \cite{Dingle:1973}, and represent the (complex) difference in heights between the critical points $j$ and $k$, and are defined by\footnote{If one keeps track of the difference in form between \eqref{eq:integral} and that of the integrals in \cite{Berry:1991, Howls:1997}, the formulae \eqref{eq:resurgence} and \eqref{eq:singulant} are consistent with those works.}:
    \begin{equation}\label{eq:singulant}
    F_{jk} = i (f(\bm{x}_j) - f(\bm{x}_k))
    \end{equation}
    and the Stokes constants $K_{jk} \in \mathbb{Z}$.  The second (formal) equality in \eqref{eq:resurgence} has been added to emphasize the clear factorial growth $(N-1)!$ of the large order terms, and its subleading exponential growth $(F_{jk})^{N}$. The Stokes constants encode whether two critical points $\bm{x}_j$ and $\bm{x}_k$ can influence other via their asymptotic series.  If that is the case, then they are said to be \textit{adjacent} to each other. Two critical points $\bm{x}_j$ and $\bm{x}_k$ are adjacent when $K_{jk} \neq 0$ and non-adjacent when $K_{jk}=0$. The sign encodes the relative orientations of the steepest descent manifolds. Building on seminal work by Dingle \cite{Dingle:1973}, Berry and Howls \cite{Berry:1990, Berry:1991,Howls:1997} used this resurgence relation to develop increasingly accurate approximations of the integral $\Psi$, known as the hyperasymptotic approximation. The hyperasymptotic expansion was further optimised and refined by Olde Daalhuis (see, for example, \cite{Daalhuis:1995, Daalhuis:1996, Daalhuis:1998, Daalhuis:1998b, Daalhuis:1998c, Daalhuis:1999}).

    Berry and Howls show that the adjacency can geometrically be identified by complexifying the parameter $k$ and studying the topology of the steepest descent manifolds while circling the origin in the complex plane  \cite{Berry:1991}. With the substituting $k \mapsto k\, e^{i\theta}$ (keeping $k$ positive),
    \begin{align}
        \Psi_\theta =\left(\frac{k}{2\pi i}\right)^{d/2} \int_{\mathbb{R}^d} e^{i k\, e^{i \theta} f(\bm{x})}\mathrm{d}\bm{x}\,,\label{eq:integral_general}
    \end{align}
    it is clear that the saddle points do not depend on the angle $\theta$. The steepest descent manifolds $\mathcal{J}_j(\theta)$ rotate when the angle $\theta$ increases from $0$ to $\pi$. Two critical points $\bm{x}_j$ and $\bm{x}_k$ are adjacent if and only if there exists an angle $\theta_s$ for which the steepest descent contour of one of the critical points includes the other saddle point, \textit{i.e.}, there exists an angle $\theta_s$ such that $\bm{x}_k \in \mathcal{J}_j(\theta_s)$. This is known as a Stokes phenomenon and marks a quantitative change in the topology of the steepest ascent and descent manifolds. Generally, the steepest descent manifold switches the valley in which it terminates. The set of relevant critical points may change at such a Stokes transition. 
    
    In the context of \eqref{eq:resurgence}, we see (e.g., \cite{Berry:1991}) that the asymptotic series of a critical point $\bm{x}_j$ is directly related to the asymptotic series of another critical point $\bm{x}_k$ if and only if there exists an angle $\theta_s$ for which the two critical points undergo a Stokes transition. 
    
    Conversely, starting from a critical point $\bm{x}_j$, the asymptotic series \eqref{eq:resurgence} may be used to detect the nearby adjacent critical points as well as their asymptotic series \cite{Daalhuis:1999} and hence to identify the adjacency relations of the associated critoval points.  
    
    The adjacency relations of a set of critical points are conveniently summarised in the adjacency graph. First, map the critical points $\bm{x}_j$ to their critical values $t_j = -i f(\bm{x}_j)$ and interpret them as the vertices of the adjacency graph. Next, connect the points $t_j$ and $t_k$ of adjacent critical points $\bm{x}_j$ and $\bm{x}_k$ with an edge and mark the edges with their Stokes constants $K_{jk}$.

    We illustrate these concepts by working out the adjacency relations of the Pearcey integral \eqref{eq:Pearcey} studied above.  For fixed values $(y,z)=(0,1)$, we rotate the phase of $k$ to find that the three critical points 1,2 and 3 are all {\it adjacent} to each other.  In \cref{fig:Pearcey_rotate} we observe that the steepest descent contours connect two saddles for the values of the phase $\theta=0$ (saddles 2 and 3), $\theta=\pi/3$ (saddles 1 and 3) and $\theta=2\pi/3$ (saddles 1 and 2). The adjacency graph this forms in the $t=-if({\bm x})$ plane is thus a triangle (see \cref{fig:Pearcey_rotate-adjacency}).  
    
    We can also read off the {\it relevant} saddles from \cref{fig:Pearcey_rotate}. A saddle will be relevant if there is a steepest ascent contour (the dashed red lines) linking the saddle to the original contour, or if the saddle itself lies within the original integration domain. Saddle 1 is relevant as it lies on the the real axis, the original integration domain.  For $0\le\theta\le\pi/3$ there is an ascent contour from saddle 3 to the real axis, making saddle 3 relevant. On the other hand the contour emanating from saddle 2 and crossing the original integration region is a steepest descent contour and so saddle 2 is {\it irrelevant}, and will not contribute to the asymptotic expansion of the integral. For $\pi/3<\theta<2\pi/3$ as shown in \cref{fig:Pearcey_rotate}, no ascent contours from saddles 2 and 3 cross the integration region, and thus these saddles now become irrelevant. The relevance of the saddles is also plotted in the adjacency graphs of \cref{fig:Pearcey_rotate-adjacency} as green dots. 
    
    \begin{figure}
        \centering
        \begin{subfigure}[b]{0.32\textwidth}
            \resizebox{\textwidth}{!}{%
            \begin{tikzpicture}
                \node[anchor=south west, inner sep=0pt, outer sep=0pt] (image) at (0,0) {\includegraphics{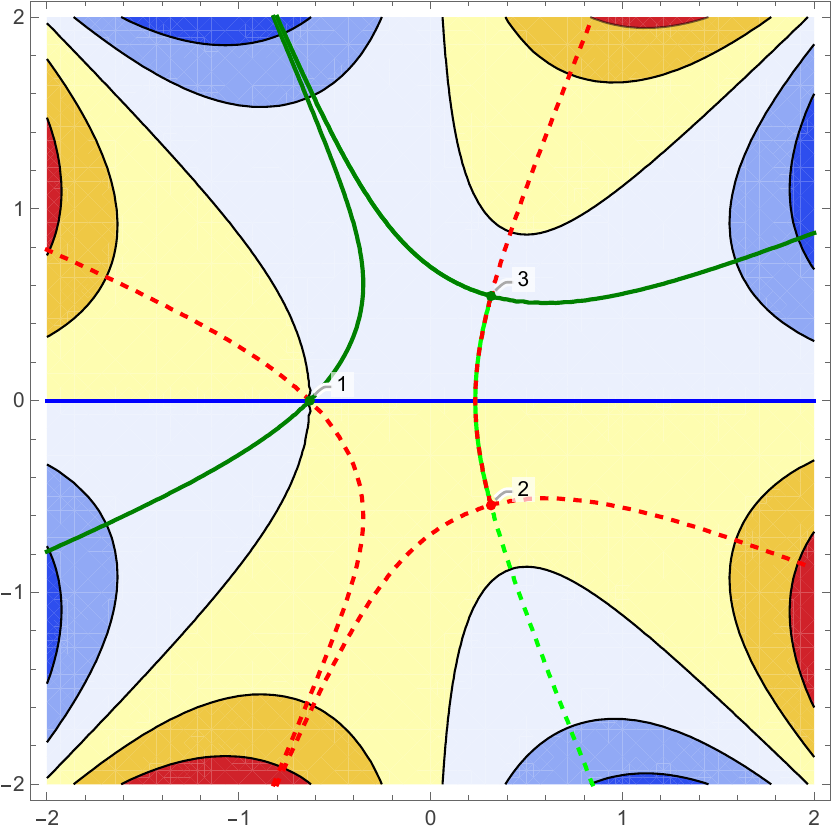}};
                \node[below right, font=\huge] at (image.south) {Re$[x]$};
                \node[above left, font=\huge, rotate = 90] at (image.west)  {Im$[x]$};
            \end{tikzpicture}
            }
            \caption{$\theta = 0$}
        \end{subfigure}
        \begin{subfigure}[b]{0.32\textwidth}
            \resizebox{\textwidth}{!}{%
            \begin{tikzpicture}
                \node[anchor=south west, inner sep=0pt, outer sep=0pt] (image) at (0,0) {\includegraphics{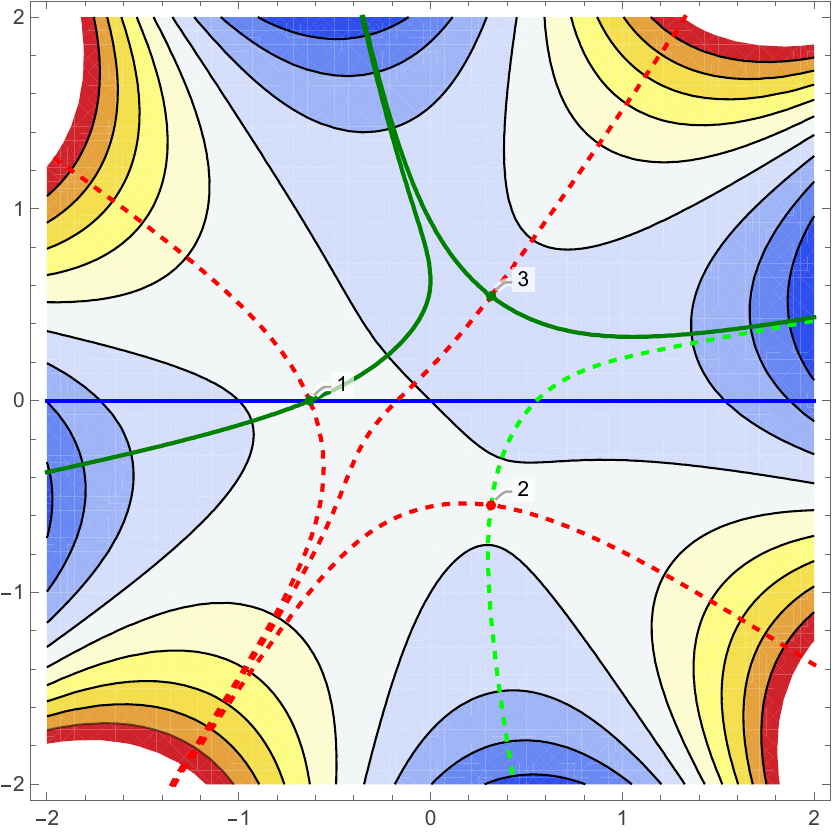}};
                \node[below right, font=\huge] at (image.south) {Re$[x]$};
                \node[above left, font=\huge, rotate = 90] at (image.west)  {Im$[x]$};
            \end{tikzpicture}
            }
            \caption{$\theta = \pi/4$}
        \end{subfigure}
        \begin{subfigure}[b]{0.32\textwidth}
            \resizebox{\textwidth}{!}{%
            \begin{tikzpicture}
                \node[anchor=south west, inner sep=0pt, outer sep=0pt] (image) at (0,0) {\includegraphics{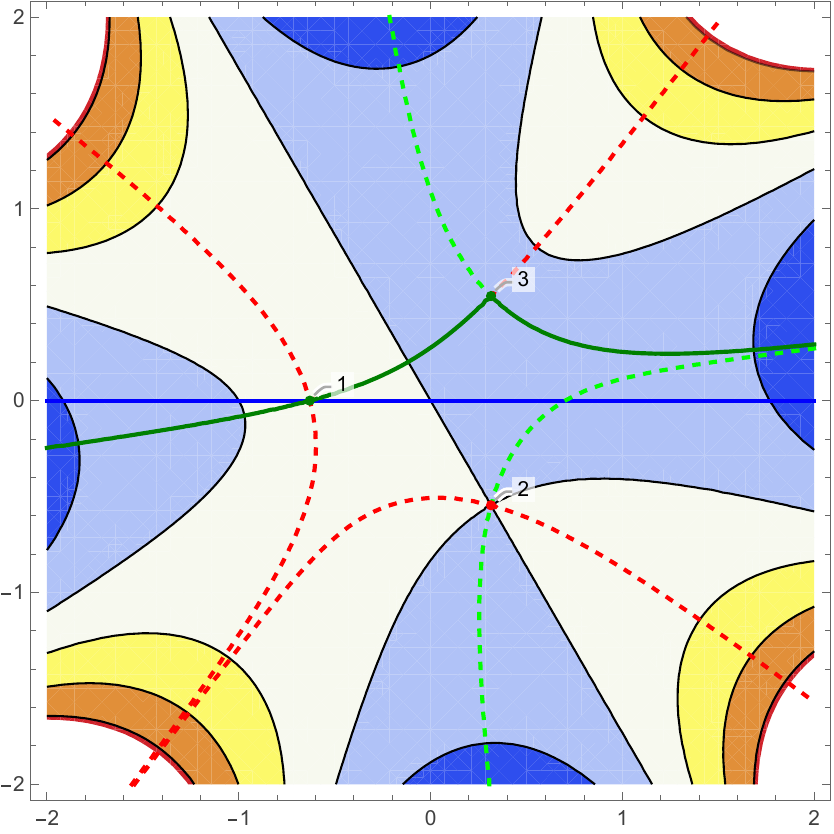}};
                \node[below right, font=\huge] at (image.south) {Re$[x]$};
                \node[above left, font=\huge, rotate = 90] at (image.west)  {Im$[x]$};
            \end{tikzpicture}
            }
            \caption{$\theta = \pi/3$}
        \end{subfigure}
        \begin{subfigure}[b]{0.32\textwidth}
            \resizebox{\textwidth}{!}{%
            \begin{tikzpicture}
                \node[anchor=south west, inner sep=0pt, outer sep=0pt] (image) at (0,0) {\includegraphics{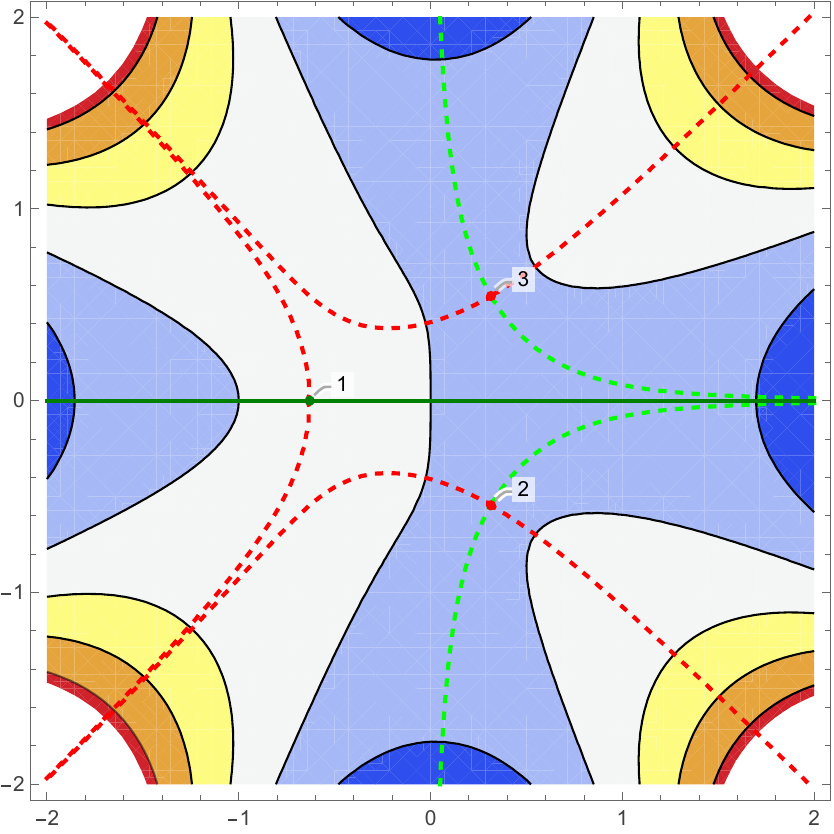}};
                \node[below right, font=\huge] at (image.south) {Re$[x]$};
                \node[above left, font=\huge, rotate = 90] at (image.west)  {Im$[x]$};
            \end{tikzpicture}
            }
            \caption{$\theta = \pi/2$}
        \end{subfigure}
         \begin{subfigure}[b]{0.32\textwidth}
            \resizebox{\textwidth}{!}{%
            \begin{tikzpicture}
                \node[anchor=south west, inner sep=0pt, outer sep=0pt] (image) at (0,0) {\includegraphics{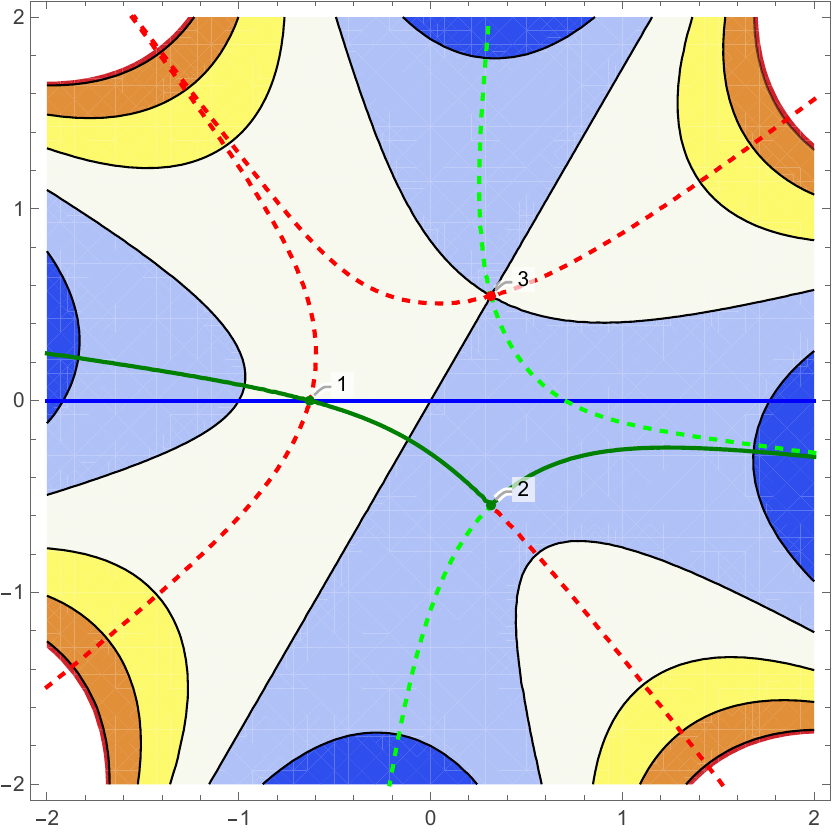}};
                \node[below right, font=\huge] at (image.south) {Re$[x]$};
                \node[above left, font=\huge, rotate = 90] at (image.west)  {Im$[x]$};
            \end{tikzpicture}
            }
            \caption{$\theta = 2\pi/3$}
        \end{subfigure}
        \caption{The analytic continuation of the Pearcey integral $\Psi_\theta(y=0,z=1)$. The Picard-Lefschetz deformation of $\Psi_{\theta}(0,1)$ for $\theta =0, \pi/4,\pi/3,\pi/2,2\pi/3$ from left to right, top to bottom following the notation outlined in \cref{fig:Pearcey}.  For each angle, the solid green contours are contours of steepest descent over the relevant saddles  which form a chain that connects the starting and ending valleys at infinity of the deformed original contour along the real axis of $x$. These contours are associated to saddles 1 and 3 for $0\le\theta\le\pi/3$ and to saddles 1 and 2 for the case $\theta=2\pi/3$. At $\theta=\pi/2$ a single steepest descent contributes, from saddle 1.  The dotted green contours are the paths of steepest descent over the saddles which do not form part of this chain and therefore do not contribute to the asymptotic expansion. These are saddle 2 up to $\theta=2\pi/3$ and saddle 3 from $\theta=\pi/3$. Stokes transitions at $\theta=0,\pi/2,2\pi/3$ show that all saddles are adjacent. The dotted red contours are paths of steepest ascent through the saddles.  In the region of interest $\theta\le\pi/2$, the steepest ascent paths through 1 and 3 intersect the original contour along the real axis, the steepest ascent path through 2 does not --  this shows the relevance of saddles 1 and 3, and irrelevance of saddle 2.      }\label{fig:Pearcey_rotate}
    \end{figure}
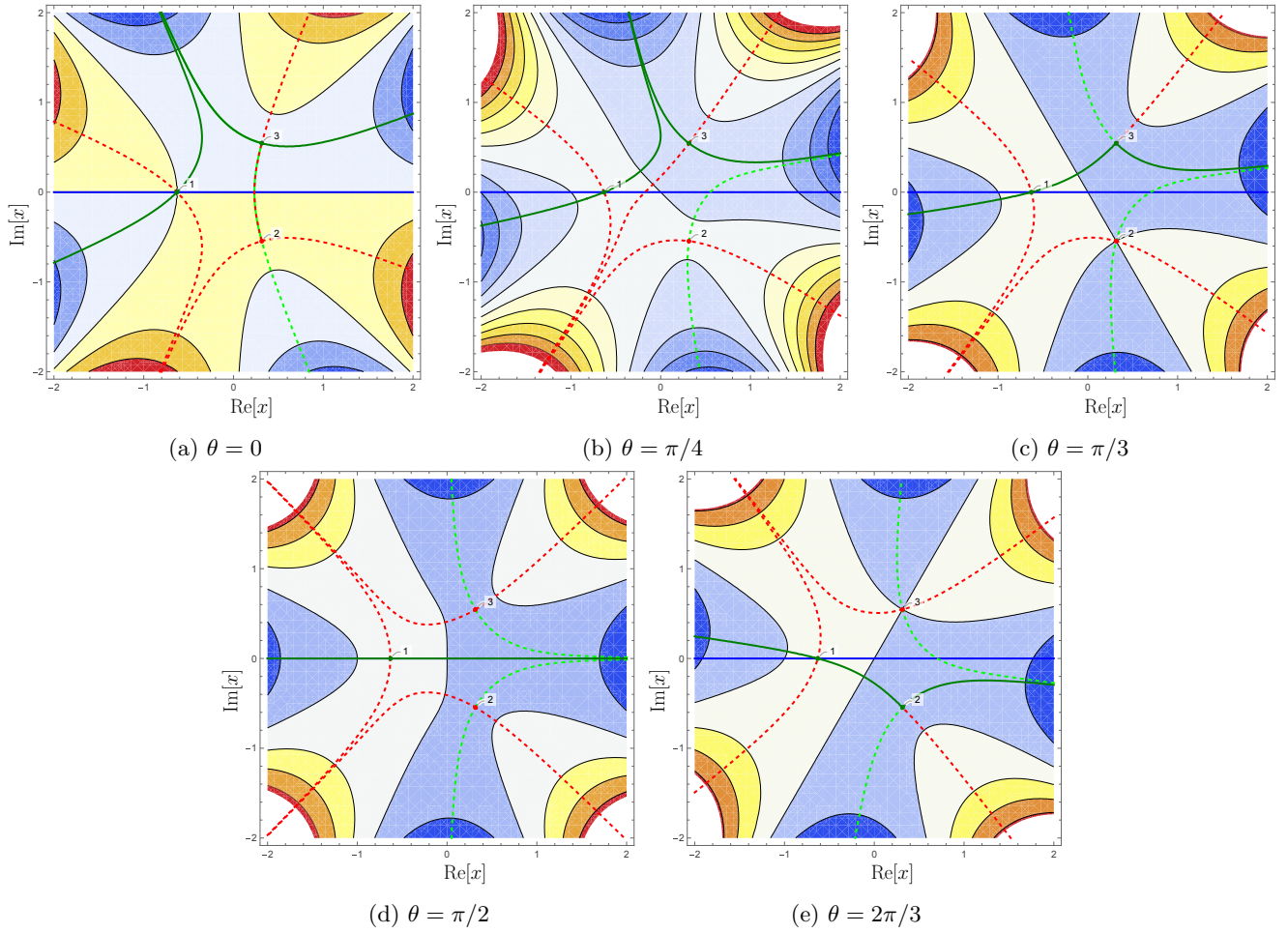
        
         \begin{figure}
        \centering
        \begin{subfigure}[b]{0.32\textwidth}
            \resizebox{\linewidth}{!}{%
                \begin{tikzpicture}[thick]
                    \draw[thick, black] (-2, -2) rectangle (2, 2);  
                    \draw[->] (-2, 0) -- (2, 0) node[above left=0.1] {$\Re(t)$};
                    \draw[->] (0, -2) -- (0, 2) node[below left=0.1] {$\Im(t)$};
                    \begin{scope}[rotate around={0:(0,0)}]
                        \draw[gray, dashed] (0, -2) -- (0, 2);
                        \draw[blue, ultra thick] (0, -2) -- (0, 0.75 );

                        \coordinate (A) at (0, 0.75);
                        \coordinate (B) at (-0.649519, - 0.375);
                        \coordinate (C) at (+0.649519, - 0.375);

                        \draw[red] (A) -- (B) -- (C) -- cycle;
                        \fill[ForestGreen](A) circle (2pt) node [right, black] (p1) {$1$}; 
                        \fill[black](B) circle (2pt) node [left, black] (p2) {$2$}; 
                        \fill[ForestGreen](C) circle (2pt) node [right, black] (p3) {$3$}; 
                    \end{scope}
                \end{tikzpicture}
            }
            \caption{$\theta = 0$}
        \end{subfigure}
        \begin{subfigure}[b]{0.32\textwidth}
            \resizebox{\linewidth}{!}{%
                \begin{tikzpicture}[thick]
                    \draw[thick, black] (-2, -2) rectangle (2, 2);  
                    \draw[->] (-2, 0) -- (2, 0) node[above left=0.1] {$\Re(t)$};
                    \draw[->] (0, -2) -- (0, 2) node[below left=0.1] {$\Im(t)$};
                    \begin{scope}[rotate around={45:(0,0)}]
                        \draw[gray, dashed] (0, -2.6) -- (0, 2.6);
                        \draw[blue, ultra thick] (0, -2.6) -- (0, 0.75 );             
                        
                        \coordinate (A) at (0, 0.75);
                        \coordinate (B) at (-0.649519, - 0.375);
                        \coordinate (C) at (+0.649519, - 0.375);

                        \draw[red] (A) -- (B) -- (C) -- cycle;
                        \fill[ForestGreen](A) circle (2pt) node [above, black] (p1) {$1$}; 
                        \fill[black](B) circle (2pt) node [left, black] (p2) {$2$}; 
                        \fill[ForestGreen](C) circle (2pt) node [right, black] (p3) {$3$}; 
                    \end{scope}
                \end{tikzpicture}
            }
            \caption{$\theta = \pi/4$}
        \end{subfigure}
        \begin{subfigure}[b]{0.32\textwidth}
            \resizebox{\linewidth}{!}{%
                \begin{tikzpicture}[thick]
                    \draw[thick, black] (-2, -2) rectangle (2, 2);  
                    \draw[->] (-2, 0) -- (2, 0) node[above left=0.1] {$\Re(t)$};
                    \draw[->] (0, -2) -- (0, 2) node[below left=0.1] {$\Im(t)$};
                    \begin{scope}[rotate around={90:(0,0)}]
                        \draw[gray, dashed] (0, -2) -- (0, 2);
                        \draw[blue, ultra thick] (0, -2) -- (0, 0.75 );     

                        \coordinate (A) at (0, 0.75);
                        \coordinate (B) at (-0.649519, - 0.375);
                        \coordinate (C) at (+0.649519, - 0.375);

                        \draw[red] (A) -- (B) -- (C) -- cycle;
                        \fill[ForestGreen](A) circle (2pt) node [above, black] (p1) {$1$}; 
                        \fill[black](B) circle (2pt) node [below, black] (p2) {$2$}; 
                        \fill[black](C) circle (2pt) node [above, black] (p3) {$3$}; 
                    \end{scope}
                \end{tikzpicture}
            }
            \caption{$\theta = \pi/2$}
        \end{subfigure}
        \caption{The analytic continuation of the Pearcey integral $\Psi_\theta(y=0,z=1)$.   The adjacency graph in the Borel plane associated to the Picard-Lefshetz plots of \cref{fig:Pearcey_rotate} for $\theta=0,\pi/4,\pi/2$.  The blue line is the Borel-plane image of the contour through saddle point 1, collapsed around the Borel branch cut.  The red lines denote saddles which are adjacent, i.e., lie on the same Riemann sheet. The green dots show which saddles are relevant to the evlauation of the integral, at each value of $\theta$.
        }\label{fig:Pearcey_rotate-adjacency}
    \end{figure}
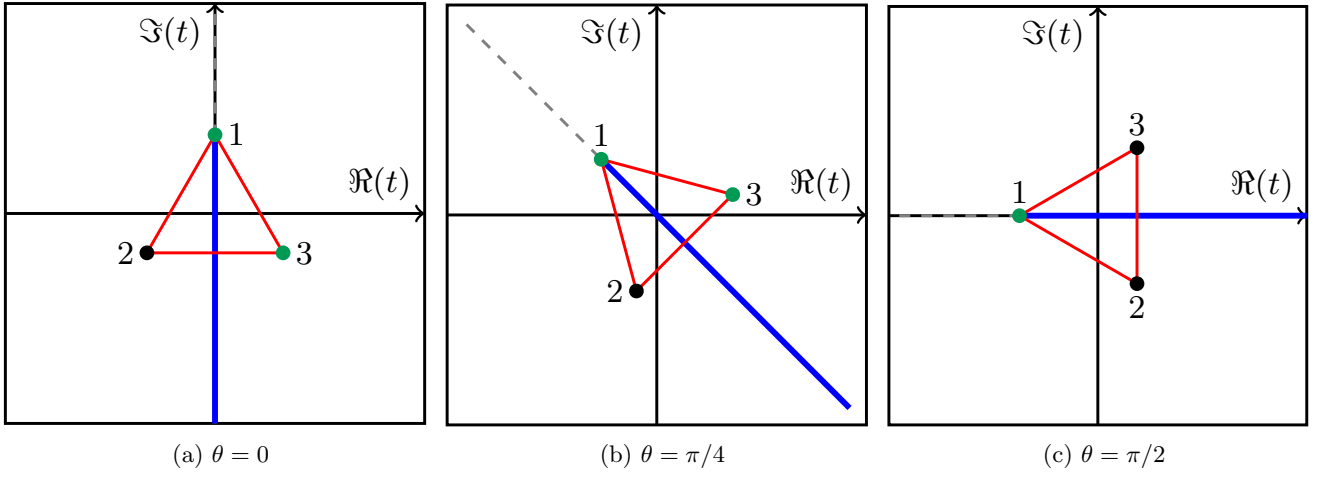

 \subsection{Transformation to the Borel Plane}\label{sec:Borel}

    The discussion above provides an algebraic interpretation of the adjacency of critical points. However, crucially, the adjacency relations and the adjacency graph also have a geometric interpretation. Let us, for simplicity, consider the one-dimensional integral 
    \begin{align}
        \Psi^{(j)} 
        = \sqrt{\frac{k}{2\pi i}} e^{i k f(x_j)}\int_{\mathcal{J}_j} e^{i k (f(x)-f(x_j))}\mathrm{d}x
        \,.
    \end{align}
    Upon changing coordinates to $t = -i(f(x) - f(x_j))$, the integral transforms in to a Laplace-type integral
    \begin{align}
        \Psi^{(j)}  
        &= \sqrt{\frac{k}{2\pi i}}  e^{i k f(x_j)}\int_{C} e^{-k t} \phi^{(j)}(t) \mathrm{d}t\,,
    \end{align}
    with the Jacobian $\phi^{(j)}(t)=\left[\frac{\mathrm{d}t}{\mathrm{d}x}\right]^{-1}$. The Jacobian is a multi-valued function with branch points at $\mathrm{d}t/\mathrm{d}x = 0$. Indeed, the non-degenerate critical points map to a square root branch point singularity. The contour $C$ starts at $t = \infty$, loops around the branch point singularity at $t=0$ and moves back to $t=\infty$ on the second Riemann sheet. Writing the difference of the Jacobian on the two Riemann sheets as $\Delta \phi^{(j)}$, which also includes the overall factor $\sqrt{\frac{k}{2\pi i}}$ above, we obtain the Laplace representation
        \begin{align}
        \Psi^{(j)}  
        &= e^{i k f(\bm{x}_j)}\,\left(T_0^{(j)}+\int_{0}^\infty e^{-k t} \Delta \phi^{(j)}(t) \mathrm{d}t\right)\,=\,e^{i k f(\bm{x}_j)} \,\sum_{m=0}^\infty \frac{T_m^{(j)}}{k^m}\,.\label{eq:borelresummation}
    \end{align}
     Importantly, the expansion of $\Psi^{(j)}$ above is unique and enables us to relate $\Delta \phi^{(j)}$ to the coefficients $T_m^{(j)}$ of the transseries,
    \begin{align}\label{eq:Borel-sec-j}
        \Delta \phi^{(j)}(t) = \sum_{m=0}^\infty \frac{T_{m+1}^{(j)}}{m!} t^{m}\,.
    \end{align}
    The function $\Delta \phi^{(j)}(t)$ obtained from the Jacobian $\phi$ is associated with the Borel transform of the asymptotic series \cite{sauzin2014intro}. Given an asymptotic series 
    \begin{equation}
    g(k) = \sum_{m=0}^\infty \frac{a_m}{k^{m+\beta}}\,,
    \end{equation}
    for some parameter $\beta\ge 1/2$, the Borel transform is defined as\footnote{The restriction on $\beta$ concerns invertiblility of the transform. If the first few terms of the series have lower powers, they can be treated separately and added at the end. This is the case in \eqref{eq:borelresummation}, for the coefficient $T_0^{(j)}$.} 
    \begin{equation}\label{eq:Borel-gen-def}
    \mathcal{B}[g](t)=\sum_{m=0}^\infty \frac{a_m}{\Gamma(m+\beta)}\, t^{m+\beta-1}\,.
    \end{equation}
    The above transform corresponds to taking an inverse Laplace transform termwise in $g(k)$. When the original asymptotic series $g(k)$ diverges factorially, the Borel transform converges. Then the Borel resummation method applied to $g(k)$ is defined as the Laplace transform of the Borel transform, \textit{i.e.},
    \begin{equation}
    \mathcal{S}g(k)=\int_0^\infty e^{-k t} \mathcal{B}[g](t)\mathrm{d}t\,.
    \end{equation}
    The Laplace integral in \cref{eq:borelresummation} Borel resums the asymptotic series $\sum_{m=0}^\infty T_m^{(j)}/k^m$ in \eqref{eq:borelresummation}. Consequently, the complex $t$-plane is often known as the Borel plane. 

    \bigskip

    When shifting coordinates to $t=-i f(x)$, we can write the original integral as 
    \begin{align}
        \Psi 
        &= \sqrt{\frac{k}{2\pi i}} \int_{\mathbb{R}} e^{i k f(x)}\mathrm{d}x\\
        &= \sqrt{\frac{k}{2\pi i}} \int_{\mathbb{R}} e^{- k t} \phi(t)\mathrm{d}x\,, \label{eq:jacobian-without-sqrt}
    \end{align}
    with the Jacobian 
    \begin{align}
        \phi(t) = \left[\frac{\mathrm{d}t}{\mathrm{d}x}\right]^{-1} = -i \left[\frac{\mathrm{d} f(x)}{\mathrm{d}x}\right]^{-1}\,.    
    \end{align}
    The Jacobian $\phi$ expressed as a function of $t$ is generally multivalued. The critical points $x_j$ map onto branch point singularities at $t_j = -i f(x_j)$ in the Borel plane. When the critical points are non-degenerate, the branch point is of the square root type. Crucially, the steepest ascent and descent contours map onto horizontal lines in the Borel plane, as they preserve the imaginary part of the exponential. In the Borel plane, the intersection of the ascent contours $\mathcal{K}_j$ with the original integration domain reduces to intersecting the map of the ascent contours $\mathcal{B}\mathcal{J}_j = -i f(\mathcal{J}_j)$ forming horizontal lines with the map of the original integration domain $D = -i f(\mathbb{R})$ running along the imaginary axes on the Riemann sheets of $\phi$. The evaluation of the intersection number is thus reduced to analysing the Riemann sheet structure of the Jacobian $\phi$. From this picture, it is clear that the relevant critical points have branch point singularities lying to the right of the domain $D$ (a necessary but not sufficient condition). The critical points with branch points outside this region are irrelevant. For an illustration, see the right panel of \cref{fig:Pearcey}. Remarkably, the geometry of the Riemann sheet structure is completely captured by the adjacency relations of the critical points. Two critical points $x_j$ and $x_k$ are adjacent if and only if there exists an angle $\theta_s$ such that $x_k \in \mathcal{J}_j(\theta_s)$. However, as the descent contour maps to straight lines in the Borel plane, it follows that $x_j$ and $x_k$ are adjacent when the associated branch points $t_j$ and $t_k$ see each other on the Riemann sheets of $\phi$. This is traditionally sometimes known as the radar method \cite{Voros:1983}. When two saddle points are not adjacent, the view of the associated branch points is obstructed by a third branch point.

    \bigskip

    For the Pearcey integral \eqref{eq:Pearcey} with $y=0$ and $z=1$, the Borel plane includes three branch points singularities associated with the critical points at $t_j = -i f(x_j)$. The original integration domain is mapped to a contour along the imaginary axes running around branch point $t_1$ (see the right panel of \cref{fig:Pearcey}). Note that the branch point $t_3$ is indeed to the right of the domain $D$. The Picard-Lefschetz deformation of the integral includes contours around branch points $t_1$ and $t_3$ in the complex $t$-plane. Both the steepest ascent and steepest descent manifolds are mapped to horizontal lines on the multivalued Riemann sheet looping around their associated branch point. For this particular example, the three critical points and associated branch points are adjacent to each other. The adjacency graph forms an equilateral triangle. When we rotate $k$ into the complex plane, or equivalently keeping $k$ real and letting $f(\bm{x}) \mapsto e^{i \theta}f(\bm{x})$, both the domain $D$ and the branch points $t_j$ rotate around the origin $t =0$ (see \cref{fig:Pearcey_rotate-adjacency}). During this rotation, critical point $3$ becomes irrelevant in a Stokes transition at $\theta = \pi/3$. At this angle, the edge between the branch points $t_1$ and $t_3$ is horizontal.

    \bigskip

    Remarkably, this geometric interpretation of the adjacency relations is not restricted to one-dimensional integrals \cite{Howls:1997}. Changing coordinates to $t = -i  f(\bm{x})$ in our original integral \eqref{eq:integral}, leads to a Laplace-type integral
    \begin{align}
        \Psi =\int_D  e^{-k t} \Delta \phi(t) \mathrm{d}t\,,
    \end{align}
    with 
    \begin{align}
        \Delta\phi(t) 
        &= \int_{\mathbb{R}^d} \delta(t + i f(\bm{x}))\mathrm{d}\bm{x}\\
        &= \int_{ f^{-1}(it) \subset \mathbb{R}^d} \frac{\mathrm{d}\bm{x}}{\|\nabla i f(\bm{x})\|}\,.
    \end{align}
    Note that the overall factor $\left(\frac{k}{2\pi i }\right)^{d/2}$ appearing in \eqref{eq:integral} was incorporated into $\Delta\phi (t)$. The integral over $t$ ranges over the image of $-i f$, \textit{i.e.},
    \begin{align}
        D= -i \{f(\bm{x})\, | \, \bm{x} \in \mathbb{R}^d\}\,.
    \end{align}
    When applied to the Picard-Lefschetz formula, the integral is written as the sum of a set of Laplace-type integrals
    \begin{align}
        \Psi = \sum_j n_j e^{i k f(\bm{x}_j)} \int_0^\infty e^{-k t}\Delta \phi^j(t)\mathrm{d}t\,,
    \end{align}
    where
    \begin{align}
        \Delta\phi^j(t) =  \int_{ f^{-1}(it) \subset \mathcal{J}_j} \frac{\mathrm{d}\bm{x}}{\|\nabla i f(\bm{x})\|}\,.
    \end{align} 
    Two critical points $\bm{x}_j$ and $\bm{x}_k$ are again adjacent when the associated branch points $t_j$ and $t_k$ see each other on the Riemann surface.
    An alternative treatment of this transformation from the original multidimensional integral in $d$-complex dimensions to the $d=1$ Borel plane in terms of complex forms may be found in \cite{Howls:1997}.

 \section{Identification of the Adjacency of Critical Points}\label{sec:Adjacency}

 We have written the solution of our integral in the form of a transseries as a sum over (potentially) all critical points, each contributing as a transseries sector. We now turn to the problem of how to systematically determine when these critical points are adjacent to each other.

The two main approaches to determining adjacency from the asymptotic behaviour of transseries sectors are hyperasymptotics \cite{Daalhuis:1999} and a Pad\'e analysis of the Borel plane  \cite{Aniceto:2018bis}. The former approach comes with rigorous error bounds in the calculation of the Stokes constants associated with adjacency.  When working with integrals, we are concerned more with a binary identification as to whether the images of the critical points in the Borel plane are visible to each other or not.  Hence, here we may focus on using the latter approach.  As described below, this approach has been used successfully in applications to gauge theories \cite{Aniceto:2018uik} and lattice models \cite{lustri2025borelpade}.

Given a transseries sector \eqref{eq:borelresummation}, the coefficients of the corresponding asymptotic series grow factorially, with a subleading exponential growth, and for the class of multidimensional integrals we are studying, this growth will be dictated by the resurgence relations \eqref{eq:resurgence}. We are interested in determining which critical points are adjacent to a particular saddle $\bm{x}_j$ -- this is equivalent to determining the non-zero Stokes constants $K_{jk}$. 
For each transseries sector associated to the critical point $\bm{x}_j$, the Borel transform is given by \eqref{eq:Borel-sec-j}\footnote{There is a slight abuse of notation here: the Borel transform is defined for asymptotic series \textit{without} the exponential part, while the $\Psi^{(j)}$ have an exponential factor. The Borel transform should be understood as acting on the asymptotic series of $\Psi^{(j)}$ alone.},
\begin{equation}
    \mathcal{B}[\Psi^{(j)}](t)=\Delta\,\phi^{(j)}(t)\,.
\end{equation}
 The Borel transform effectively removes the factorial growth of the coefficients, leaving a Taylor series centred at $t=0$ with a finite radius of convergence, representing an analytic function with branch cuts starting at the critical points $t_{jk}=-i(f(\bm{x}_k)-f(\bm{x}_j))$. In the complex Borel $t$-plane the resurgence relations \eqref{eq:resurgence} are encoded in these branch cuts, and the Stokes constants $K_{jk}$ can be obtained from the local behaviour at the branch point $t_{jk}$. 

Importantly, by knowing the exact factorial growth of the large order coefficients, we can control the type of branch cuts appearing on the Borel $t-$plane. As seen in \eqref{eq:resurgence}, the coefficients of the asymptotic series grow as $T_N^{(j)}\sim \Gamma(N)$. The Borel transform \eqref{eq:Borel-sec-j} removes this factorial growth \textit{exactly}, in the sense that each coefficient $T_m^{(j)}$ is divided by $\Gamma(m)$, leading to logarithmic branch cuts in the Borel plane. A simple multiplication of the original asymptotic series \eqref{eq:borelresummation} by an overall factor $k^{-\alpha}$ with $\alpha \ge 1/2$ will change the nature of the Borel singularities (see section 4 of \cite{Aniceto:2018bis})\footnote{A multiplication of two functions in the physical $k-$variable leads to the convolution of their respective Borel transforms in the $t-$variable. This is why the Borel plane is also known as the convolutive plane.}. In particular for $\alpha=1/2$, then using the definition of Borel transform \eqref{eq:Borel-gen-def} we obtain
\begin{equation}\label{eq:borel-sqrts}
    \mathcal{B}\left[k^{-1/2}\Psi^{(j)}\right](t)=\frac{1}{t^{1/2}}\sum_{m=0}^\infty \frac{T_{m}^{(j)}}{\Gamma(m+1/2)} t^{m}\,.
\end{equation}
One can go between the two Borel transforms above through the actions of a half derivative \cite{Aniceto:2018bis}, which transforms logarithmic branch cuts into square root branch cuts -- in both cases the adjacency of critical points can be obtained from the local behaviour at the respective branch points\footnote{The expectation of square root branch cuts when one keeps a $\sqrt{k}$ factor out was already discussed for the Jacobian in \eqref{eq:jacobian-without-sqrt}}.

To retrieve the adjacency information, we first note that the resurgence large-order relations \eqref{eq:resurgence} can be naturally rewritten as a consequence of Darboux's theorem in the Borel plane \cite{Dingle:1973}. In the case of the Borel transform \eqref{eq:borel-sqrts}, the relation \eqref{eq:resurgence} between the large order coefficients of the regular expansion at $t=0$ and the coefficients of the singular expansions around nearest singularities can be rewritten as: 
\begin{equation}\label{eq:analytic-at-sing}
    \left.\mathcal{B}\left[k^{-1/2}\,\Psi^{(j)}\right](t)\right|_{t=F_{jk}}=\frac{K_{jk}}{2}\,\mathcal{B}\left[k^{-1/2}\,\Psi^{(k)}\right](t-F_{jk})+\mathrm{regular}\,.
\end{equation}
The above relation should be seen as the analytic continuation of the Borel transform associated to critical point $\bm{x}_j$, $\mathcal{B}\left[k^{-1/2}\,\Psi^{(j)}\right](t)$, evaluated at the singular branch point $t=F_{jk}$, where $F_{jk}$ is the singulant \eqref{eq:singulant}, is determined by the regular series of the Borel transform associated to critical point $\bm{x}_k$, weighted by half the Stokes constant $K_{jk}$. 

At this point, the Borel transform \eqref{eq:borel-sqrts} has a square root branch point at $t=0$, and consequently so does the behaviour of the r.h.s. of \eqref{eq:analytic-at-sing} at $t=F_{jk}$. To have a regular expansion at $t=0$ we will multiply the Borel transform $\mathcal{B}\left[k^{-1/2}\,\Psi^{(j)}\right](t)$ by $t^{1/2}$ before analysing the square root behaviour of the function at $t=F_{jk}$:
\begin{equation}
    \Phi^{(j)}(t)\equiv t^{1/2}\,\mathcal{B}\left[k^{-1/2}\,\Psi^{(j)}\right](t)\,.
\end{equation}
We can further simplify this square root behaviour by applying a conformal map \cite{jentschura2001improved,caliceti2007useful,costin2021conformal}, allowing us to transform the original square root branch point singularity into a simple pole and to determine the Stokes constant via a residue calculation. We perform the transformation
\begin{equation}\label{eq:conf-map}
    t=F_{jk}+e^{i\pi}\,F_{jk}(s-1)^2,
\end{equation}
noting that the branch point $t=F_{jk}$ gets mapped to the singular point $s=1$. The leading behaviour at $s=1$ is now a simple pole:
\begin{equation}
    \left.\Phi^{(j)}(s)\right|_{s=1}=K_{jk}\,\frac{T^{(k)}_0}{2i\sqrt{\pi}}\frac{1}{(s-1)}+\mathrm{regular}\,.
\end{equation}

Clearly, we can then obtain the value of the Stokes constant $K_{jk}$ via a residue calculation of $\Phi^{(j)}$, as long as we know the latter around the singular point. To have an effective procedure to determine adjacency, \textit{i.e.}, the non-zero Stokes constants, we need to determine the analytic continuation of $\Phi^{(j)}$ from the known convergent expansion \eqref{eq:borel-sqrts}. A very efficient procedure is using rational approximations through (diagonal) Pad\'{e} approximants \cite{graves1981pade}. 

The Padé approximant algorithmically approximates the series around $t=0$ (or the respective point for $s$ which is $s=0$) by a ratio of two polynomials of order $M$ where $M$ is half of the total number of terms we have on the expansion \eqref{eq:borel-sqrts} (the number of coefficients $T_n^{(j)}$ we have calculated). The poles of the denominator will carry the information of the singularities of the original Borel transform. In the case of the simple pole of $\Phi^{(j)}(s)$, the Pad\'{e} approximant of order $M$,
\begin{equation}
\mathrm{P}_M\Phi^{(j)}(s)\quad\mathrm{determined\,\,at}\quad s=0\,,
\end{equation}
 will have a simple pole\footnote{The expansion around the point $s=0$ corresponds to the original expansion around $t=0$.} at $s=1$. We can then determine the residue of this Pad\'{e} approximant and obtain the relevant Stokes constant
 \begin{equation}
     K_{jk}=\frac{2i\sqrt{\pi}}{T_0^{(k)}}\,\mathrm{Res}_{s=1}\left(\mathrm{P_M\Phi^{(j)}}\right)\,.
 \end{equation}

The Stokes constant has a sign ambiguity, which is related to the choice of orientation of the steepest manifolds, or in other words to the definition of the contour around the branch cut in the Borel plane. Nevertheless, the combination $K_{jk}\,T^{(k)}_0$ will be unambiguous, as the choice of orientation will change the sign of the coefficients $T^{(k)}_\ell$. The non-zero intersection numbers appearing in the transseries decomposition \eqref{eq:Useries} will inherit the same ambiguity, but they will always appear multiplied by the relevant expansion coefficients such that the combination will be unambiguous.

 If we determine the Pad\'{e} approximant $\mathrm{P}_M\mathcal{B}[\Psi^{(j)}]$ of the original expansion \eqref{eq:Borel-sec-j} at $t=0$, we can plot the poles of the approximant (the zeros of the denominator) and this provides a very useful visual aid to check for adjacent saddles: the function is expected to have branch cuts and the Pad\'{e} approximant, which has only pole singularities, mimics the branch cuts by seeding its poles along the locus of these cuts. 
 
\begin{figure}
    \centering
 \begin{subfigure}[b]{0.46\textwidth}
            \includegraphics[width=\textwidth]{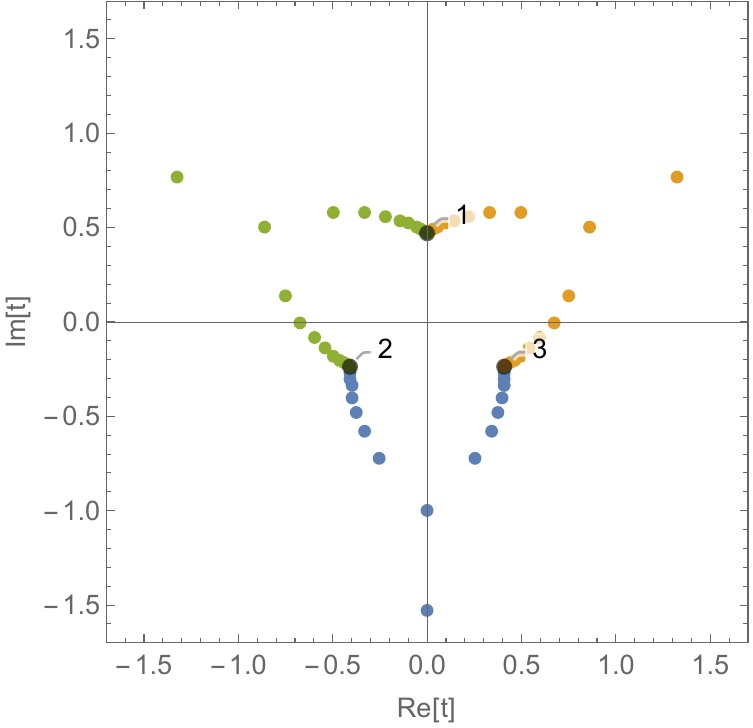}
        \end{subfigure}
        \hspace{1cm}
        \begin{subfigure}[b]{0.46\textwidth}
            \includegraphics[width=\textwidth]{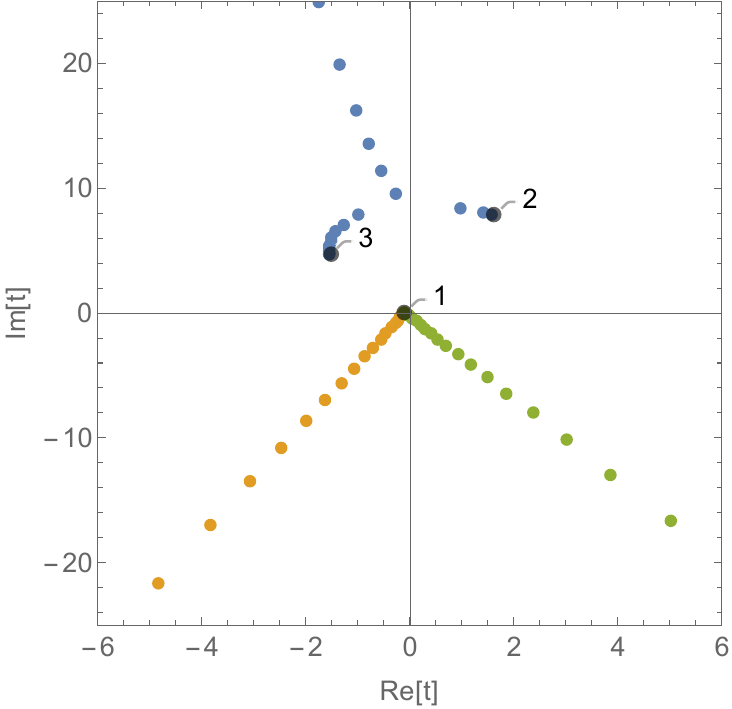}
        \end{subfigure}
        \caption{Superposition of the singularity structure of the Borel planes for each of the 3 critical points of the Pearcey integral \eqref{eq:Pearcey}. \textit{Left}: case $y=0,\,z=1$. The three Borel singularities $t_j=-i \,f(x_j)$ are shown in grey. The pole singularities of the Pad\'{e} approximant $\mathrm{P}_{30}\mathcal{B}[\Psi^{(j)}]$ (using 60 terms of the Borel transform) are shown in green ($j=1$), orange ($j=2$) and blue ($j=3$). The Borel plane of saddle $1$ has two cuts starting at saddles $2$ and $3$ (shown as a condensation of poles in green), evidence of adjacency between these saddles. The same is true for saddles $2$ and $3$. \textit{Right:} case $y=5,\,z=1+i$. This corresponds to a complex exponent; in this case, we can see that saddle $1$ sees both saddles $2$ and $3$ (in blue), while saddles $2$ and $3$ see saddle $1$ (orange and green) but do not see each other.
        }
        \label{fig:Pearcey-pade}
    \end{figure}

 On the left panel of \cref{fig:Pearcey-pade}, we can see a superposition of the Borel singularities associated with each of the 3 critical points of the Pearcey integral \eqref{eq:Pearcey} for $y=0$, $z=1$, which were determined via a Pad\'{e} approximant of the respective Borel transform. It is clear that every critical point is adjacent to the other two, as it predicts branch cuts starting at the respective Borel singularities\footnote{The Borel transform for each critical point is always defined at $t=0$. In \cref{fig:Pearcey-pade} we have shifted the Borel plane for each critical point $x_j$ from $t=0$ to the respective critical value $t=-i\,f(x_j)$}. This is in agreement with the adjacency plot in \cref{fig:Pearcey_rotate-adjacency}. The right panel of \cref{fig:Pearcey-pade} shows the adjacency plot for a different configuration of the Pearcey integral, with $y=5$ and $z=1+i$ -- for this choice of parameters, we see that while saddle $1$ sees both saddles $2,3$, the latter only see saddle $1$ but not each other. Note that this case corresponds to a complex exponent in the integral \eqref{eq:Pearcey}, and for this case, the original integration along the real axis would diverge. Nevertheless, the adjacency structure depends solely on the steepest descent contours associated with each critical point and is independent of the convergence of the original integral.

  \section{Relevance from Adjacency: the ``South-East Rule"}\label{sec:Relevance}

    We are now in a position to derive the main result of the paper, namely that it is possible to use a simple geometric rule, based on the disposition of images of the critical points in the Borel plane, which, when combined with adjacency information, will determine the relevance of the critical points.  We term this the "South-East Rule", for reasons that will become apparent. 
    
\begin{figure}
        \centering
        \includegraphics[width=0.8 \textwidth]{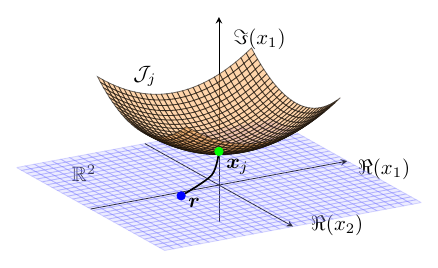}
        \caption{A sketch of the steepest descent manifold $\mathcal{J}_j$ (the yellow surface) of a relevant complex critical point $\bm{x}_j$ (the green point). The critical point is relevant, as there exists a steepest descent contour (the black curve) connecting the critical point to the original integration domain (the blue plane) via the intersection point $\bm{r}$ (the blue point).}\label{fig:Sketch}
    \end{figure}

    Picard–Lefschetz theory relates the weight $n_j$ to the intersection of the steepest ascent manifold $\mathcal{K}_j$ with the original integration domain $\mathbb{R}^d$. As both $\mathcal{K}_j$ and $\mathbb{R}^d$ are $d$-dimensional manifolds embeded in $\mathbb{C}^d$, they generically intersect in points. Consequently, given an intersection point $\bm{r}_j \in \mathbb{R}^d$, there exists a steepest descent line from $\bm{r}_j$ to $\bm{x}_j$ connecting the critical point to the original integration domain (see \cref{fig:Sketch}). Direct evaluation of the intersections of the ascent manifold with the real plane is computationally expensive for high-dimensional integrals \cite{Weber:2025, Shoji:2026}. However, note that only the existence of the intersection points and not their locations plays a role in the Picard-Lefschetz formula.

    As illustrated in \cref{sec:Borel}, the Borel plane simplifies Picard-Lefschetz analysis. The steepest descent and ascent manifolds $\mathcal{J}_j$ and $\mathcal{K}_j$ map to horizontal curves on the Riemann sheet of the Jacobian $\phi$, looping around the branch point singularity $t_j$ associated with the critical points $\bm{x}_j$. The original integration domain maps to a curve on the Riemann sheet along the imaginary axes. The Riemann sheet structure is fully encoded in the adjacency graph. Two branch point singularities are connected by an edge if and only if they can see each other on the Riemann sheet. This corresponds to the existence of a value of $\theta$ for which a Stokes phenomenon takes place between the two. We use this perspective to derive an algorithm to evaluate the intersection numbers from the adjacency graph. We first consider integrals for which the exponent $f$ is bounded from below. Next, we extend the analysis to general analytic exponentials.

    \subsection{Bounded exponentials}
    
\begin{figure}
        \centering
        \begin{subfigure}[b]{0.32\textwidth}
            \includegraphics[width=\textwidth]{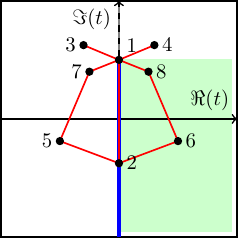}
            \caption{$\theta=0$}
        \end{subfigure}
        \begin{subfigure}[b]{0.32\textwidth}
            \includegraphics[width=\textwidth]{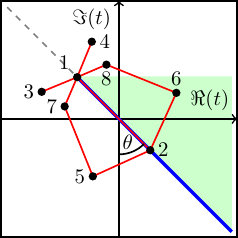}
            \caption{$\theta=\pi/4$}
        \end{subfigure}
                \begin{subfigure}[b]{0.32\textwidth}
            \includegraphics[width=\textwidth]{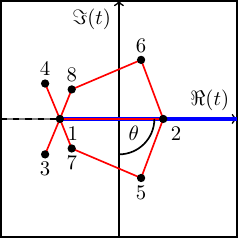}
            \caption{$\theta=\pi/2$}
        \end{subfigure}
        \caption{A sketch of the Borel plane for a constructed example, for illustrative purposes, for  different angles $\theta=0, \pi/4,$ and $\pi/2$ (left to right), with the original integration domain maps to the line segment $D$ (the blue line segment) running from $-M e^{i \theta}$ to infinity in the $ e^{i \theta}$ direction. The green region consists of the points to the right of interval $D$, marking the space in which relevant complex critical points reside. The relevance of the complex critical points is governed by the adjacency graph (the red lines) connecting the branch point singularities (the black points).}\label{fig:BorelPlane}
    \end{figure}

    Let us assume that the function $f$ is bounded by the constant $\mathcal{M}$, \textit{i.e.}, $\mathcal{M}$ is the largest number such that $f(\bm{x}) \geq \mathcal{M}$ for all $\bm{x} \in \mathbb{R}^d$. Following the geometric definition of the adjacency criteria, we generalise the integral by introducing a phase in $k$,
    \begin{align}
        \Psi_\theta =\left(\frac{k}{2\pi i}\right)^{d/2} \int_{\mathbb{R}^d} e^{i k e^{i \theta} f(\bm{x})}\mathrm{d}\bm{x}\,.
    \end{align}
    While the critical points are independent of the angle $\theta$, the steepest descent and ascent manifolds perform a half rotation when increasing $\theta$ from $0$ to $2\pi$. First, note that for $\theta=0$, the original integration domain is mapped to the interval $D$ running from $t=-\mathcal{M}i e^{i \theta}$ to $t=-\infty i e^{i \theta}$ in the Borel plane (the blue line in the \cref{fig:BorelPlane}(a)). As the ascent contours map onto horizontal lines running from the branch point singularities to the  left, at constant $\mathrm{Im}[-if(\bm{x}_j)]$, it follows that the relevant saddle points need to map to branch points on the right of the interval $D$ (see the shaded green regions in the sketch  \cref{fig:BorelPlane}). In terms of Picard-Lefschetz theory, the green region follows from the observation that the real part of the exponent increases while the imaginary part is preserved by the gradient ascent flow, corresponding with a horizontal move to the left in the Borel plane.
    
    Suppose now that $f({\bm x})$ possesses 8 critical points labelled $j=1,2, \cdots 8$, where we assume that $1$ and $2$ are real critical points. The choice of 8 singularities here is only for illustrative purposes, more or fewer do not change the essence of the following argument.
    From the sketch of the Borel plane in \cref{fig:BorelPlane}(a), it is clear that the singularities corresponding to the critical points $1$ and $2$ are relevant as they lie on the original integration domain.  Since the singularities corresponding to critical points $3, 5$ and $7$ lie to the left of $D$, they are irrelevant. Critical point $4$ is also irrelevant because it lies above the contour $D$ and its horizontal (in the Borel plane) steepest ascent contour can never intersect with the original integration domain.

    The critical points $ 6$ and $8$ lie to the right of $D$ at $\theta=0$, and so are candidates to be relevant. To identify which of these (if any) are indeed relevant, we rotate $\theta$ from $0$ to $\pi/2$, \cref{fig:BorelPlane}(c). 
    The interval $D$ then rotates to the real axis, the images of the critical points in the Borel plane also rotate about the origin, and the shaded green region shrinks to a horizontal line. The net effect is that at $\theta=\pi/2$ we have a non-oscillatory convergent integral
     \begin{align}
        \Psi_{\theta=\pi/2} = \left(\frac{k}{2\pi i}\right)^{d/2}\int_{\mathbb{R}^d} e^{- k  f(\bm{x})}\mathrm{d}\bm{x}\,.\label{eq:Euclidean}
    \end{align}
   as shown in  \cref{fig:BorelPlane}, and the only relevant critical points are the critical points for which $f(\bm{x}_j)$ is real. Moreover, as the original integration domain coincides with its downward flow, \textit{i.e.}, the plane $\mathbb{R}^d$ is a fixed point of the downward flow, we can express the real plane as the union of descent manifolds of the real critical points. Consequently, at $\theta=\pi/2$, the real critical points are the relevant critical points.
   
    The next step is crucial.  We now rotate $\theta$ back towards $\theta=0$, and take note of whether any of the Borel singularities corresponding to the potentially relevant critical points (the ones in the green region at $\theta=0$) line up horizontally with the branch points of any critical point that is relevant at that angle $\theta$.  Whenever two branch point singularities are aligned horizontally in the Borel plane, there is the possibility that a Stokes phenomenon may occur. At such a Stokes transition a relevant critical points changes the relevance of another critical point if turns out that they are adjacent. When such a Stokes transition occurs, the Borel singularity of the critical point that changes intersection number, crosses the branch cut of the relevant saddle moving into another sheet, see for example \cite{Howls:1997}.
   
    At $\theta=0$, critical points that undergo a Stokes transition under this rotation are connected by an edge running in the \textit{South-East} direction in the adjacency graph. When the North-West critical point is relevant, it will change the relevance of the adjacent critical point pointing in the South-East direction.

    In our illustration, critical point $6$ is the first to enter the green region. However, as it is not adjacent to critical point $1$, its relevance does not change. Next, critical point $8$ undergoes a Stokes transition with critical point $1$, making it relevant to the integral. Note that critical points $6$ and $8$ do undergo a Stokes transition, but not while either one is relevant. This transition does not change the relevance of either critical points. We conclude that critical points $1,2$ and $8$ are the relevant critical points to the integral $\Psi_{\theta=0}$.

    In summary, a complex critical point $\bm{x}_k$ switches from being irrelevant to relevant (or vice versa) when $\bm{x}_j$ is relevant, and there exists a horizontal edge in the adjacency graph from $t_j$ to $t_k$. Taking into account the orientation of the descent manifolds through the Stokes constants $K_{jk}$, when $\bm{x}_j$ is relevant the intersection numbers change as 
    \begin{align}
        n_k \mapsto n_k + n_j K_{jk}\,,
    \end{align} 
    at the Stokes transition of $\bm{x}_k$ with $\bm{x}_j$. We can formalise this process as an algorithm.

    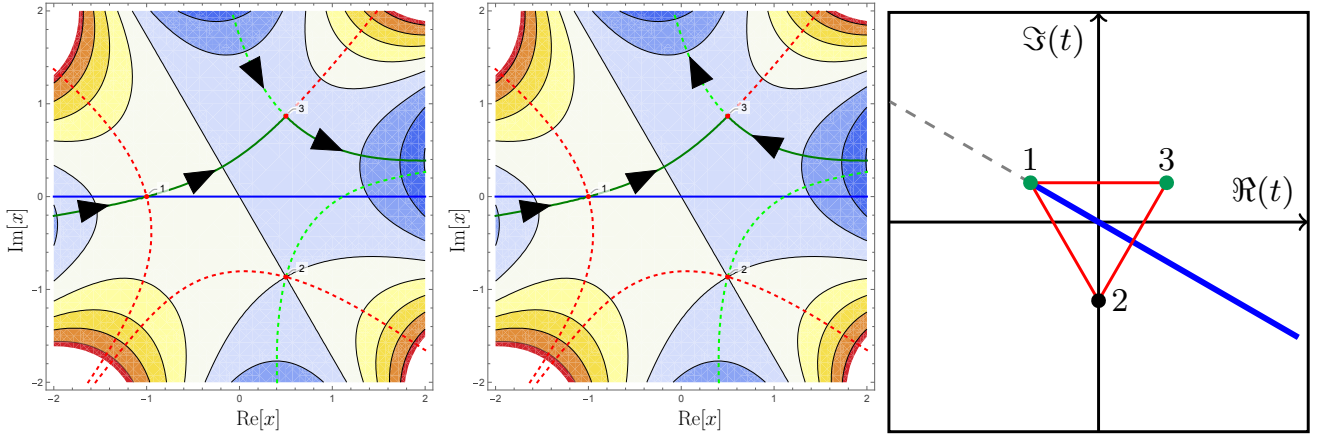
\begin{figure}
        \centering
        \begin{subfigure}[b]{0.32\textwidth}
            \resizebox{\textwidth}{!}{%
            \begin{tikzpicture}
                \node[anchor=south west, inner sep=0pt, outer sep=0pt] (image) at (0,0) {\includegraphics{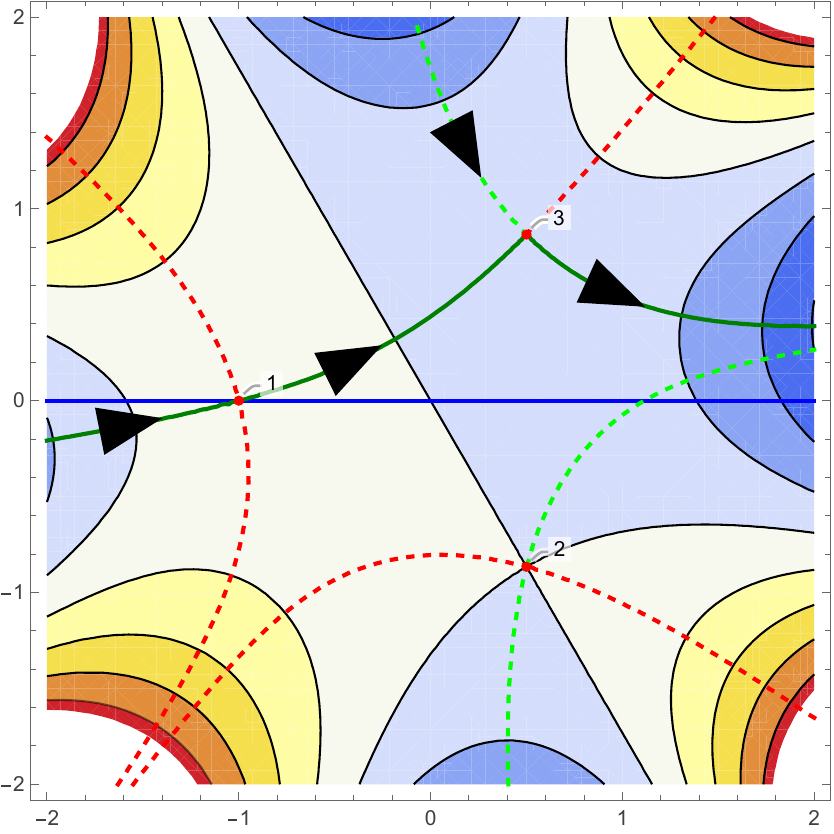}};
                \node[below right, font=\huge] at (image.south) {Re$[x]$};
                \node[above left, font=\huge, rotate = 90] at (image.west)  {Im$[x]$};
            \end{tikzpicture}
            }
        \end{subfigure}
        \begin{subfigure}[b]{0.32\textwidth}
            \resizebox{\textwidth}{!}{%
            \begin{tikzpicture}
                \node[anchor=south west, inner sep=0pt, outer sep=0pt] (image) at (0,0) {\includegraphics{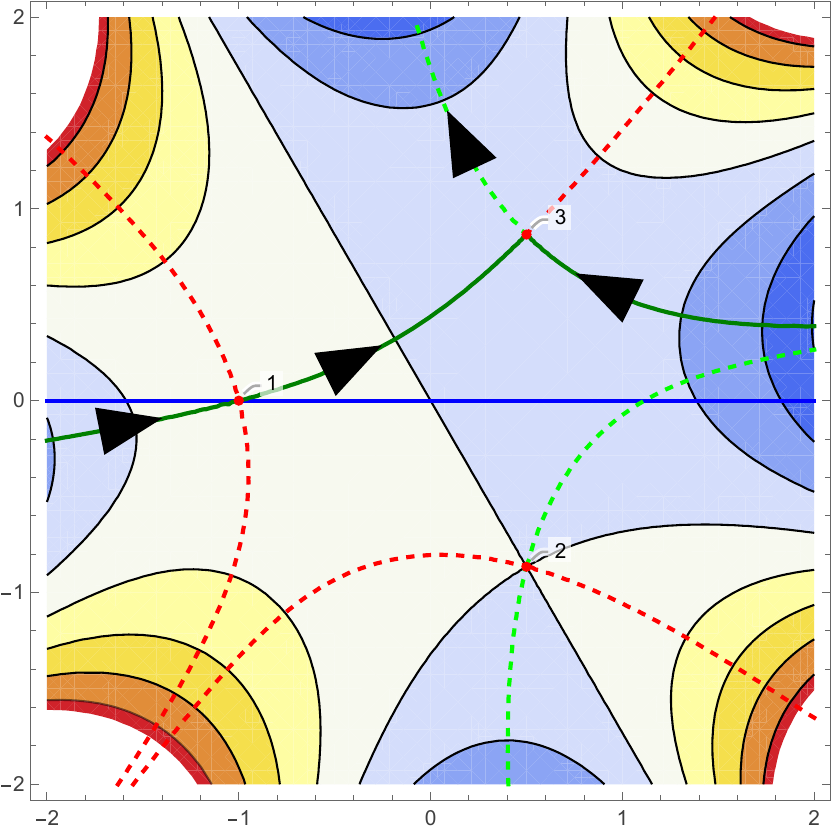}};
                \node[below right, font=\huge] at (image.south) {Re$[x]$};
                \node[above left, font=\huge, rotate = 90] at (image.west)  {Im$[x]$};
            \end{tikzpicture}
            }
        \end{subfigure}
        \begin{subfigure}[b]{0.32\textwidth}
            \resizebox{\linewidth}{!}{%
                \begin{tikzpicture}[thick]
                    \draw[thick, black] (-2, -2) rectangle (2, 2);  
                    \draw[->] (-2, 0) -- (2, 0) node[above left=0.1] {$\Re(t)$};
                    \draw[->] (0, -2) -- (0, 2) node[below left=0.1] {$\Im(t)$};
                    \begin{scope}[rotate around={60:(0,0)}]
                        \draw[gray, dashed] (0, -2.2) -- (0, 2.3);
                        \draw[blue, ultra thick] (0, -2.2) -- (0, 0.75 );  
                        
                        \coordinate (A) at (0, 0.75);
                        \coordinate (B) at (+0.649519, - 0.375);
                        \coordinate (C) at (-0.649519, - 0.375);

                        \draw[red] (A) -- (B) -- (C) -- cycle;
                        \fill[ForestGreen](A) circle (2pt) node [above, black] {$1$}; 
                        \fill[ForestGreen](B) circle (2pt) node [above, black] {$3$}; 
                        \fill[black](C) circle (2pt) node [right, black] {$2$}; 
                    \end{scope}
                \end{tikzpicture}
            }
        \end{subfigure}
        
        \caption{This is the Pearcey integral for $y=0,\,z=1$ at the Stokes transition $\theta = \pi/3$. \textit{Left and centre:} The relative orientation of the descent manifolds in the complex $x$-plane. \textit{Right:} The associated adjacency graph in the complex $t$-plane. We follow the notation outlined in \cref{fig:Pearcey}.}\label{fig:Stokes}
    \end{figure}

    \begin{algorithm}[H]
        \caption{The "South-East rule" for the evaluation of the intersection numbers for bounded exponentials.}
        \label{alg:asymptoticTerms}
        \begin{algorithmic}[1]
            \State Select the edges $(t_j, t_k)$ in the adjacency graph going in the South-East direction and order these edges by their slope 
            \begin{align}
                \theta_{jk} = \arg(t_k - t_j)\in [-\pi/2,0]\,.
            \end{align}
            from steep to shallow. Explicitly, we order pairs $(t_j,t_k)$ such that $-\pi/2 \leq \theta_{j_1 k_1} < \theta_{j_2 k_2}< \dots < \theta_{j_N k_N} \leq 0$. 
            \State Set $n_j$ to be the relative orientations of $\mathcal{J}_j$ and $\mathbb{R}^d$ for real critical points (often set to $1$). Set $n_j = 0$ for the complex critical points. This corresponds to the rotated integral $\Psi_{\theta = \pi/2}$ where the relevant critical points are the real critical points.
            \State Loop over the ordered list of edges. An edge between $t_j$ and $t_k$, change the intersection number 
            \begin{align}
                n_k \mapsto n_k + n_j K_{jk}
            \end{align} when the critical point $\bm{x}_j$ is relevant. This corresponds to smoothly rotating clockwise back from $\Psi_{\theta=\pi/2}$ to $\Psi_{\theta=0}$.
        \end{algorithmic}
        \label{alg}
    \end{algorithm}

    Note that for the vast majority of oscillatory integrals in physics, the exponent $f$ is bounded from below. In the next section, we consider unbounded exponents.

    \subsection{Unbounded exponentials}\label{sec:unbounded}
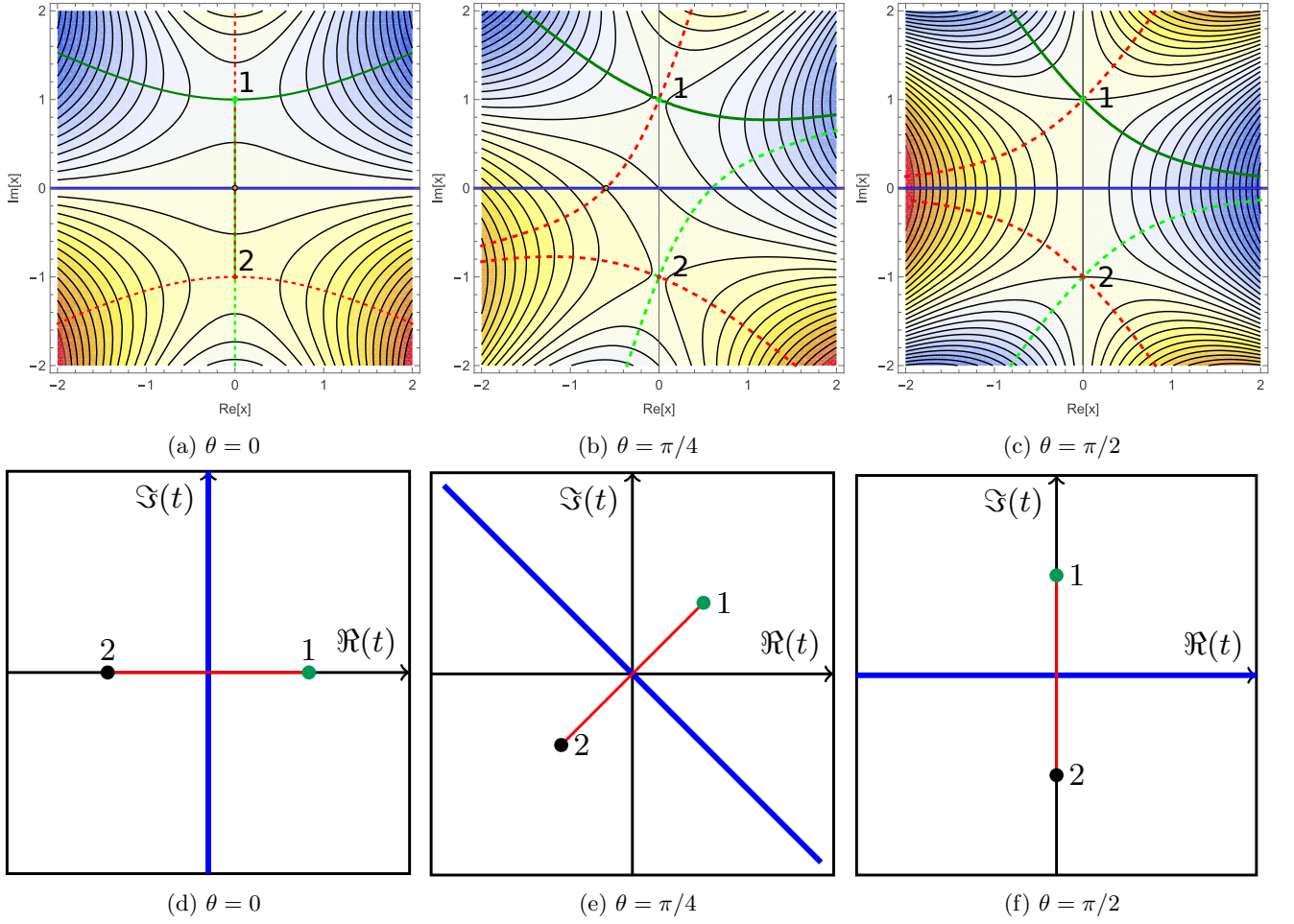
\begin{figure}
        \centering
        \begin{subfigure}[b]{0.32\textwidth}
            \includegraphics[width=\textwidth]{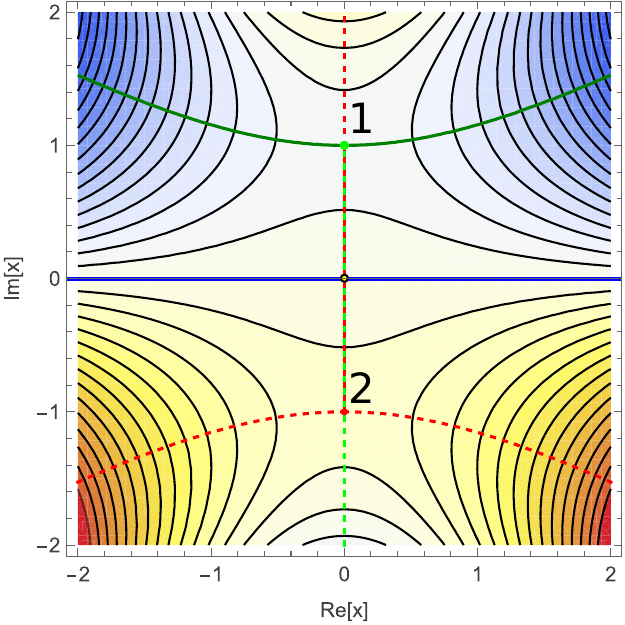}
            \caption{$\theta = 0$}
        \end{subfigure}
        \begin{subfigure}[b]{0.32\textwidth}
            \includegraphics[width=\textwidth]{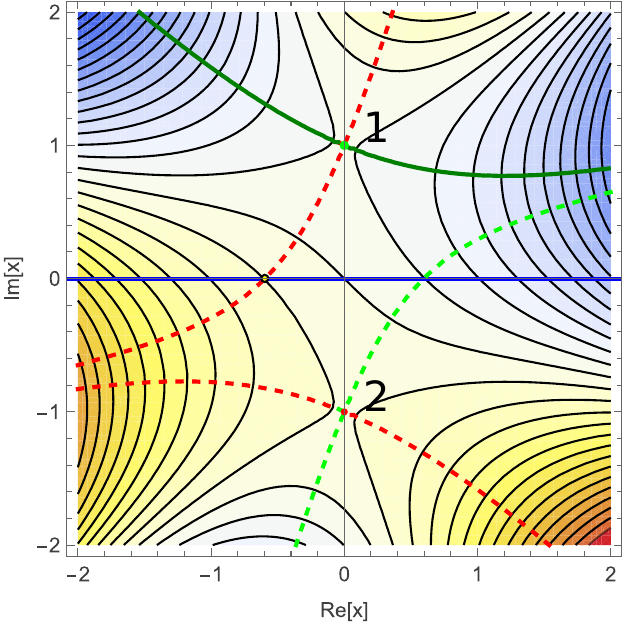}
            \caption{$\theta = \pi/4$}
        \end{subfigure}
        \begin{subfigure}[b]{0.32\textwidth}
            \includegraphics[width=\textwidth]{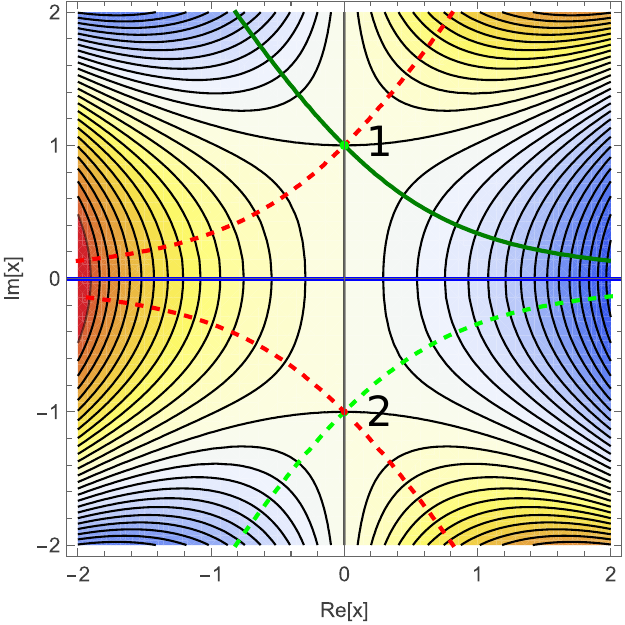}
            \caption{$\theta = \pi/2$}
        \end{subfigure}
        \begin{subfigure}[b]{0.32\textwidth}
            \resizebox{\linewidth}{!}{%
                \begin{tikzpicture}[thick]
                    \draw[thick, black] (-2, -2) rectangle (2, 2);  
                    \draw[->] (-2, 0) -- (2, 0) node[above left=0.1] {$\Re(t)$};
                    \draw[->] (0, -2) -- (0, 2) node[below left=0.1] {$\Im(t)$};
                    \begin{scope}[rotate around={0:(0,0)}]
                        
                        \draw[blue, ultra thick] (0, -2) -- (0, 2. );

                        \coordinate (A) at (+1,0);
                        \coordinate (B) at (-1,0);

                        \draw[red] (A) -- (B);

                        \fill[ForestGreen](A) circle (2pt) node [above, black] {$1$}; 
                        \fill[black](B) circle (2pt) node [above, black] {$2$}; 
                    \end{scope}
                \end{tikzpicture}
            }
            \caption{$\theta = 0$}
        \end{subfigure}
        \begin{subfigure}[b]{0.32\textwidth}
            \resizebox{\linewidth}{!}{%
                \begin{tikzpicture}[thick]
                    \draw[thick, black] (-2, -2) rectangle (2, 2);  
                    \draw[->] (-2, 0) -- (2, 0) node[above left=0.1] {$\Re(t)$};
                    \draw[->] (0, -2) -- (0, 2) node[below left=0.1] {$\Im(t)$};
                    \begin{scope}[rotate around={45:(0,0)}]
                        \draw[blue, ultra thick] (0, -2.65) -- (0, 2.65);

                        \coordinate (A) at (+1,0);
                        \coordinate (B) at (-1,0);

                        \draw[red] (A) -- (B);

                        \fill[ForestGreen](A) circle (2pt) node [right, black] {$1$}; 
                        \fill[black](B) circle (2pt) node [right, black] {$2$}; 
                    \end{scope}
                \end{tikzpicture}
            }
            \caption{$\theta = \pi/4$}
        \end{subfigure}
        \begin{subfigure}[b]{0.32\textwidth}
            \resizebox{\linewidth}{!}{%
                \begin{tikzpicture}[thick]
                    \draw[thick, black] (-2, -2) rectangle (2, 2);  
                    \draw[->] (-2, 0) -- (2, 0) node[above left=0.1] {$\Re(t)$};
                    \draw[->] (0, -2) -- (0, 2) node[below left=0.1] {$\Im(t)$};
                    \begin{scope}[rotate around={90:(0,0)}]
                        \draw[blue, ultra thick] (0, -2) -- (0, 2. );

                        \coordinate (A) at (+1,0);
                        \coordinate (B) at (-1,0);

                        \draw[red] (A) -- (B);

                        \fill[ForestGreen](A) circle (2pt) node [right, black] {$1$}; 
                        \fill[black](B) circle (2pt) node [right, black] {$2$}; 
                    \end{scope}
                \end{tikzpicture}
            }
            \caption{$\theta = \pi/2$}
        \end{subfigure}
        \caption{The analytic continuation of Airy function at $y=1$ for $\theta =0, \pi/4$ and $\pi/2$. \textit{Top:} The Picard-Lefschetz deformation of the original integration domain in the complex $x$-plane. \textit{Bottom:} The associated adjacency graph in the complex $t$-plane. We follow the notation outlined in \cref{fig:Pearcey}.}\label{fig:fold_unfolding_Borel}
    \end{figure}

    When the exponent $f$ is unbounded, the adjacency graph still governs the Stokes phenomena which determines the intersection numbers. However, the situation is more subtle as the original integration domain $\mathbb{R}^d$ is no longer a steepest descent manifold for the integral $\Psi_{\theta=\pi/2}$. The integrand $e^{- kf(\bm{x})}$ diverges when $f$ approaches $-\infty$. Consequently, the rotated integral $\Psi_{\theta=\pi/2}$ may have a relevant complex critical point. To see this, let us consider the Airy function
    \begin{align}
        \Psi_\theta = \sqrt{\frac{k}{2 \pi i}} \int_{-\infty}^\infty e^{i k e^{i \theta} \left[\frac{x^3}{3} + y x\right]}\mathrm{d}x\,,
    \end{align}
    which includes two critical points located at $x_{\pm} = \pm i\sqrt{y}$. For positive $y$, the integral has two complex critical points, one of which is relevant and one of which is irrelevant to $\Psi_{\theta =0}$ (see the upper left panel of \cref{fig:fold_unfolding_Borel}). In the Borel plane, the interval $D$ extends from $-i \infty$ to $+i \infty$ (see the lower left panel of \cref{fig:fold_unfolding_Borel}). As we rotate to $\theta = \pi/2$, the intersection of $\mathbb{R}$ and $\mathcal{K}_1$ moves to infinity, ensuring the relevance of critical point $1$. Complex critical points can remain relevant for $\theta=\pi/2$ as the domain $D$ extends to $+i \infty$. This reflects the fact that the integral $\Psi_{\theta}$ diverges for any angle $0<\theta\leq \pi/2$. When considering the analytic continuation of the Airy function, not only the integrand but also the integration domain is deformed in the integral representation (see the upper panels of \cref{fig:fold_unfolding_Borel}). The relevant critical points associated with the analytic continuation of the Airy function are correctly identified by the intersection algorithm. After finding the relevant critical points at $\theta = \pi/2$, we may, analogously to the bounded case, rotate back to $\theta =0$ while tracking the Stokes transitions to infer the intersection numbers for $\Psi_{\theta =0}$ using \cref{alg}. In practice, we can identify the relevant complex critical points of $\Psi_{\theta = \pi/2}$ by regulating the integral. After adding a sufficiently quickly rising analytic regulator $r_L$ with $\lim_{L\to \infty}r_L(\bm{x})=0$ such that the sum $f(\bm{x}) + r_L(\bm{x})$ is bounded from below, we may define the integral in the limit 
    \begin{align}
        \Psi_\theta
        \equiv
        \lim_{L \to \infty }\int_{\mathbb{R}^d} e^{i k e^{i \theta} (f(\bm{x}) + r_L(\bm{x}))}\mathrm{d}\bm{x}\,.
    \end{align}
    The dominated convergence theorem guarantees that the integral converges to the Picard-Lefschetz deformation and is independent of the regulator of choice (for a detailed discussion, see \cite{Feldbrugge:2023}). At finite $L$, the regulator introduces a set of auxiliary critical points on which we can apply \cref{alg}. An auxiliary critical point may undergo a Stokes transition, turning on a complex critical point of the unregulated integral. In the limit $L\to \infty$, these auxiliary critical points move to infinity and do no longer contribute; however, the complex critical points they turned on will stay relevant.

    \bigskip 

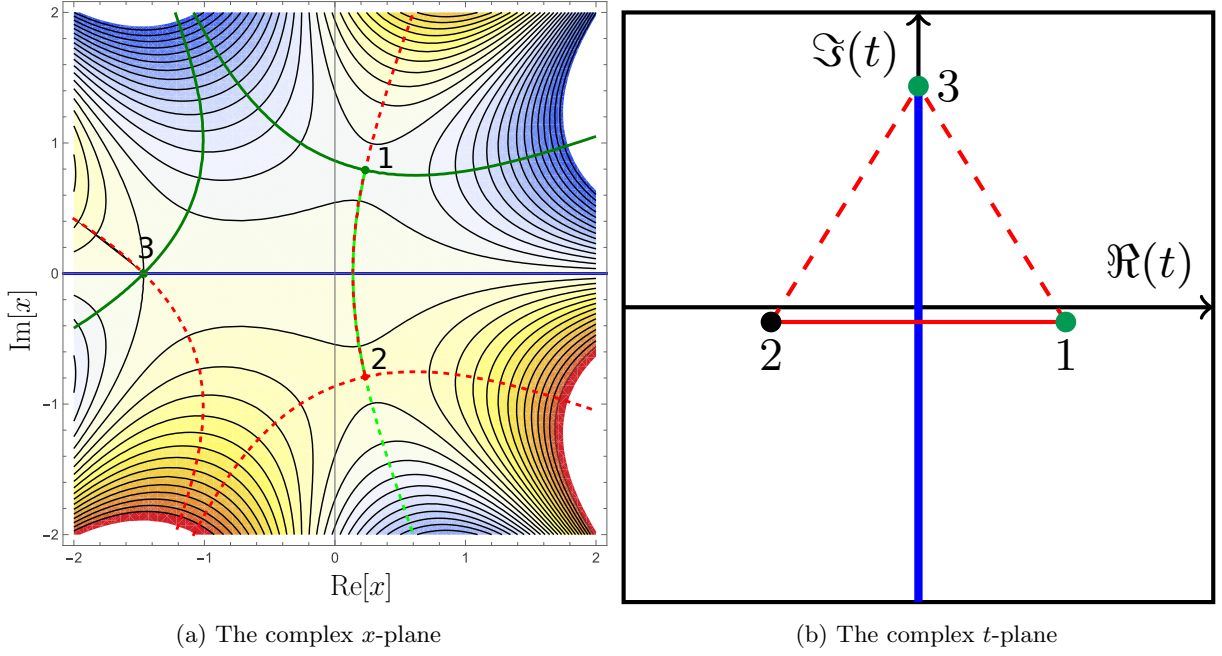
\begin{figure}
        \centering
        \begin{subfigure}[b]{0.45 \textwidth}
            \resizebox{\textwidth}{!}{%
            \begin{tikzpicture}
                \node[anchor=south west, inner sep=0pt, outer sep=0pt] (image) at (0,0) {\includegraphics{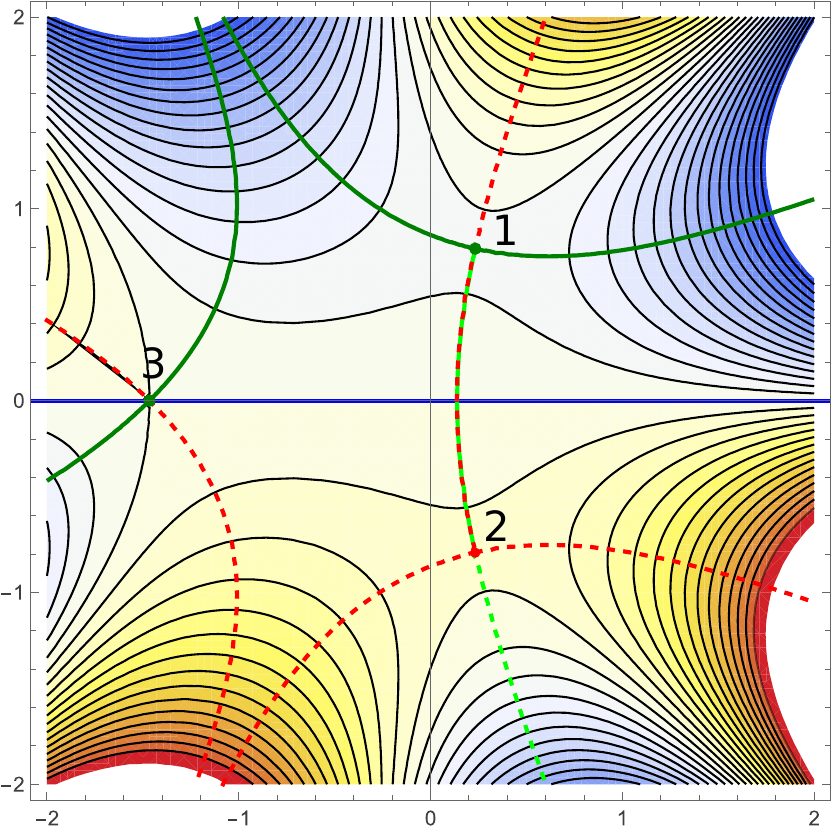}};
                \node[below right, font=\huge] at (image.south) {Re$[x]$};
                \node[above left, font=\huge, rotate = 90] at (image.west)  {Im$[x]$};
            \end{tikzpicture}
            }
            \caption{The complex $x$-plane}
        \end{subfigure}
        \begin{subfigure}[b]{0.45\textwidth}
            \resizebox{\linewidth}{!}{%
                \begin{tikzpicture}[thick]
                    \draw[thick, black] (-2, -2) rectangle (2, 2);  
                    \draw[->] (-2, 0) -- (2, 0) node[above left=0.1] {$\Re(t)$};
                    \draw[->] (0, -2) -- (0, 2) node[below left=0.1] {$\Im(t)$};
                    \begin{scope}[rotate around={0:(0,0)}]
                        
                        \draw[blue, ultra thick] (0, -2) -- (0, 1.5 );

                        \coordinate (A) at (+1,-0.1);
                        \coordinate (B) at (-1,-0.1);
                        \coordinate (C) at (0, 1.5);

                        \draw[red] (A) -- (B);
                        \draw[red, dashed] (A) -- (C);
                        \draw[red, dashed] (B) -- (C);

                        \fill[ForestGreen](A) circle (2pt) node [below, black] {$1$}; 
                        \fill[black](B) circle (2pt) node [below, black] {$2$}; 
                        \fill[ForestGreen](C) circle (2pt) node [right, black] {$3$}; 
                    \end{scope}
                \end{tikzpicture}
            }
            \caption{The complex $t$-plane}
        \end{subfigure}
        \caption{The regulated Airy function for $L=4$. \textit{Left:} The Picard-Lefschetz deformation in the complex $x$-plane. \textit{Right:} The associated adjacency graph in the complex $t$-panel with the adjacency relations with the auxiliary critical point represented by the red dashed edges. We follow the notation outlined in \cref{fig:Pearcey}.}\label{fig:fold_regulated}
    \end{figure}

    To illustrate this, let us consider the regulated Airy function
    \begin{align}
        \Psi = \lim_{L \to \infty}\sqrt{\frac{k}{2 \pi i}} \int_{-\infty}^\infty e^{i k\left[\frac{x^3}{3} + y x + \frac{x^4}{L}\right]}\mathrm{d}x\,.
    \end{align}
    The regulator makes the exponent $f$ bounded from below and introduces a real auxiliary critical point that moves to $t=+i \infty$ as $L$ increases to $\infty$ (see the left panel of \cref{fig:fold_regulated}). As the three critical points are adjacent, it follows that critical point $1$ undergoes a Stokes transition with critical point $3$, making it relevant to the integral. In the limit $L \to \infty$, the thimbles of the regulated Airy function reduce to the ones of the unregulated Airy function.

    \bigskip
    The observation that the analytic continuation of an integral generally not only involves the deformation of the integrand but also the integration domain is central to the famous conformal factor problem in quantum gravity \cite{Gibbons:1978, Hawking:1979, Mazur:1990} and the sign problem in many-particle systems \cite{Loh:1990, Kieu:1994, Pan:2022}. When the potential and action in the path integral is bounded from below, the analytic continuation of the real-time path integral to the Euclidean path integral is well-defined. The original integration domain coincides with the steepest descent manifolds of the relevant critical points. In these situations, no sign problem emerges. When the potential is not bounded from below, the Picard-Lefschetz deformation of the analytically continued path integral no longer coincides with the space of real-valued paths. The naive formulation of the Euclidean path integral diverges, as we observed for the Airy function.

    \bigskip
    
    Throughout this work, we consider oscillatory integrals for which the exponent $if(\bm{x})$ is imaginary for $x\in {\mathbb R}^d$. In the case that $f$ is instead complex, the original integration domain no longer maps to the imaginary line in the Borel plane. Rather, the original integration domain will generically map to a region in that plane. While it is still true that the relevant complex critical points map to branch point singularities residing to the right of this region, the rotated integral for $\theta=\pi/2$ may now already have contributions from complex critical points. Once we rotate the integral to $\theta=\pi/2$ we can define a horizontal strip which surrounds the rotated region of integration in the Borel plane. If some of the critical points map to branch points $t_j$ that lie within this horizontal strip for the rotated integral $\theta=\pi/2$, our present methods are inconclusive . These critical points may be relevant and influence the relevance of other critical points while rotating back to $\theta=0$. If this is the case, one may need to deform the integral further to identify the relevant critical points at $\theta=\pi/2$ before rotating back to the original integral, keeping track of how such a deformation changes the adjacency graph.
    
    \section{Examples}\label{sec:Examples}

    We now demonstrate the ``South-East" rule for determining relevant critical points in action. To do so, we evaluate the adjacency relations and relevant saddles for a set of canonical diffraction integrals. We can then easily read off the intersection numbers appearing in the final transseries decomposition \eqref{eq:Useries}, from the residue calculation of the Stokes constants as explained in \cref{sec:Adjacency}. 
    
    Note that in the class of examples we discuss, whenever a complex critical point becomes relevant, it will not turn on further complex critical points, and thus the non-zero intersection numbers will be $n_{jk}=\pm 1$. As discussed in \cref{sec:Adjacency}, this ambiguity is related to the orientation of the steepest descent manifold, but the quantity $K_{jk}\,T^{(k)}_0$ is unambiguous and only combinations $n_{jk}\,T^{(k)}_0$ will be relevant for the final transseries \eqref{eq:Useries}.

    \subsection{The cusp integral (\texorpdfstring{$A_3$}{})}
    
\begin{figure}
        \begin{subfigure}[t]{0.5 \textwidth}
            \centering
            \begin{tikzpicture}
                \node (img) {\includegraphics[width=\textwidth]{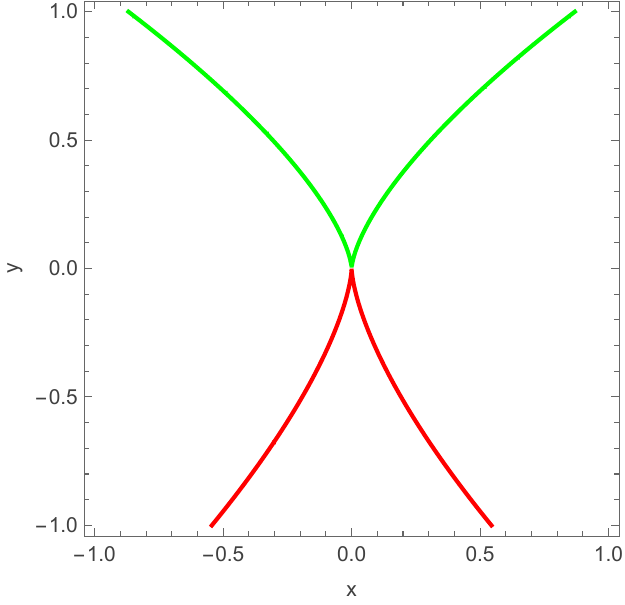}};
                \node[scale=2] at (0.55,-1.5) {$\Circled{1}$};
                \node[scale=2] at (-1, 0.3) {$\Circled{2}$};
                \node[scale=2] at (2.05, 0.3) {$\Circled{2}$};
                \node[scale=2] at (0.55, 2) {$\Circled{3}$};
            \end{tikzpicture}
        \end{subfigure}
        \caption{The unfolding of the cusp caustic, with the fold caustic (the red curve) and the Stokes line (the green curve). The caustic and Stokes lines partition the external parameter space into regions with different sets of relevant critical points. \textit{Region 1:} Three relevant real critical points, all real. \textit{Region 2:} Two relevant critical points, one real and one complex. \textit{Region 3:} One relevant real critical point.}\label{fig:cusp_unfolding}
        \centering
        \begin{subfigure}[b]{0.32\textwidth}
            \resizebox{\linewidth}{!}{%
                \begin{tikzpicture}[thick]
                    \draw[thick, black] (-2, -2) rectangle (2, 2);  
                    \draw[->] (-2, 0) -- (2, 0) node[above left=0.1] {$\Re(t)$};
                    \draw[->] (0, -2) -- (0, 2) node[below left=0.1] {$\Im(t)$};

                    \coordinate (A) at (0,1);
                    \coordinate (B) at (0,-0.5);
                    \coordinate (C) at (0,-1);

                    
                    \draw[blue, ultra thick] (0, -2) -- (A);
                    \draw[red] (A) -- (B);
                    \draw[red] (A) -- (C);
                    \draw[red] (B) -- (C);
                    
                    \fill[ForestGreen](A) circle (2pt) node [right, black] (p1) {$1$}; 
                    \fill[ForestGreen](B) circle (2pt) node [right, black] (p2) {$2$}; 
                    \fill[ForestGreen](C) circle (2pt) node [right, black] (p3) {$3$}; 
                \end{tikzpicture}
            }
            \caption{Region $1$}
        \end{subfigure}
        \begin{subfigure}[b]{0.32\textwidth}
            \resizebox{\linewidth}{!}{%
                \begin{tikzpicture}[thick]
                    \draw[thick, black] (-2, -2) rectangle (2, 2);  
                    \draw[->] (-2, 0) -- (2, 0) node[above left=0.1] {$\Re(t)$};
                    \draw[->] (0, -2) -- (0, 2) node[below left=0.1] {$\Im(t)$};

                    \coordinate (A) at (0,1.3);
                    \coordinate (B) at (-0.4,-0.5);
                    \coordinate (C) at (+0.4,-0.5);

                    \draw[blue, ultra thick] (0, -2) -- (A);
                    \draw[red] (A) -- (B);
                    \draw[red] (A) -- (C);
                    \draw[red] (B) -- (C);
                    
                    \fill[ForestGreen](A) circle (2pt) node [right, black] (p1) {$1$}; 
                    \fill[black](B) circle (2pt) node [left, black] (p2) {$2$}; 
                    \fill[ForestGreen](C) circle (2pt) node [right, black] (p3) {$3$}; 
                \end{tikzpicture}
            }
            \caption{Region $2$}
        \end{subfigure}
        \begin{subfigure}[b]{0.32\textwidth}
            \resizebox{\linewidth}{!}{%
                \begin{tikzpicture}[thick]
                    \draw[thick, black] (-2, -2) rectangle (2, 2);  
                    \draw[->] (-2, 0) -- (2, 0) node[above left=0.1] {$\Re(t)$};
                    \draw[->] (0, -2) -- (0, 2) node[below left=0.1] {$\Im(t)$};

                    \coordinate (A) at (0,0.2);
                    \coordinate (B) at (-0.4,+1.3);
                    \coordinate (C) at (+0.4,+1.3);

                    \draw[blue, ultra thick] (0, -2) -- (A);
                    \draw[red] (A) -- (B);
                    \draw[red] (A) -- (C);
                    \draw[red] (B) -- (C);
                    
                    \fill[ForestGreen](A) circle (2pt) node [right, black] (p1) {$1$}; 
                    \fill[black](B) circle (2pt) node [left, black] (p2) {$2$}; 
                    \fill[black](C) circle (2pt) node [right, black] (p3) {$3$}; 
                \end{tikzpicture}
            }
            \caption{Region $3$}
        \end{subfigure}
        \caption{The adjacency graph of the Pearcey integral associated with the unfolding of the cusp catastrophe in the complex $t$-plane in regions 1, 2 and 3 of \cref{fig:cusp_unfolding}. We follow the notation outlined in \cref{fig:Pearcey}.}\label{fig:cusp_unfolding_Borel}
    \end{figure}

    Our first example is the cusp integral 
    \begin{align}
        \Psi(x,y) = \sqrt{k} \int_{-\infty}^\infty e^{i k f(u)}\mathrm{d}u\,,
    \end{align}
    with the exponential
    \begin{align}
        f(u) = u^4 + y u^2 + x u\,.
    \end{align}
    This is the Pearcey integral \eqref{eq:Pearcey} we used to exemplify the Borel plane analysis and adjacency relations in \cref{sec:resurgence}.  This integral has $3$ critical points solving the cubic equation 
    \begin{align}
        4 u^3 + 2y u + x =0\,.
    \end{align}
    The relevant critical points vary as a function of the external parameters $x$ and $y$. At the configuration $x=y=0$, the three critical points coalesce at $u=0$. This is a degenerate critical point, known as a cusp catastrophe. As we move away from the cusp, we identify three qualitatively distinct regions separated by the fold caustic -- where two of the three critical points coalesce -- and the Stokes lines (see the red and green curves in \cref{fig:cusp_unfolding}). In region $1$, the integral has three real critical points that are all relevant. When crossing the fold curve, two of the three critical points become complex, forming a complex conjugate pair. In region $2$, the integral has one real and one complex relevant critical point. In region $3$, while crossing the Stokes line from region $2$, only the real critical point remains relevant to the integral. For a more detailed exposition, see \cite{NIST:DLMF}.

    Next, we analyse the adjacency graph for the three regions (see \cref{fig:cusp_unfolding_Borel}). In region $1$, the real critical points map to three branch point singularities on the imaginary axes. The three critical points are clearly relevant to the integral. In region $2$, we observe that critical point $3$ is relevant, as there exists an edge in the adjacency graph going in the lower right direction connecting branch points associated to critical points $1$ and $3$. While rotating from $\theta = \pi/2$ to $\theta =0$, critical point $3$ undergoes a Stokes transition with critical point $1$ making critical point $3$ relevant to the integral. Finally, in region $3$, branch point $t_1$ is still connected with branch point $3$. However, as this connection runs in the upper right direction, the rotation $\theta = \pi/2$ to $\theta =0$ does not lead to a Stokes transition. Consequently, critical point $3$ remains irrelevant. When moving from region $3$ to region $2$, the branch points associated to critical points $2$ and $3$ move down relative to critical point $1$. The three branch points lie on a horizontal line while crossing the Stokes line (the green line in \cref{fig:cusp_unfolding}).

    \bigskip

    In \cref{fig:cusp_unfolding_Borel}, we observe that for the Pearcey integral, the complex critical points lying to the right of the blue line are the relevant critical points. While this is a necessary condition, it is not a sufficient condition. Indeed, for more intricate integrals, this will generally not be true. To see this, imagine taking the Pearcey integral in region $3$, and deforming $f$ to create a new global minimum away from the region $u=0$, while leaving the original complex structure around $u=0$ unchanged. This deformation has the effect of adding a set of critical points to the original integral and corresponding branch points to the adjacency graph in the Borel plane, without changing the branch points in the right panel of \cref{fig:cusp_unfolding_Borel}. When the global minimum of the deformed integral is sufficiently low, the deformation will add a new branch point on the imaginary axes above the points $2$ and $3$. However, even though point $3$ now resides to the right of the blue contour is still irrelevant, as the deformation did not introduce an adjacency between the new critical point and
    point $3$.

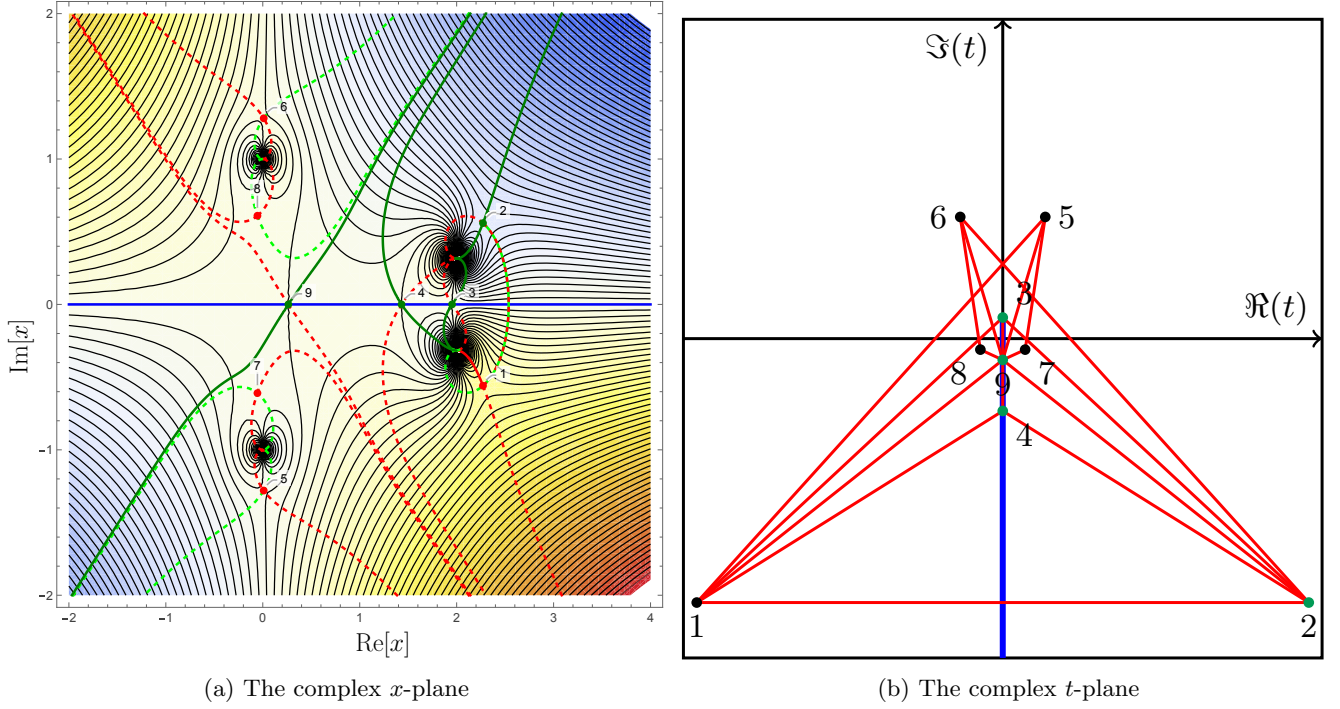
\begin{figure}
        \centering
        \begin{subfigure}[b]{0.49\textwidth}
            \resizebox{\textwidth}{!}{%
            \begin{tikzpicture}
                \node[anchor=south west, inner sep=0pt, outer sep=0pt] (image) at (0,0) {\includegraphics{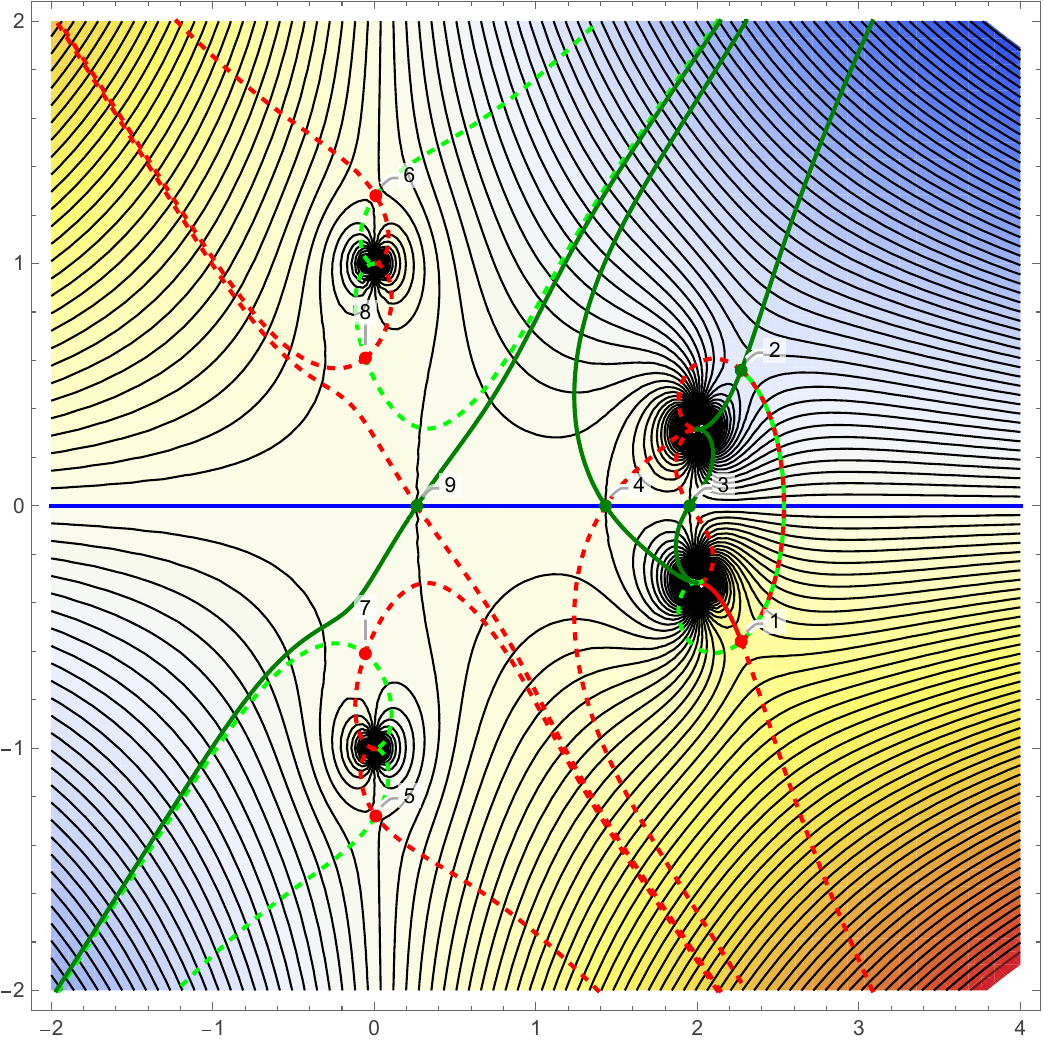}};
                \node[below right, font=\huge] at (image.south) {Re$[x]$};
                \node[above left, font=\huge, rotate = 90] at (image.west)  {Im$[x]$};
            \end{tikzpicture}
            }
            \caption{The complex $x$-plane}
        \end{subfigure}
        \begin{subfigure}[b]{0.49\textwidth}
            \resizebox{\linewidth}{!}{%
                \begin{tikzpicture}[thick]
                    \draw[thick, black] (-3, -3) rectangle (3, 3);  
                    \draw[->] (-3, 0) -- (3, 0) node[above left=0.1] {$\Re(t)$};
                    \draw[->] (0, -3) -- (0, 3) node[below left=0.1] {$\Im(t)$};

                    \pgfmathsetmacro{\alpha}{2.5}
                    \pgfmathsetmacro{\radius}{0.05pt}

                    \coordinate (A) at (\alpha * -1.15, -2.4794);
                    \coordinate (B) at (\alpha * 1.15, -2.4794);
                    \coordinate (C) at (\alpha * 0., 0.198432);
                    \coordinate (D) at (\alpha * 0., -0.479439 -0.2);
                    \coordinate (E) at (\alpha * 0.159769, 1.1432);
                    \coordinate (F) at (\alpha * -0.159769, 1.1432);
                    \coordinate (G) at (\alpha * 0.0841148, -0.103323);
                    \coordinate (H) at (\alpha * -0.0841148, -0.103323);
                    \coordinate (I) at (\alpha * 0., -0.2);

                    \draw[blue, ultra thick] (0, -3) -- (C);

                    \draw[red] (A) -- (B);
                    \draw[red] (A) -- (C);
                    \draw[red] (A) -- (D);
                    \draw[red] (A) -- (E);
                    \draw[red] (A) -- (I);
                    \draw[red] (B) -- (C);
                    \draw[red] (B) -- (D);
                    \draw[red] (B) -- (F);
                    \draw[red] (B) -- (I);
                    \draw[red] (C) -- (I);
                    \draw[red] (D) -- (I);
                    \draw[red] (E) -- (G);
                    \draw[red] (E) -- (I);
                    \draw[red] (F) -- (H);
                    \draw[red] (F) -- (I);
                    \draw[red] (G) -- (I);
                    \draw[red] (H) -- (I);
                    
                    \fill[black](A) circle (\radius) node [below, black] {$1$}; 
                    \fill[ForestGreen](B) circle (\radius) node [below, black]  {$2$}; 
                    \fill[ForestGreen](C) circle (\radius) node [above right, black] {$3$}; 
                    \fill[ForestGreen](D) circle (\radius) node [below right , black] {$4$}; 
                    \fill[black](E) circle (\radius) node [right, black] {$5$}; 
                    \fill[black](F) circle (\radius) node [left, black] {$6$}; 
                    \fill[black](G) circle (\radius) node [below right, black] {$7$}; 
                    \fill[black](H) circle (\radius) node [below left, black] {$8$}; 
                    \fill[ForestGreen](I) circle (\radius) node [below, black] {$9$}; 
                \end{tikzpicture}
            }
            \caption{The complex $t$-plane}
        \end{subfigure}
        \caption{The Picard-Lefschetz deformation of the one-dimensional double Lorentzian lens model in the complex $x$-plane (the left panel) and the adjacency graph in the complex $t$-plane (the right panel). We follow the notation outlined in \cref{fig:Pearcey}.}\label{fig:doubleLorentzian}
    \end{figure}

    To illustrate this phenomenon, we consider the  Kirchhoff-Fresnel integral of a thin lens system
    \begin{align}
        \psi(y) = \sqrt{\frac{k}{2\pi i}} \int_{-\infty}^\infty e^{i \omega T(x,y)}\mathrm{d}x\,,
    \end{align}
    with the dimensionless frequency of the radiation $\omega$ and the dimensionless time delay function 
    \begin{align}
        T(x,y) = \frac{(x-y)^2}{2} +
        \frac{1}{5 \left(1+x^2\right)}-\frac{2}{1 + 10 (x-2)^2}\,,
    \end{align}
    consisting of two Lorentzian lenses. For $y=1/10$, the Kirchhoff-Fresnel integral has four relevant critical points (see the left panel of \cref{fig:doubleLorentzian}). Critical points $3$, $4$ and $9$ are real (classical rays) and consequently relevant. Critical point $2$ is the only relevant complex critical point. When evaluating the adjacency graph in the Borel plane, we remark that critical point $7$ lies to the right of the original integration domain. Its irrelevance becomes only apparent when we note that critical point $3$ is not adjacent to critical point $7$. Critical points $7$ and $9$ are adjacent. However, the associated edge in the Borel plane does not lie in the South-East direction and does not lead to a Stokes phenomenon when rotating from $\theta = \pi/2$ to $\theta = 0$. It turns out that this phenomenon is increasingly common for multidimensional integrals when the geometry of the function $f$ tends to become more intricate. The phenomenon, for example, occurs in the two-dimensional lens integral 
    \begin{align}
        \Psi(\bm{y}) = \frac{k}{2\pi i}\int_{\mathbb{R}^2}e^{i \omega \left(\frac{(\bm{x}-\bm{y})^2}{2} + \frac{2}{1+x^2+2y^2}\right)}\mathrm{d}x\mathrm{d}y\,,
    \end{align}
    with $\bm{x}=(x,y)$.

    \subsection{The swallowtail integral (\texorpdfstring{$A_4$}{})}

    We now turn to the case of an unbounded potential, the swallowtail integral 
    \begin{align}
        \Psi(x,y,z) = \sqrt{k} \int_{-\infty}^\infty e^{i k f(u)}\mathrm{d}u\,,
    \end{align}
    with the exponential 
    \begin{align}
        f(u) = u^5 + z u^3 + y u^2 + x u\,.
    \end{align}
    This integral has $4$ critical points solving the quartic equation 
    \begin{align}
        5 u^4 + 3z u^2 + 2 y u + x =0\,.
    \end{align}

\begin{figure}
        \begin{subfigure}[b]{0.5 \textwidth}
            \begin{tikzpicture}
                \node (img) {\includegraphics[width=\textwidth]{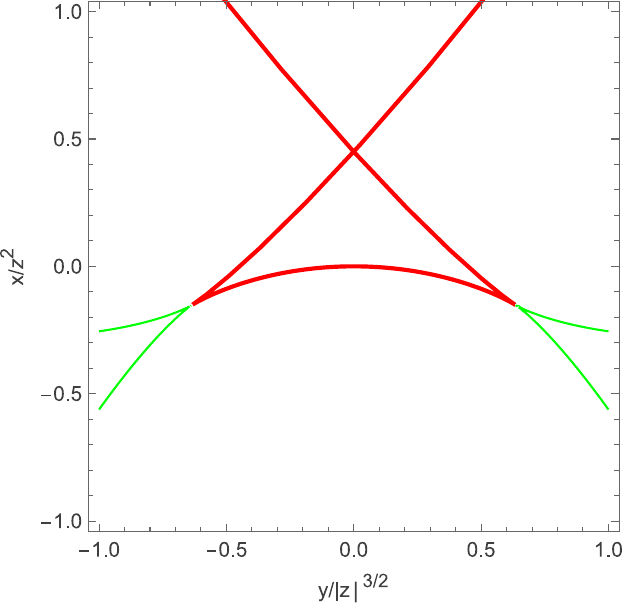}};
                \node[scale=2] at (0.55,-1.5) {$\Circled{1}$};
                \node[scale=2] at (0.55, 1.1) {$\Circled{2}$};
                \node[scale=2] at (0.55, 3.3) {$\Circled{3}$};
                \node[scale=2] at (-1., 2.2) {$\Circled{4}$};
                \node[scale=2] at (2.05, 2.2) {$\Circled{4}$};
                \node[scale=2] at (-2.7, -0.6) {$\Circled{5}$};
                \node[scale=2] at (3.8, -0.6) {$\Circled{5}$};
            \end{tikzpicture}
        \end{subfigure}
        \caption{The unfolding of the swallowtail caustic for $z=-1$, with the fold caustic line (the red curve) and the Stokes line (the green curve). The caustic and Stokes lines partition the external parameter space into regions with different sets of relevant critical points. \textit{Region 1:} Three relevant critical points. Two real and one complex. \textit{Region 2:} Four relevant real critical points. \textit{Region 3:} Two relevant complex critical points. \textit{Region 4:} Three relevant critical points. Two real and one complex. \textit{Region 5:} Two relevant real critical points.}\label{fig:swallowtail_unfolding}
        \centering
        \begin{subfigure}[B]{0.45\textwidth}
            \resizebox{\textwidth}{!}{%
            \begin{tikzpicture}
                \node[anchor=south west, inner sep=0pt, outer sep=0pt] (image) at (0,0) {\includegraphics{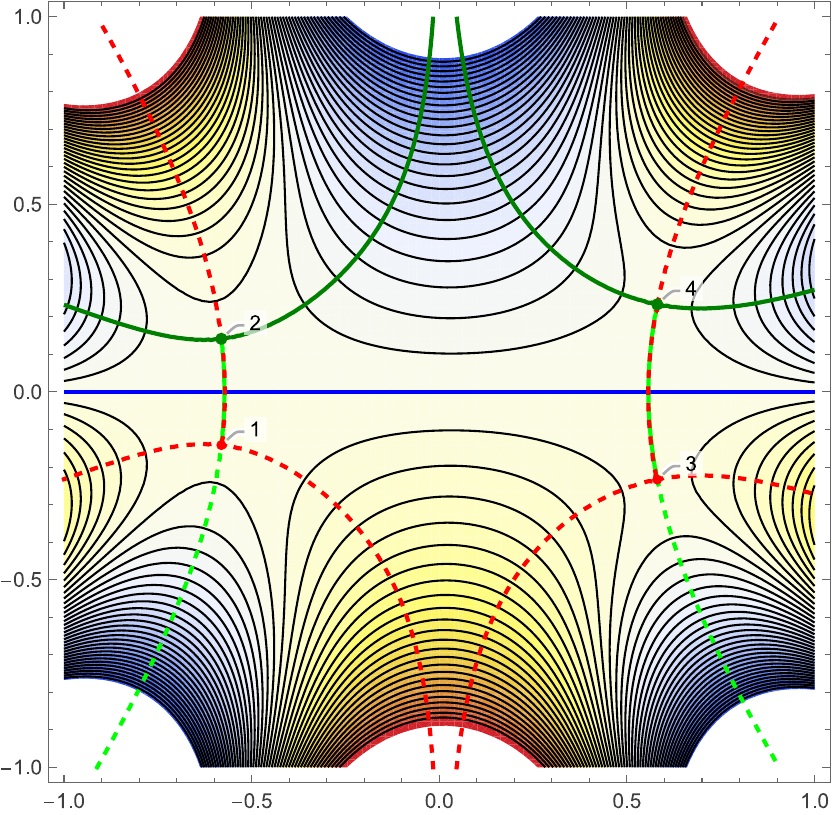}};
                \node[below right, font=\huge] at (image.south) {Re$[x]$};
                \node[above left, font=\huge, rotate = 90] at (image.west)  {Im$[x]$};
            \end{tikzpicture}
            }
            \caption{The complex $u$-plane.}
        \end{subfigure}
        \begin{subfigure}[b]{0.45\textwidth}
            \resizebox{\linewidth}{!}{%
                \begin{tikzpicture}[thick]
                    \draw[thick, black] (-2, -2) rectangle (2, 2);  
                    \draw[->] (-2, 0) -- (2, 0) node[above left=0.1] {$\Re(t)$};
                    \draw[->] (0, -2) -- (0, 2) node[below left=0.1] {$\Im(t)$};

                    \coordinate (A) at (-0.5,1);
                    \coordinate (B) at (+0.5,1);
                    \coordinate (C) at (-1,-1);
                    \coordinate (D) at (+1,-1);

                    
                    \draw[blue, ultra thick] (0, -2) -- (0,2);
                    \draw[red] (A) -- (B);
                    \draw[red] (A) -- (C);
                    \draw[red] (A) -- (D);
                    \draw[red] (B) -- (C);
                    \draw[red] (B) -- (D);
                    \draw[red] (C) -- (D);
                    
                    \draw[red, dashed] (A) -- (-0.5, 2);
                    \draw[red, dashed] (B) -- (+0.5, 2);
                    
                    \fill[black](A) circle (2pt) node [left, black] {$1$}; 
                    \fill[ForestGreen](B) circle (2pt) node [right, black] {$2$}; 
                    \fill[black](C) circle (2pt) node [left, black] {$3$}; 
                    \fill[ForestGreen](D) circle (2pt) node [right, black] {$4$}; 
                \end{tikzpicture}
            }
            \caption{The complex $t$-plane}
        \end{subfigure}
        \caption{The regulated swallowtail diffraction integral \eqref{eq:swallowtail-reg} in region $3$: the Picard-Lefschetz deformation in the complex $x$-plane (the left panel) and the associated adjacency graph in the complex $t$-plane with the adjacency relations to the auxiliary critical points represented with the red dashed edges (the right panel). We follow the notation outlined in \cref{fig:Pearcey}.}\label{fig:Swallowtail_unfolding_Borel}
    \end{figure}

    The relevant critical points vary as a function of the external parameters $x, y$ and $z$. In \cref{fig:swallowtail_unfolding}, we illustrate the relevant regions for $z=-1$.

    The exponent for the swallowtail integral is not bounded, making the rotation $\Psi_\theta$ diverge for $\theta \neq 0$. To analyse the swallowtail integral, we work with the regulated integral 
    \begin{align}\label{eq:swallowtail-reg}
        \Psi(x,y,z) = \sqrt{k} \lim_{L\to \infty} \int_{-\infty}^\infty e^{ik\left(f(u) + \frac{u^6}{L}\right)}\mathrm{d}u\,.
    \end{align}
    The regulator adds a fifth real critical point, playing the role of a global minimum. As an illustration, we consider region $3$, where we find two relevant complex critical points. As it turns out, this auxiliary critical point is adjacent to critical points $1$ and $2$ (see the right panel of \cref{fig:Swallowtail_unfolding_Borel}). An analysis of the adjacency graph explains the relevance of critical points $2$ and $4$: The auxiliary critical point makes critical point $2$ relevant at $\theta=\pi/2$. While rotating back to $\theta=0$, critical points $2$ and $4$ undergo a Stokes transition, making critical point $4$ relevant to the integral.
    
    \subsection{The hyperbolic integral (\texorpdfstring{$D_4^+$}{})}
    
\begin{figure}
        \begin{subfigure}[b]{0.45\textwidth}
            \begin{tikzpicture}
                \node (img) {\includegraphics[width=\textwidth]{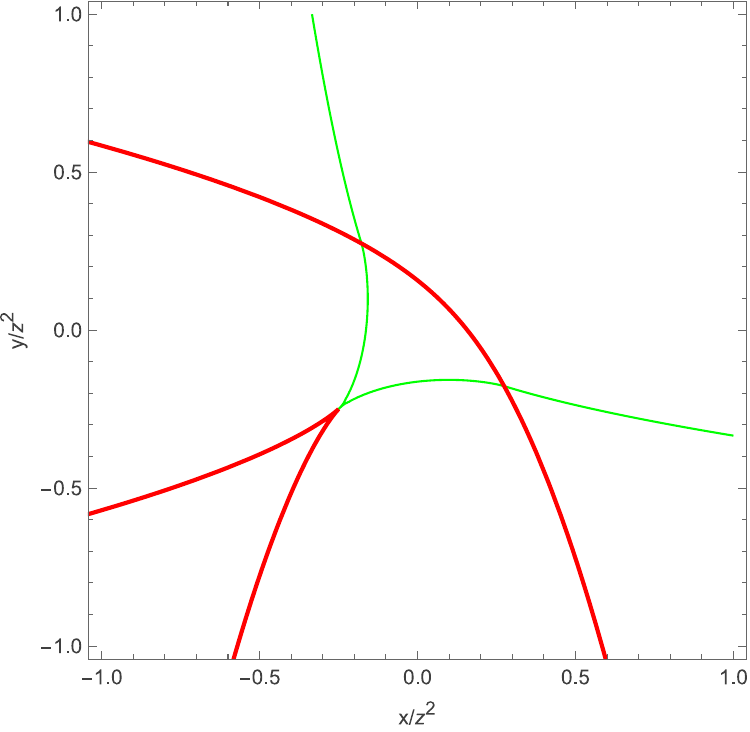}};
                \node[scale=2] at (0.3,0.3) {$\Circled{1}$};
                \node[scale=2] at (0.3,-2) {$\Circled{2}$};
                \node[scale=2] at (-1.7,0.3) {$\Circled{2}$};
                \node[scale=2] at (3,3) {$\Circled{3}$};
                \node[scale=2] at (3,-2) {$\Circled{4}$};
                \node[scale=2] at (-1.7,3) {$\Circled{4}$};
            \end{tikzpicture}
            
        \end{subfigure}
        \caption{The unfolding of the hyperbolic umbilic caustic, with the fold caustic (the red curve) and the Stokes line (the green curve). The caustic and Stokes lines partition the external parameter space into regions with different sets of relevant critical points. \textit{Region 1:} Two real relevant critical points. \textit{Region 2:} Three relevant critical points. Two real and one complex. \textit{Region 3:} One relevant complex critical point. \textit{Region 4:} Two relevant complex critical points.}\label{fig:hyperbolic_unfolding}
    \centering
        \begin{subfigure}[b]{0.24\textwidth}
            \resizebox{\linewidth}{!}{%
                \begin{tikzpicture}[thick]
                    \draw[thick, black] (-2, -2) rectangle (2, 2);  
                    \draw[->] (-2, 0) -- (2, 0) node[above left=0.1] {$\Re(t)$};
                    \draw[->] (0, -2) -- (0, 2) node[below left=0.1] {$\Im(t)$};

                    \coordinate (A) at (0,-0.2);
                    \coordinate (B) at (0,-1.7);
                    \coordinate (C) at (-0.5,-0.5);
                    \coordinate (D) at (+0.5,-0.5);

                    \draw[blue, ultra thick] (0, -2) -- (0, 2);

                    \draw[red] (A) -- (B);
                    \draw[red] (A) -- (C);
                    \draw[red] (A) -- (D);

                    \draw[red,dashed] (A) -- (0, 2);
                    \draw[red,dashed] (C) -- (-0.5, 2);
                    \draw[red,dashed] (D) -- (+0.5, 2);
                    
                    \fill[ForestGreen](A) circle (2pt) node [right, black] {$1$}; 
                    \fill[ForestGreen](B) circle (2pt) node [right, black] {$2$}; 
                    \fill[black](C) circle (2pt) node [left, black] {$3$}; 
                    \fill[black](D) circle (2pt) node [right, black] {$4$}; 
                \end{tikzpicture}
            }
            \caption{Region $1$}
        \end{subfigure}
        \begin{subfigure}[b]{0.24\textwidth}
            \resizebox{\linewidth}{!}{%
                \begin{tikzpicture}[thick]
                    \draw[thick, black] (-2, -2) rectangle (2, 2);  
                    \draw[->] (-2, 0) -- (2, 0) node[above left=0.1] {$\Re(t)$};
                    \draw[->] (0, -2) -- (0, 2) node[below left=0.1] {$\Im(t)$};

                    \coordinate (A) at (0,-1);
                    \coordinate (B) at (0,-1.7);
                    \coordinate (C) at (-0.5,+0.5);
                    \coordinate (D) at (+0.5,+0.5);

                    \draw[blue, ultra thick] (0, -2) -- (0, 2);

                    \draw[red] (A) -- (B);
                    \draw[red] (A) -- (C);
                    \draw[red] (A) -- (D);
                    \draw[red] (C) -- (D);

                    \draw[red, dashed] (A) -- (0, 2);
                    \draw[red, dashed] (C) -- (-0.5, 2);
                    \draw[red, dashed] (D) -- (+0.5, 2);
                    
                    \fill[ForestGreen](A) circle (2pt) node [right, black] {$1$}; 
                    \fill[ForestGreen](B) circle (2pt) node [right, black] {$2$}; 
                    \fill[black](C) circle (2pt) node [left, black] {$3$}; 
                    \fill[ForestGreen](D) circle (2pt) node [right, black] {$4$}; 
                \end{tikzpicture}
            }
            \caption{Region $2$}
        \end{subfigure}
        \begin{subfigure}[b]{0.24\textwidth}
            \resizebox{\linewidth}{!}{%
                \begin{tikzpicture}[thick]
                    \draw[thick, black] (-2, -2) rectangle (2, 2);  
                    \draw[->] (-2, 0) -- (2, 0) node[above left=0.1] {$\Re(t)$};
                    \draw[->] (0, -2) -- (0, 2) node[below left=0.1] {$\Im(t)$};

                    \coordinate (A) at (-0.4,-1.6);
                    \coordinate (B) at (0.4,-1.6);
                    \coordinate (C) at (-1.4,+1.1);
                    \coordinate (D) at (+1.4,+1.1);

                    \draw[blue, ultra thick] (0, -2) -- (0, 2);

                    \draw[red] (A) -- (C);
                    \draw[red] (A) -- (D);
                    \draw[red] (B) -- (C);
                    \draw[red] (B) -- (D);
                    \draw[red] (C) -- (D);

                    \draw[red, dashed] (C) -- (-1.4, 2);
                    \draw[red, dashed] (D) -- (+1.4, 2);
                    
                    \fill[black](A) circle (2pt) node [left, black] {$3$}; 
                    \fill[black](B) circle (2pt) node [right, black] {$4$}; 
                    \fill[black](C) circle (2pt) node [left, black] {$1$}; 
                    \fill[ForestGreen](D) circle (2pt) node [right, black] {$2$}; 
                \end{tikzpicture}
            }
            \caption{Region $3$}
        \end{subfigure}
        \begin{subfigure}[b]{0.24\textwidth}
            \resizebox{\linewidth}{!}{%
                \begin{tikzpicture}[thick]
                    \draw[thick, black] (-2, -2) rectangle (2, 2);  
                    \draw[->] (-2, 0) -- (2, 0) node[above left=0.1] {$\Re(t)$};
                    \draw[->] (0, -2) -- (0, 2) node[below left=0.1] {$\Im(t)$};

                    \coordinate (A) at (-0.4,-1.1);
                    \coordinate (B) at (0.4,-1.1);
                    \coordinate (C) at (-1.3,+0.8);
                    \coordinate (D) at (+1.3,+0.8);

                    \draw[blue, ultra thick] (0, -2) -- (0, 2);

                    \draw[red] (A) -- (B);
                    \draw[red] (A) -- (C);
                    \draw[red] (B) -- (D);

                    \draw[red, dashed] (A) -- (-0.4, 2);
                    \draw[red, dashed] (B) -- (+0.4, 2);
                    \draw[red, dashed] (C) -- (-1.3, 2);
                    \draw[red, dashed] (D) -- (+1.3, 2);
                    
                    \fill[black](A) circle (2pt) node [left, black] {$1$}; 
                    \fill[ForestGreen](B) circle (2pt) node [right, black] {$2$}; 
                    \fill[black](C) circle (2pt) node [left, black] {$3$}; 
                    \fill[ForestGreen](D) circle (2pt) node [right, black] {$4$}; 
                \end{tikzpicture}
            }
            \caption{Region $4$}
        \end{subfigure}
        \caption{The adjacency graph (the red graph) in the complex $t$-plane of the unfolding of the hyperbolic umbilic diffraction integral in the four regions outlined in \cref{fig:hyperbolic_unfolding}. We follow the notation outlined in \cref{fig:Pearcey}.}\label{fig:hyperoblic_unfolding_Borel}

\centering
    \begin{subfigure}[b]{0.24\textwidth}
    \includegraphics[width=\textwidth]{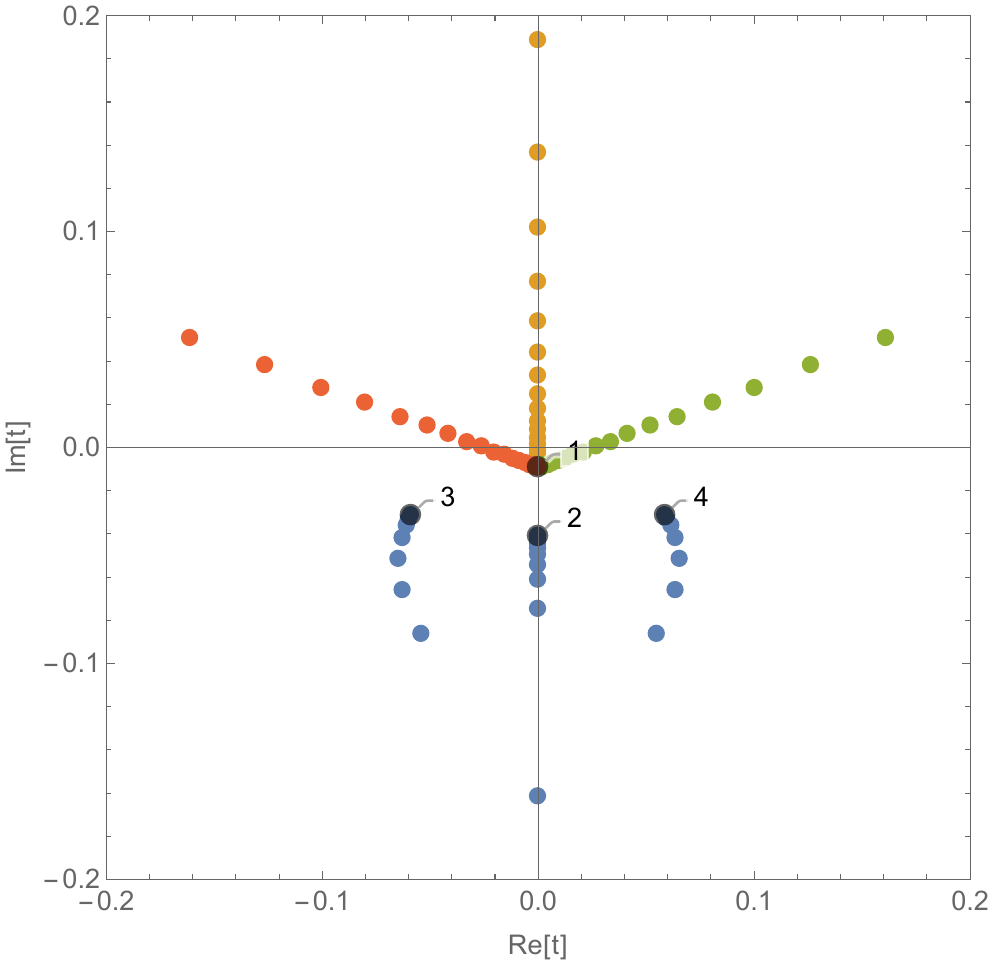}
            \caption{Region 1}
    \end{subfigure}
    \begin{subfigure}[b]{0.24\textwidth}
    \includegraphics[width=\textwidth]{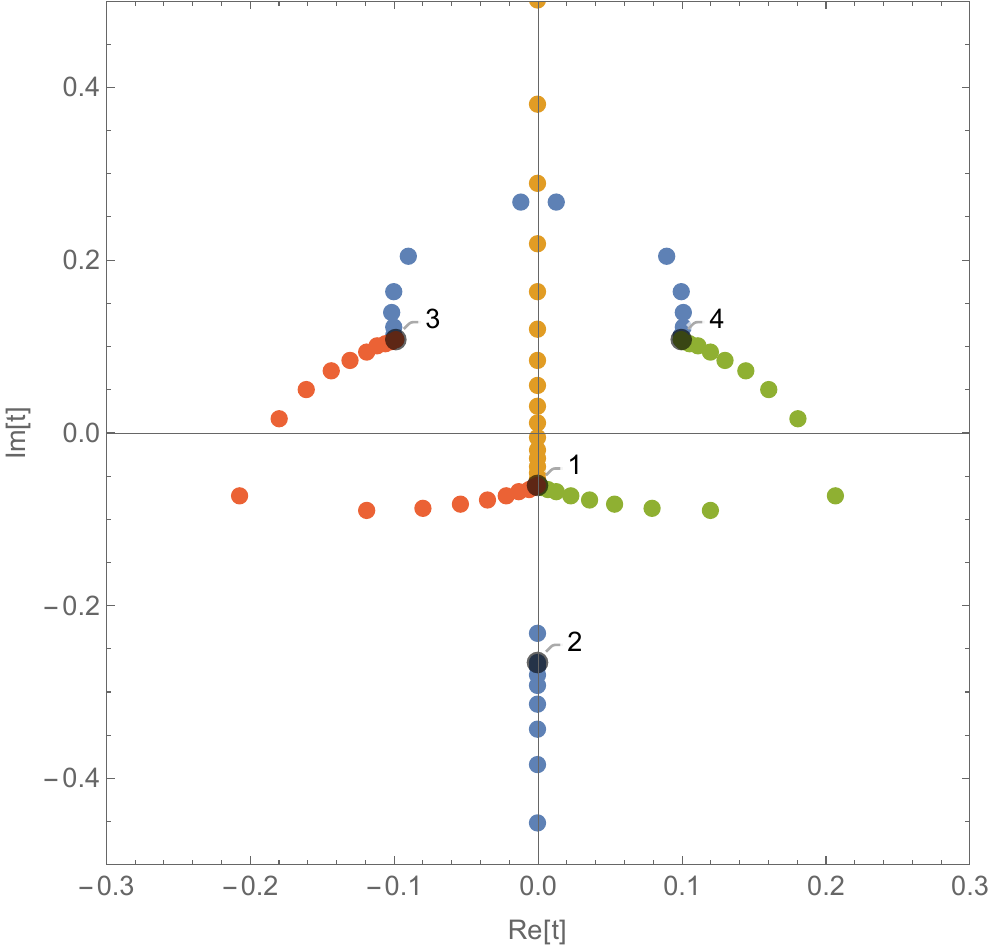}
            \caption{Region 2}
    \end{subfigure}
    \begin{subfigure}[b]{0.24\textwidth}
    \includegraphics[width=\textwidth]{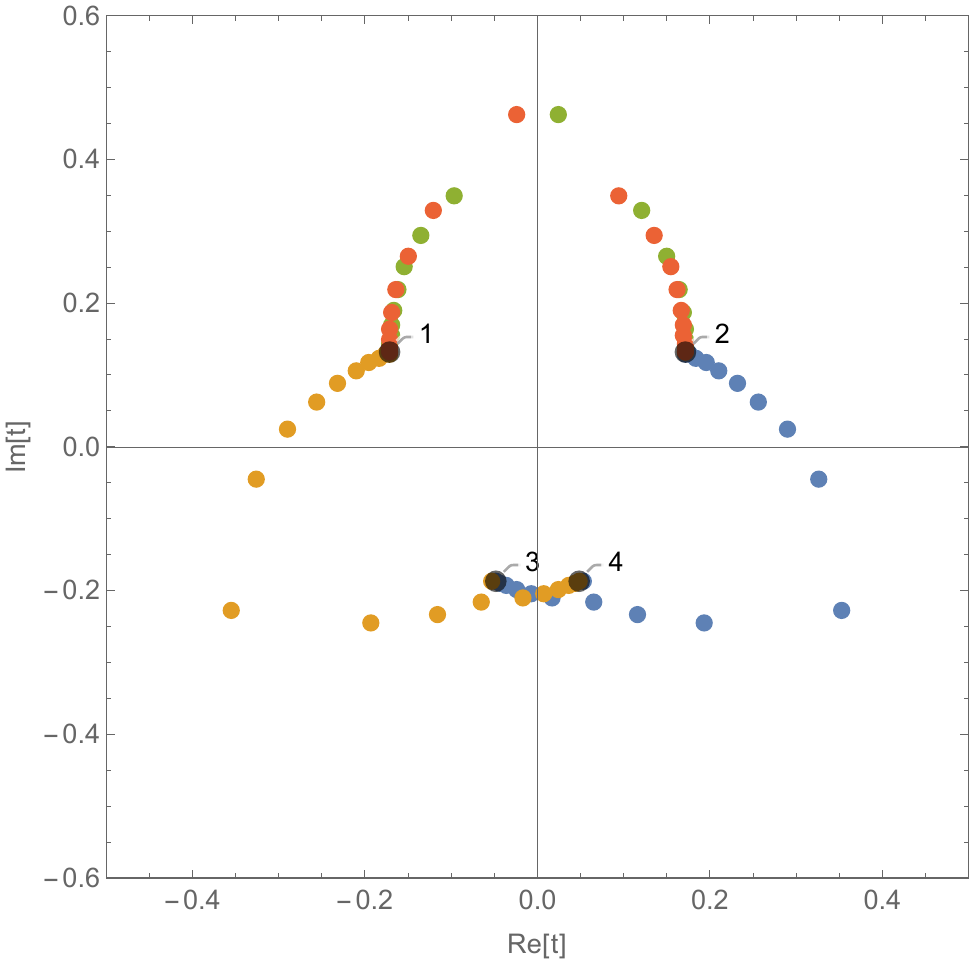}
            \caption{Region 3}
    \end{subfigure}
    \begin{subfigure}[b]{0.24\textwidth}
    \includegraphics[width=\textwidth]{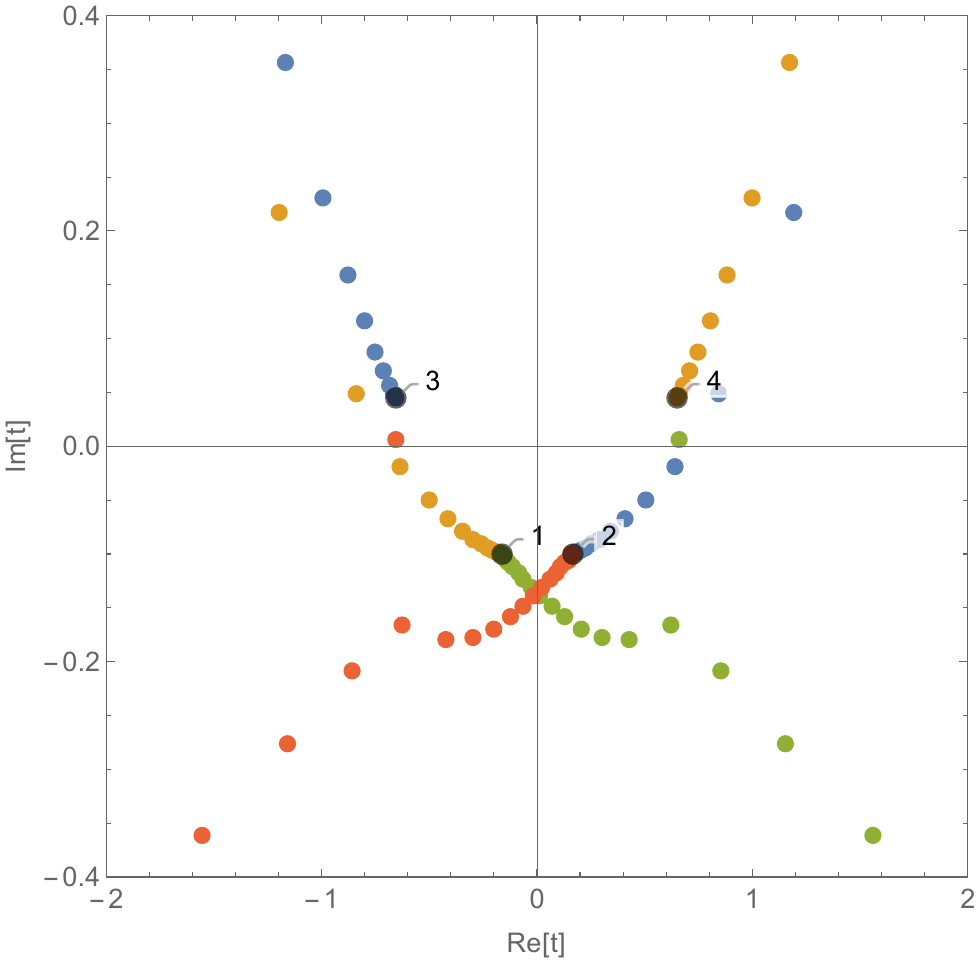}
            \caption{Region 4}
    \end{subfigure}
        \caption{Borel-Pad\'{e} prediction of the adjacency graph in the complex $t$-plane of the unfolding of the hyperbolic umbilic diffraction integral in the four regions outlined in \cref{fig:hyperbolic_unfolding}. We followed the procedure outlined in \cref{sec:Adjacency}. The values of the external parameters used were $z=1$ and: region $1$ $x=1/10,\,y=-1/10$, region $2$ $x=0,\,y=-1/2$, region $3$ $x=1/2,\,y=2/5$ and region $4$ $x=1,\,y=-1/2$.}\label{fig:hyperbolic_pade}
    \end{figure}

    Let us consider the two-dimensional representation of the hyperbolic integral
    \begin{align}\label{eq:hyperbolic-integral2d}
        \Psi(x,y, z) = k \int_{\mathbb{R}^2} e^{i k f(u,v)}\mathrm{d}u \mathrm{d}v\,,
    \end{align}
    with the exponential
    \begin{align}\label{eq:hyperbolic-potential}
        f(u,v) = u^3+v^3+x u + y v + z u v\,.
    \end{align}
    The hyperbolic integral has four critical points $(s_c,t_c)$ satisfying the equations 
    \begin{align}
        3 u^2 + x + z v & = 0\,,\\
        3 v^2 + y + z u & = 0\,.
    \end{align}
    The relevant critical points vary as a function of the external parameters $x, y$ and $z$. An illustration of the relevant regions can be found in \cref{fig:hyperbolic_unfolding} for $z=1$. We can see the fold caustic (in red), where two of the critical points coalesce, and Stokes lines (in green) where one expects adjacency graphs to change.

    We can study the adjacency graphs in each of the four regions shown in \cref{fig:hyperbolic_unfolding}, and this is summarised in \cref{fig:hyperoblic_unfolding_Borel}, where adjacency lines between critical points are shown in red,  In region $1$ we find $2$ real critical points that are mapped to two branch point singularities on the imaginary axes (labelled $1$ and $2$), and which are adjacent to each other, and two complex critical points mapped to the complex Borel plane ($3$ and $4$). These complex critical points are adjacent only to the top real critical point ($1$). To cross from region $1$ to region $2$ the branch points associated to critical points $3$ and $4$ move upwards, becoming collinear to points $1$ and $2$ when we cross a Stokes line. Because of this collinearity the Stokes transition turns on the adjacency between the two complex critical points ($3$ and $4$) through a process called \textit{higher order Stokes phenomenon} \cite{Howls:2004}. We can reach region $3$ by crossing the fold caustic line, at which point points $1$ and $2$ move upwards, coalesce and then become complex critical points. Thus in region $3$ all critical points are mapped to complex branch points in the Borel plane. Critical points $1$ and $2$ are adjacent to all others, while critical points $3$ and $4$ stop being adjacent to each other. Finally, to get to region $4$ from region $3$ branch points associated to critical points $1$ and $3$ move downwards while $3$ and $4$ move upwards. They all become collinear when we cross the Stokes line between the two regions, after which the adjacency between $1$ and $4$ as well as between $2$ and $3$ get turned off.

    The adjacency graphs sketched in \cref{fig:hyperoblic_unfolding_Borel} have been determined via the calculation of Stokes constants and large order relations as described in \cref{sec:Adjacency} and the Pad\'{e} prediction of the adjacency graph can be found in \cref{fig:hyperbolic_pade} for various values of the external parameters covering the four regions.

    Following \cite{Berry:1990b}, the hyperbolic integral \eqref{eq:hyperbolic-potential} can also be expressed as a one-dimensional integral, upon completing the square and performing a Gaussian integral:
    \begin{align}
        \Psi(x,y,z) 
        &=  \sqrt{\frac{8 \pi k }{3}} e^{i k \left[\frac{1}{6}  (x+y)z+\frac{z^3}{27}\right]+\frac{i \pi}{4}}
        \int_{\infty \exp \left(\frac{5 \pi  i}{12}-\frac{\text{arg} (k)}{6}\right)}^{\infty \exp \left(\frac{\pi  i}{12} - \frac{\text{arg}(k)}{6}\right)} e^{
            i k\left[2 u^6+2 z u^4+  \left(\frac{1}{2}z^2+x+y\right)u^2-\frac{ (x-y)^2}{24  u^2}\right]} \mathrm{d}u\,.
    \end{align}
    The same adjacency relations can be obtained by analysing the steepest descent contours for this one-dimensional representation. However, each critical point of the two-dimensional integral is mapped to 2 separate critical points due to symmetry, and both need to be taken into consideration when determining the adjacency graphs. There are critical points that one would na\"{i}vely consider as adjacent but where contributions from the two symmetric points in fact cancel, resulting in a Stokes constant that is zero (not adjacent).

    We now turn to analyse the relevance of critical points in each region. First, notice that the exponent for the hyperbolic integral is not bounded, and to make the rotation $\Psi_\theta$ converge for $\theta \neq 0$ we need to introduce the regulated integral (analogous to the swallowtail example above):
    \begin{align}
        \Psi(x,y,z)  = k \lim_{L\to \infty} \int_{\mathbb{R}^2} e^{i k\left( f(u,v) +\frac{u^4+v^4}{L}\right)}\mathrm{d}u \mathrm{d}v\,.
    \end{align}
    This regulator introduces 5 extra critical points, one of which will effectively work as a global minimum (depending on the external parameters), and the other are auxiliary, playing an irrelevant role. Consider region $1$ in left panel of \cref{fig:hyperoblic_unfolding_Borel}: we have two real critical points $1$ and $2$ (mapped to the imaginary axis in the Borel plane) and the critical point $4$ is both adjacent to $1$ and within its "South-East" area in the Borel plane, so we could be lead to believe that $1$ would make $4$ relevant when rotation from $\theta=\pi/2$ back to $\theta=0$. However, the auxiliary critical point is also adjacent to $4$, and consequently, during the rotation $\theta=\pi/2$ to $\theta=0$, the auxiliary critical point turns on $4$ and then the critical point $1$ turns $4$ back off, making it irrelevant. Now consider region $2$ instead. From \cref{fig:hyperoblic_unfolding_Borel}, second panel, we can see that critical point $4$ is not in the "South-East" area of the relevant real saddles $1$ and $2$. It is nevertheless adjacent to the auxiliary critical point and gets turned on when we rotate from $\theta=\pi/2$ to $\theta=0$ -- in this region, critical point $4$ is relevant. In regions $3$ and $4$ (as right panels of \cref{fig:hyperoblic_unfolding_Borel}), all critical points are complex, and the relevant ones are dictated solely by the adjacency to the auxiliary critical point: critical point $2$ in region $3$ and critical points $2$ and $4$ in region $4$.

   In \cref{fig:hyperbolicEvaluated}, we compare the numerical results with the \textit{resummed} asymptotic transseries approximation of the hyperbolic integral \eqref{eq:hyperbolic-integral2d} with potential \eqref{eq:hyperbolic-potential}, computed in regions 1 and 2. Region 1 has only two relevant real critical points, both included in the plot. In region 2 we present the asymptotic approximation both with and without the relevant (complex) critical point identified by the South-East rule. The excellent improvement in agreement obtained when contributions from this critical point are included, especially for relatively small values of the asymptotic parameter $|k|$, provides strong evidence that the South-East rule correctly identifies the relevant critical points. The resummation of each transseries sector associated to the relevant critical points was performed using standard Borel-Pad\'{e} resummation methods (see \cite{Aniceto:2018bis} for further details).
    
    \section{Implications for Path Integrals and Wick rotations}\label{sec:Discussion}
    
    The theory of resurgence is gaining increasing application within the field of mathematical physics. Standard perturbation theory in quantum mechanics and quantum field theory leads to formal power series in the coupling constants that typically diverge. Using resurgence, one may resum these asymptotic series and find exponential terms encoding non-perturbative effects that traditional perturbation theory misses. In particular, resurgence has found applications in the study of  energy spectra, analytic behaviour of quasi-normal modes, novel non-perturbative phenomena and phase transitions, see for example \cite{Dunne:2012, Basar:2017, Serone:2017, Marino:2019wra,Fujimori:2021oqg, Bajnok:2022rtu,Schiappa:2023ned,Spaendonck:2024,Aniceto:2026zmc}. However, we propose that the connection between adjacency and relevance described above establishes a fundamentally new avenue for the application of resurgence in theoretical physics.

    The Feynman path integral expresses quantum evolution as a highly oscillatory functional integral \cite{Feynman:1948, Feynman:1965}. The transition amplitude between two states is expressed as a sum over histories. For example, in quantum mechanics, the Green's function of a particle in a potential $V$ evolving from $x_0$ to $x_1$ in time $T$ is given by the path integral 
    \begin{align}
        G(x_1,x_0,T) = \int_{x(0)=x_0}^{x(T)=x_1} e^{i S[x]/\hbar}\mathcal{D}x\,,
    \end{align} 
    with the action 
    \begin{align}
        S[x] = \int_0^T \left[\frac{1}{2} m \dot{x}(t)^2 - V(x(t))\right]\mathrm{d}t\,,
    \end{align}
    the mass of the particle $m$ and the reduced Planck constant $\hbar$. Evolution arises through constructive interference \cite{Feynman:1985} around the critical points
    \begin{align}
        \frac{\delta S}{\delta x(t)} = 0:\quad m \ddot{x}(t) = -V''(x(t))\,.
    \end{align} 
    Indeed, in the Wentzel–Kramers–Brillouin (WKB) approximation \cite{Wentzel1926, Kramers1926, Brillouin1926}, the path integral is approximated by a sum over the relevant (complex) classical paths
    \begin{align}
        G(x_1,x_0,T) \approx \Theta(T) \sqrt{\frac{i}{2 \pi \hbar}} \sum_{j} n_j \sqrt{\frac{\partial^2 S[x_j]}{\partial x_0 \partial x_1}} e^{i S[x_j]/\hbar}\,,
    \end{align}
    where the classical path $x_j$ is a solution to the classical boundary value problem. The complex classical paths, sometimes known as instantons, have elucidated many quantum phenomena \cite{Coleman:1978ae} and may be the key to a rigorous definition of the real-time path integral \cite{Feldbrugge:2023}. Yet while real classical paths always contribute, finding the relevant complex classical paths has remained a fundamental problem in real-time quantum physics. In practice, we can only use Picard-Lefschetz theory to identify the relevant instantons for a limited set of problems. See, for example, the first application of Picard-Lefschetz theory to the path integral for gravity \cite{Feldbrugge:2017}. In this example, the path integral is Gaussian, allowing the propagator to be expressed as a one-dimensional integral over the lapse. In more general theories, the relevant classical paths can be inferred from a numerical evaluation of the real-time path integral \cite{Feldbrugge:2023arXiv230912427F, Feldbrugge:2026Step}.

    When interpreting the path integral as the limit of the time-discretised path integral \cite{Feynman:1965}, the South-East rule solves this problem. Expanding the path integral around the classical paths to higher order gives rise to the transseries expansion 
        \begin{align}
        G(x_1,x_0,T) \sim \Theta(T) \sqrt{\frac{i}{2 \pi \hbar}} \sum_{j} n_j  e^{i S[x_j]/\hbar} \sum_{m=0}^\infty T^{(j)}_m \hbar^m\,,
    \end{align}
    with coefficients $T^{(j)}_m$. Analogous to the finite-dimensional integrals, this expansion is formal in nature, as the sums $\sum_{m=0}^\infty T_m^{(j)}\hbar^m$ generally diverge. The way this divergence occurs tells us about the adjacency relations between the classical paths through the resurgence relations. Finally, the South-East rule translates these adjacency relations into the relevance of the complex classical paths. In a future paper, we will perform this analysis in the context of quantum mechanics. 
    
    Note that when the potential diverges to $+\infty$ for any $x$, the action is unbounded. The identification of the relevant complex classical paths through the South-East rule requires the smooth regulator discussed above. Alternatively, when the potential is bounded from below but diverging -- like, for example, the famous double-well potential $V(x)=(x^2-1)^2$ -- we propose to identify the relevant complex classical paths with a \textit{smooth} Wick rotation $t \mapsto e^{-i \vartheta}t$, like proposed in \cite{Cherman:2014}. Under this deformation, the path integral can be written as
    \begin{align}
        G_\vartheta(x_1,x_0;T) = \int_{x(0)=x_0}^{x(T)=x_1} e^{iS_\vartheta/\hbar}\mathcal{D}x\,,
    \end{align}
    with the generalised action 
    \begin{align}
        i S_{\vartheta} = ie^{i \vartheta} \int_{0}^T \left[\frac{1}{2} m \dot{x}^2(t) - e^{-i2 \vartheta} V(x(t))\right]\mathrm{d}t\,.
    \end{align}
    This deformation transforms the real-time path integral, at $\vartheta=0$, into the Euclidean path integral, at $\vartheta = \pi/2$, with Euclidean action $iS_{\vartheta = \pi/2} = -S_E$ where
    \begin{align}
        S_E = \int_{0}^T \left[\frac{1}{2} m \dot{x}^2(t) + V(x(t))\right]\mathrm{d}t\,.
    \end{align}
    Now, importantly, as the potential is real and bounded from below, so is the Euclidean action. The space of real paths interpolating between $x_0$ and $x_1$ is mapped to a horizontal line segment in the Borel plane. Consequently, the Euclidean path integral 
    \begin{align}
        G_E(x_1,x_0,T) = \int_{x(0)=x_0}^{x(T)=x_1} e^{-S_E/\hbar}\mathcal{D}x\,.\label{eq:Wick}
    \end{align}
    can only have relevant classical paths for which the action is real. Moreover, as the integral converges (as described by the Feynman-Kac formula), the original integration domain consists of a set of steepest descent manifolds. Consequently, only the real classical paths contribute (this is generally not true when the potential is unbounded). When rotating back to the real-time path integral, moving from $\vartheta = \pi/2$ to $\vartheta=0$, we propose that identifying the Stokes transitions governed by the adjacency relations allows us to identify the relevant complex classical paths of the real-time path integral. Some relevant classical paths of the Euclidean action may become irrelevant, and some irrelevant classical paths can become relevant while we rotate back. Note that the classical paths depend on the angle $\vartheta$, and the adjacency relations may change as a function of $\vartheta$ via higher-order Stokes phenomena \cite{Howls:2004}. We will fully explore this idea in a future publication.

    Finally, the Wick rotation famously cannot be applied to problems for which the potential $V$ is unbounded, as the Euclidean path integral diverges. This is known as the conformal factor problem in quantum gravity \cite{Gibbons:1978, Hawking:1979, Mazur:1990} and the sign problem in many-particle systems \cite{Loh:1990, Kieu:1994, Pan:2022}. The analysis in section \cref{sec:unbounded} demonstrates that a continuous deformation of an integral should not be restricted to the integrand but generally requires the simultaneous deformation of the integration contour. When the potential is not bounded from below, the Picard-Lefschetz deformation of the analytically continued path integral no longer coincides with the space of real-valued paths. The naive formulation of the Euclidean path integral diverges. When defining the Wick-rotated path integral along the relevant descent manifolds, the Euclidean path integral will remain finite and well-defined. As illustrated by the regulator, the Picard-Lefschetz deformation follows directly from the decision to define the conditionally convergent integral using analyticity (for more details, see \cite{Feldbrugge:2023}).

    \section{Conclusion}\label{sec:conclusion}
    In this paper, we have identified a direct link between the adjacency relations from resurgence theory and the intersection numbers of the Picard-Lefschetz representation of multidimensional highly oscillatory integrals over the real plane. The adjacency relations may be inferred from the diverging behaviour of the asymptotic series associated with the different critical points in the transseries of the integral. They provide a complete description of the topology of the integration cycles in the complex plane and govern whether a critical point is relevant or irrelevant to the integral. We propose an elementary algorithm to evaluate the intersection numbers from the geometry of the adjacency diagram in the Borel plane. The real critical points are always relevant. The relevance of complex critical points follows from the South-East rule introduced in\cref{sec:Relevance}. 
    
\begin{figure}
        \centering
        \begin{subfigure}[b]{0.49\textwidth}
            \includegraphics[width=\textwidth]{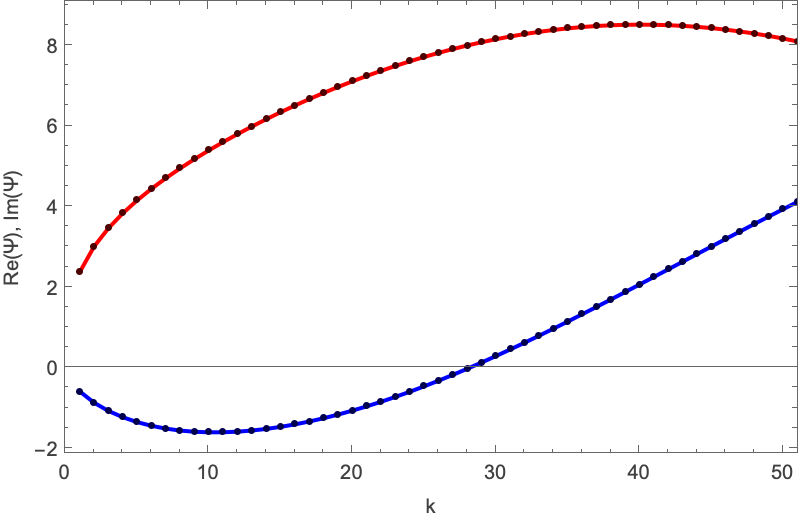}
            \caption{Region 1}
        \end{subfigure}
        \begin{subfigure}[b]{0.49\textwidth}
            \includegraphics[width=\textwidth]{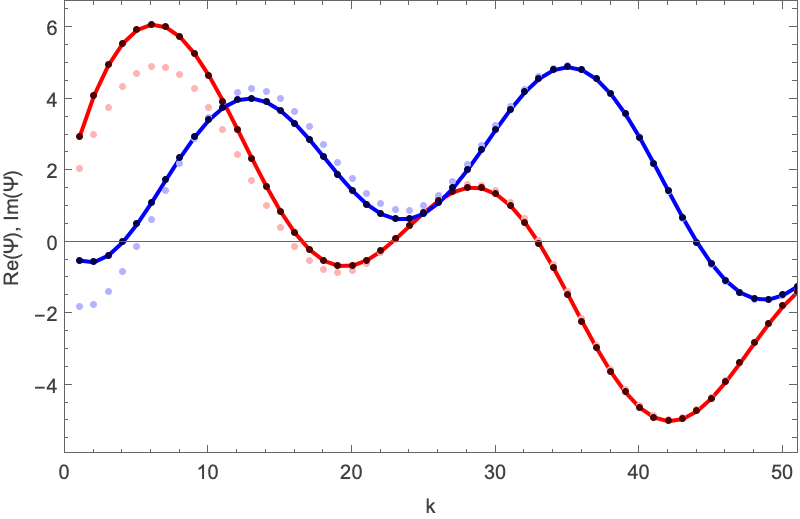}
            \caption{Region 2}
        \end{subfigure}
        \caption{The hyperbolic integral and the relevance of complex critical points. Left: The hyperbolic integral in region $1$ (see \cref{fig:hyperoblic_unfolding_Borel}) with $x = 1/10$, $y = -1/10$ and $z = 1$, as a function of $k$ with the red and blue curves being the real and imaginary parts of the hyperbolic integral. The dark red and blue points represent the integral along the steepest descent manifold associated to the two real relevant critical points. This is evaluated directly from the asymptotic series using the Borel-Pad\'e resummation method involving expansions about critical points 1 and 2. Right: The hyperbolic integral in region 2 (as in \cref{fig:hyperoblic_unfolding_Borel}) with $x = 0$, $y = -1/2$ and $z = 1$, as a function of $k$, with the red and blue curves being the real and imaginary parts of the hyperbolic integral. The dark red and blue points represent the integral along the steepest descent manifolds of the relevant critical points, two real (1 and 2) and one complex (4). The light red and blue points are the real and imaginary parts of the integral obtained when only including the integrals associated with the real critical points in the asymptotic approximation.}\label{fig:hyperbolicEvaluated}
    \end{figure}

    Using the South-East rule in combination with either the hyperasymptotic method or the Borel-Pad\'{e} resummation technique, we can recover highly oscillatory integrals without the need for numerical evaluations (see \cref{fig:hyperbolicEvaluated}). The South-East rule is crucial as it singles out the relevant critical points that contribute. This is a particularly exciting prospect for high-dimensional oscillatory integrals, where the numerical evaluation becomes increasingly expensive. Moreover, this is a significant step towards the identification of instanton contributions in real-time path integrals governing quantum evolution.
    
    The South-East rule and the resurgent approximation of the integral ultimately rely on the evaluation of the asymptotic series of the critical points. Dingle famously demonstrated that the asymptotic series can be expressed in terms of the derivatives of the exponent in the critical point \cite{Dingle:1973}. Recently, this calculation was generalized to multidimensional integrals, yielding an efficient algorithmic evaluation of asymptotic series used in this paper \cite{Feldbrugge:2026}. This method will be further analysed and extended in an upcoming paper. Another future direction we would like to pursue is to extend these techniques to oscillatory integrals with boundaries.

    We believe that the South-East rule has applications in many fields in mathematics and theoretical physics. We in particular expect that the here proposed method will be beneficial to the study of laser physics and lensing in wave optics. Ultimately, we hope that this method will impact the study of real-time path integrals in quantum physics. In an upcoming paper, we will verify the South-East rule for a set of non-trivial solvable models in quantum mechanics. If successful, this method may support new investigations in quantum mechanics, quantum field theory and quantum cosmology.
    
    \acknowledgments
    We thank Samuel Crew for introducing JF to resurgence theory and for many discussions on the interface of Picard-Lefschetz theory and resurgence theory. IA would like to thank Tm\'{a}s Reis for pointing out a simplification in the conformal maps used. IA and JF thank the International Centre for Mathematical Sciences at the University of Edinburgh for supporting this research with a Research-in-Groups grant.

    The work of JF is supported by the STFC Consolidated Grant ‘Particle Physics at the Higgs Centre,’ and a Higgs Fellowship at the University of Edinburgh. IA was supported in part by UKRI EPSRC Early Career Fellowship EP/S004076/1.

    \bibliographystyle{apsrev4-1}
    \bibliography{library.bib}

    \appendix

    \section{Asymptotic series of canonical diffraction integrals}\label{ap:asymptotic}

    For completeness, to help with the example calculations above, we explicitly evaluate the asymptotic series of the canonical diffraction integral associated with the seven elementary catastrophes. For an elementary introduction to catastrophe theory and its applications we refer to \cite{Saunders:1980, Poston:1996, Nye:1999}.

    \subsection{The canonical diffraction integrals of the \texorpdfstring{$A$}{}-type catastrophes}
    Starting with  canonical diffraction integral of the fold ($A_2$), cusp ($A_3$), swallowtail ($A_4$) and butterfly catastrophe ($A_5$) along the steepest descent manifold $\mathcal{J}_j$,
    \begin{align}
        \Psi^{(j)}_{A_n} = \sqrt{k} \int_{\mathcal{J}_j} e^{i k f_{A_n}(u)}\mathrm{d}u    
    \end{align}
    with the exponential
    \begin{align}
        f_{A_2}(u) &=  u^3 + x u\,,\\
        f_{A_3}(u) &=  u^4 + y u^2 + x u\,,\\
        f_{A_4}(u) &=  u^5 + z u^3 + y u^2 + x u\,,\\
        f_{A_5}(u) &=  u^6 + w u^4 + z u^3 + y u^2  + x u\,,
    \end{align}
    we expand around the critical point $u_j$,
    \begin{align}
        f_{A_2}(u_j + \Delta u) &= f_{A_2}(u_j) 
        + \frac{1}{2} f_{A_2}''(u_j)\Delta u^2 
        +\Delta u^3\,,\\
        f_{A_3}(u_j + \Delta u) &= f_{A_3}(u_j) 
        + \frac{1}{2} f_{A_3}''(u_j)\Delta u^2 
        + 4 u_j \Delta u^3 + \Delta u^4\,,\\
        f_{A_4}(u_j + \Delta u) &= f_{A_4}(u_j) 
        + \frac{1}{2} f_{A_4}''(u_j)\Delta u^2 
        + (10 u_j^2 + z)\Delta u^3
        + 5 u_j \Delta u^4
        + \Delta u^5\,,\\
        f_{A_5}(u_j + \Delta u) &= f_{A_5}(u_j) 
        + \frac{1}{2} f_{A_5}''(u_j)\Delta u^2 
        +(20 u_j^3+4 u_j w +z)\Delta u^3 + (15 u_j^2 + w)\Delta u^4 + 6 u_j \Delta u^5 + \Delta u^6\,,
    \end{align}
    with $u_j$ a root of the polynomial $f_{A_n}'(u)=0$.
    Upon expanding the cubic and higher-order terms in the integral, we obtain the asymptotic series
    \begin{align}
        \Psi_{A_2}^{(j)}
        &\sim
        e^{i k f_{A_2}(u_j)} 
        \sum_{m=0}^\infty
        \frac{(-1)^m}{(2m)! k^{m}}
        \int_{-\infty}^\infty e^{\frac{i }{2}f_{A_2}''(u_j) \Delta u^2 }
         \Delta u^{6m}
        \mathrm{d} \Delta u\,,\\
        \Psi^{(j)}_{A_3}   
        &\sim
        e^{i k f_{A_3}(u_j)} 
        \sum_{m=0}^\infty
        \frac{(ik)^m}{m!}
        \int_{-\infty}^\infty e^{\frac{i }{2}f_{A_3}''(u_j) \Delta u^2 }
         \left[4 u_j \frac{\Delta u^3}{k^{3/2}} + \frac{\Delta u^4}{k^2}\right]^m
        \mathrm{d} \Delta u\,,\\
        \Psi^{(j)}_{A_4}  
        &\sim
        e^{i k f_{A_4}(u_j)} 
        \sum_{m=0}^\infty
        \frac{(ik)^m}{m!}
        \int_{-\infty}^\infty e^{\frac{i }{2}f_{A_4}''(u_j) \Delta u^2 }
         \left[(10 u_j^2 + z)\frac{\Delta u^3}{k^{3/2}}
        + 5 u_j \frac{\Delta u^4}{k^2}
        + \frac{\Delta u^5}{k^{5/2}}\right]^m
        \mathrm{d} \Delta u\,,\\
        \Psi^{(j)}_{A_5}  
        &\sim
        e^{i k f_{A_5}(u_j)} 
        \sum_{m=0}^\infty
        \frac{(ik)^m}{m!}
        \int_{-\infty}^\infty e^{\frac{i }{2}f_{A_5}''(u_j) \Delta u^2 }
         \left[
            (20 u_j^3+4 u_j w +z)\frac{\Delta u^3}{k^{3/2}} + (15 u_j^2 + w)\frac{\Delta u^4}{k^2} + 6 u_j \frac{\Delta u^5}{k^{5/2}} + \frac{\Delta u^6}{k^3}
         \right]^m
        \mathrm{d} \Delta u\,,
    \end{align}
    which we write as $\Psi_{A_n}^{(j)} \sim e^{i k f(u_j)} \sum_{m=0}^\infty \frac{T_{A_n,m}^{(j)}}{k^m}$, where we evaluate the coefficients $T_{A_n,m}$ with the Gaussian integral
    \begin{align}
        \int_{-\infty}^\infty e^{\frac{i}{2} \lambda u^2}
        u^{p}\mathrm{d}u = 
        \begin{cases}
            \sqrt{\frac{2 \pi i}{\lambda}} \left(\frac{2 i}{ \lambda}\right)^{p/2}
            \frac{(p-1)!!}{2^{p/2}}
            & \text{if $p$ is even,}\\
            0 & \text{if $p$ is odd.}
        \end{cases}
        \label{eq:gauss}
    \end{align}
    For example, for the fold catastrophe, we obtain the coefficients
    \begin{align}
        T_{A_2,m}^{(j)} 
        &= 
        \frac{(-1)^{m}}{(2m)! }\frac{(6m-1)!!}{2^{3m}}
        \sqrt{\frac{2 \pi i}{f_{A_2}''(u_j)}} \left(\frac{2i}{f_{A_2}''(u_j)}\right)^{3m}\,.
    \end{align}

    \subsection{The canonical diffraction integrals of the \texorpdfstring{$D$}{}-type catastrophes}\label{ap:hyperbolic}
    Starting with the canonical diffraction integral of the the elliptic ($D_4^-$), the hyperbolic ($D_4^+$), and the parabolic umbilic ($D_5$) catastrophes along the steepest descent manifold $\mathcal{J}_j$,
    \begin{align}
        \Psi_{D_n}^{(j)} = k \int_{\mathcal{J}_j} e^{i k f_{D_n}(\bm{u})}\mathrm{d}\bm{u}\,,
    \end{align}
    with $\bm{u}=(u,v)$ and the exponentials
    \begin{align}
        f_{D_4^-}(\bm{u}) &= u^3 - 3 u v^2 + z(u^2 + v^2) + y v + x u\,,\\
        f_{D_4^+}(\bm{u}) &= u^3+v^3 + z u v + y v + x u\,,\\
        f_{D_5}(\bm{u})   &= u^4 + u v^2 + w u^2 + z v^2 + y v + x u \,,
    \end{align}
    we expand around the critical point $\bm{u}_j = (u_j,v_j)$ (solving $\nabla f_{D_n}(\bm{u}_j)=\bm{0}$) 
    \begin{align}
        f_{D_4^-}(\bm{u} + \Delta\bm{u}) &= f_{D_4^-}(\bm{u}_j) + 
        \frac{1}{2}\Delta\bm{u}^T \mathcal{H}f_{D_4^-}\Delta\bm{u}
        + \Delta u^3 - 3 \Delta u \Delta v^2\,,\\
        f_{D_4^+}(\bm{u} + \Delta\bm{u}) &= f_{D_4^+}(\bm{u}_j) + 
        \frac{1}{2}\Delta\bm{u}^T \mathcal{H}f_{D_4^+}\Delta\bm{u}
        + \Delta u^3 + \Delta v^3\,,\\
        f_{D_5}(\bm{u} + \Delta\bm{u}) &= f_{D_5}(\bm{u}_j) + 
        \frac{1}{2}\Delta\bm{u}^T \mathcal{H}f_{D_5}\Delta\bm{u}
         + 4 u_j \Delta u^3+ \Delta u \Delta v^2  +\Delta u^4\,,
    \end{align}
    with the displacement $\Delta \bm{u}=(\Delta u, \Delta v)$ and the Hessian operator $\mathcal{H}$.
    Upon expanding the cubic term, we obtain the expansion 
    \begin{align}
        \Psi_{D_4^-}^{(j)} &\sim  e^{i k f_{D_4^-}(\bm{u}_j)} 
        \sum_{m=0}^\infty \frac{(-1)^m}{(2m)!k^{m}}
        \int_{\mathbb{R}^2} 
        e^{\frac{i}{2} \Delta\bm{u}^T \mathcal{H}f_{D_4^-}(\bm{u}_j)\Delta\bm{u}}
        \left[\Delta u^3 - 3 \Delta u \Delta v^2\right]^{2m}
        \mathrm{d}\Delta\bm{u}\,,\\
        \Psi_{D_4^+}^{(j)} &\sim  e^{i k f_{D_4^+}(\bm{u}_j)} 
        \sum_{m=0}^\infty \frac{(-1)^m}{(2m)!k^{m}}
        \int_{\mathbb{R}^2} 
        e^{\frac{i}{2} \Delta\bm{u}^T \mathcal{H}f_{D_4^+}(\bm{u}_j)\Delta\bm{u}}
        \left[\Delta u^3 + \Delta v^3\right]^{2m}
        \mathrm{d}\Delta\bm{u}\,,\\
        \Psi_{D_5}^{(j)} &\sim  e^{ i k f_{D_5}(\bm{u}_j)} 
        \sum_{m=0}^\infty \frac{(ik)^m}{m!}
        \int_{\mathbb{R}^2} 
        e^{\frac{i}{2} \Delta\bm{u}^T \mathcal{H}f_{D_5}(\bm{u}_j)\Delta\bm{u}}
        \left[
            \frac{ 4 u_j \Delta u^3 + \Delta u \Delta v^2}{k^{3/2}} +\frac{\Delta u^4}{k^2}
        \right]^{m}
        \mathrm{d}\Delta\bm{u}\,.
    \end{align}
    We write the series as $\Psi_{D_n}^{(j)} \sim e^{i k f(u_j, v_j)} \sum_{m=0}^\infty T_{D_n, m}^{(j)}/k^m$, where the coefficients $T_{D_n, m}^{(j)}$ follow from the Gaussian integral \eqref{eq:gauss}.

\end{document}